\documentclass[article]{aa}  

\usepackage{graphicx}
\usepackage{txfonts}

\usepackage{xfrac, amsmath, bm, multirow}
\usepackage[dvipsnames, table]{xcolor}

\newcommand{\figpath}{}

\usepackage{multicol}
\usepackage{siunitx}
\usepackage{empheq}

\graphicspath{{figures/}}
\usepackage{hyperref}
\hypersetup{colorlinks, linkcolor={blue},citecolor={blue},urlcolor={magenta}}  

\usepackage[labelfont=bf]{caption, subcaption}
\usepackage[compact]{titlesec}
 
 \allowdisplaybreaks

 \definecolor{colortbd}{rgb}{0.44, 0.16, 0.39}
 
\begin{document} 

\title{Performance analysis of extragalactic classifications \\in \textit{Gaia} Data Release 4}


\author{
	Sara Jamal \inst{1} 
  	\and
	Coryn A. L.\ Bailer-Jones \inst{1}
    \and
	Ruth Carballo \inst{2}
    \and
	Orlagh L. Creevey  \inst{3} 
  }

\institute{
	Max-Planck Institute f\"ur Astronomy,
	K\"onigstuhl 17, 69117 Heidelberg, Germany
	\and
	Dpto. de Matemática Aplicada y Ciencias de la Computación, Univ. de Cantabria, 
	ETS Ingenieros de Caminos, Canales y Puertos, Avda. de los Castros s/n, 39005 Santander, Spain
	\and
	Université Côte d’Azur, Observatoire de la Côte d’Azur, CNRS, Laboratoire Lagrange, 
	Bd de l’Observatoire, CS 34229, 06304 Nice Cedex 4, France\\
	}

\date{Received xxx. Accepted xxx}

\abstract 
{
The Discrete Source Classifier (DSC) provides probabilistic classifications of sources in \textit{Gaia} Data Release 4 (GDR4) based on empirically-trained Bayesian classifiers. 
Using the \textit{Gaia} astrometry, photometry, and low-resolution spectra (`XP'), DSC classifies all sources into one of three classes: quasar, galaxy, or star. 
DSC comprises three trained neural networks and three different combinations of their probabilities. 
When evaluated as a function of brightness and sky position on a test set that excludes the Magellanic Clouds, the DSC purity in GDR4 has improved for a small loss in completeness. 
The average performance of the best classifiers at magnitudes brighter than $G$=20 is at least 88\% completeness and 96\% purity for the extragalactic classes, namely the quasar and galaxy classes.
At fainter magnitudes, the performance is lower due to the increased noise.
The average performance at magnitudes of 20$\leq$$G$<20.5 is a minimum of 55\% completeness and 71\% purity for the extragalactic classes. At $G$>20.5\,mag, the completeness is considerably reduced, primarily for the models that depend on the XP spectra.
Furthermore, we train additional models on the \textit{Gaia} optical data together with mid-infrared photometry from the CatWISE2020 catalogue. 
Inclusion of infrared photometry increases the completeness of extragalactic samples at $G$>20\,mag between 9 and 29 percentage points, at the cost of reducing the purity between 1 and 9 percentage points.
In the GDR4, the best DSC-combined classifier prioritising completeness identifies three million quasars and two million galaxies, but with expected high contamination among fainter sources.
In contrast, the combined classifiers prioritising purity identify approximately two million quasars and 1.3 million galaxies with an expected lower level of contamination.
Finally, we provide recommendations for enhancing the purity of the DSC extragalactic selection by applying quality cuts to the \textit{Gaia} photometry and astrometry.
}

\keywords{surveys/galaxy: general / stars: general / quasars: general / methods: data analysis/methods: statistical}
\maketitle
\nolinenumbers

\section{Introduction}\label{sec:1_introduction}

Over the last few decades, numerous ground- and space-based surveys have compiled extensive catalogues of extragalactic sources, including quasars and galaxies, such as 
the Sloan Digital Sky Survey Quasar catalogue (SDSS DR16Q; \citealt{york_sloan_2000, lyke_sloan_2020}), 
the Large Quasar Astrometric Catalogue (LQAC-6, \citealt{souchay_lqac-6_2024}), 
the Wide-field Infrared Survey Explorer (WISE) Active Galactic Nuclei (AGN) catalogues (R90 and C75; \citealt{assef_wise_2018}),  
the 2dF QSO Redshift Survey (2QZ, \citealt{croom_2df_2004}),  
the AllBRICQS Survey \citep{onken_allbricqs_2023},  
the Milliquas Catalogue \citep{flesch_million_2023},  
and the \textit{Gaia} Data Release 3 extragalactic candidates \citep{bailer-jones_gaia_2023}. 
The extragalactic sky will be mapped in even greater detail, reaching fainter magnitudes, thanks to various other surveys, such as 
	the Vera C. Rubin Observatory Legacy Survey of Space and Time (LSST; \citealt{ivezic_lsst_2016,ivezic_lsst_2019}), 
	the Dark Energy Spectroscopic Instrument survey (DESI; 
	\citealt{desi_collaboration_desi_2025-1,desi_collaboration_desi_2025,desi_collaboration_validation_2025}), 
	the \textit{Euclid} ESA mission \citep{mellier_euclid_2025}, 
	as well as, the \textit{Gaia} ESA mission \citep{vallenari_gaia_2023}.
    
A quasar -- a quasi-stellar object (QSO) -- is an extremely luminous object driven by an accreting supermassive black hole (SMBH) located at the centre of its host galaxy.
The identification of quasars and galaxies is essential to 
	constrain galaxy evolutionary models \citep{hopkins_unified_2006},
	characterise the growth of SMBHs \citep{di_matteo_energy_2005,hopkins_black_2005,volonteri_quasars_2006}, 
	and trace the evolution history of structures in the early Universe and the epoch of reionisation \citep{fan_constraining_2006, mortlock_luminous_2011, banados_800-million-solar-mass_2018}, to mention a few examples.
Furthermore, cosmological probes with detectable signatures in the distribution of quasars and galaxies $-$ such as distortions in the redshift space (RSDs) and baryon acoustic oscillations (BAOs) -- are used to investigate the growth history of large-scale structures and the role of dark energy in the late-time acceleration of the expansion of the Universe. 
RSDs emerge as an anisotropic clustering in the spatial distribution of galaxies caused by the expansion of the Universe and the peculiar velocities of galaxies. 
Their observation provides a direct measurement of the linear growth rate of large-scale structures \citep{beutler_6df_2012,zarrouk_clustering_2018}. 
Conversely, the imprint from the BAOs on the late-time clustering of galaxies in the early universe is used to measure the distance scale and trace the expansion history, through measurements of the angular comoving distances and the Hubble parameter as functions of the redshift \citep{eisenstein_baryonic_1998,seo_probing_2003,linder_baryon_2003,blake_cosmology_2005,eisenstein_detection_2005,dolney_baryon_2006, beutler_6df_2011}. 
Likewise, the Ly$\alpha$ forest in quasars' spectra traces the matter distribution and clustering, and serves as a powerful probe to infer the expansion history of the Universe \citep{busca_baryon_2013,slosar_measurement_2013,font-ribera_quasar-lyman_2014,du_mas_des_bourboux_completed_2020,bautista_measurement_2017}.
Therefore, the construction of high-purity quasar and galaxy catalogues lies at the core of several cosmological studies.

Primarily designed to map stars in the Milky Way, the \textit{Gaia} ESA satellite has observed more than two billion sources across the entire sky down to very faint magnitudes, including millions of quasars and galaxies \citep{bailer-jones_gaia_2023}.
Several modules within the \textit{Gaia} Data Processing and Analysis Consortium (DPAC) perform classification using machine learning (ML) techniques. 
One of these modules is the Discrete Source Classifier (DSC), as part of the Coordination Unit CU8 for the Apsis (Astrophysical ParameterS Inference System) chain \citep{bailer-jones_gaia_2013, andrae_gaia_2018, creevey_gaia_2023-1, fouesneau_gaia_2023, delchambre_gaia_2023}.
DSC uses supervised classification on \textit{Gaia} photometry, astrometry, and low-resolution spectroscopy to identify a source as a quasar, a galaxy, or a star.
DSC computes posterior probabilities across the target classes and uses a prior function that represents our expectation of the overall distribution of classes in the sky \citep{delchambre_gaia_2023}.
In \textit{Gaia} Data Release 3 (GDR3), DSC achieved high completeness of 92\% but low purity of 22\% for the quasar and galaxy classes. 
For the upcoming \textit{Gaia} DR4 (GDR4), we aim to improve the classifications and produce quasar and galaxy catalogues for DSC of a higher purity than was achieved in GDR3.

In literature, sources are classified using information from spectra, photometry and astrometry to determine their type.
Quasars' spectra typically exhibit a flat continuum alongside broad emission lines such as H$\alpha$, Ly$\alpha$, as well as narrow emissions due to ionisation [OII] and [OIII] \citep{alexander_desi_2023}.
Spectral features of quasars can reveal evidence of 
	strong star-formation activity \citep{lacy_large_2007,zakamska_mid-infrared_2008},
	outflows of ionised gas \citep{villar-martin_ionized_2016},  
	as well as streams or tidal effects resulting from interactions and major mergers \citep{bessiere_importance_2012,pierce_galaxy_2023}.
Quasar signatures can be detected across multiple wavelengths, including ultraviolet (UV), optical and infrared (IR). However, these key signatures in the spectrum shift towards longer wavelengths with increasing redshift. This result means that very distant quasars at redshifts greater than $z$$\approx$6.5 can only be detected via their Ly$\alpha$ line in the infrared or radio spectrum \citep{mortlock_probabilistic_2012}.
Galaxies' spectra are dominated by a strong continuum component superimposed on absorption lines characteristic of the stellar population, as well as emission lines from ionised gas that indicate whether a galaxy is undergoing active star formation, is a host of an active galactic nucleus (AGN), or is in a quiescent phase.
Stellar spectra show a flat continuum with absorption lines informing about the star's evolutionary phase and key properties such as effective temperature, luminosity, and element abundances. 
In the colour space, the different classes occupy distinct regions. 
For instance, quasars appear bluer than stars or galaxies, and stars tend to be concentrated along a distinct locus \citep{richards_colors_2001,richards_spectroscopic_2002}.

Historical methods for the quasar-galaxy-star classification have relied on colour cuts in UV, optical or IR photometry 
\citep{newberg_three-dimensional_1997,fan_simulation_1999,richards_spectroscopic_2002,lacy_obscured_2004,trammell_uv_2007,hutchings_catalog_2010,wu_quasar_2010,stern_mid-infrared_2012,wang_survey_2016}.  
Nevertheless, these approaches are limited by the low dimensionality of the data and the arbitrary selection of thresholds applied to a small sample of sources. 
An improved classification requires exploring large, highly dimensional data sets.
To this end, ML methods have been recognised as powerful tools for processing large data sets and detecting patterns and outliers.
ML has been applied extensively to the quasar-galaxy-star classification using various methods, such as 
	tree-based estimators \citep{weir_automated_1995,carrasco_photometric_2015,jin_efficient_2019,schindler_extremely_2017,bai_machine_2018,nakazono_discovery_2021,zhang_classification_2021,golob_classifying_2021,li_identification_2021,baqui_minijpas_2021,hughes_quasar_2022,rimoldini_gaia_2023,stoppa_autosourceid-classifier_2023,delchambre_gaia_2023,cunha_identifying_2024,fu_catnorth_2024,fu_catsouth_2025},
	Bayesian kernel density estimators \citep{richards_efficient_2004,kirkpatrick_simple_2011,peters_quasar_2015,richards_bayesian_2015,bailer-jones_quasar_2019},
	Support Vector Machines 
		(SVMs; e.g. \citealt{gao_support_2008,peng_selecting_2012,malek_vimos_2013,wang_survey_2016,nakazono_discovery_2021}),
	and neural networks 
		(NNs; \citealt{yeche_artificial_2010,tuccillo_neural-network_2015,kim_stargalaxy_2017,burke_deblending_2019,he_deep_2021,guo_unsupervised_2022,martinez-solaeche_minijpas_2023,stoppa_autosourceid-classifier_2023,merz_detection_2023,rodrigues_minijpas_2023,chaini_photometric_2023,cunha_identifying_2024}).
Several studies have used a combination of multi-wavelength observations to improve quasar selection \citep{wu_quasar_2010,bovy_photometric_2012,dipompeo_quasar_2015,carrasco_photometric_2015,richards_bayesian_2015,cunha_identifying_2024,fu_catnorth_2024,fu_catsouth_2025}.

In GDR3, DSC has used a combination of a tree-based algorithm, called Extratrees, and a density-based method, called Gaussian Mixture Models, to classify sources as quasars, galaxies, single stars, white dwarfs, or binaries based on \textit{Gaia} astrometry, photometry, and low-resolution spectroscopy.
For the GDR4, we have explored various classification algorithms and refined the construction of the labelled data sets used to train the models. For DSC, we now use a combination of neural network models, known as multi-layer perceptrons, which have proven highly efficient at rejecting stellar false detections, albeit at a reduced completeness.
Moreover, DSC in GDR4 provides independent classifications combining \textit{Gaia} optical data with mid-infrared photometry from the CatWISE2020 catalogue \citep{marocco_catwise2020_2021}.  
Throughout this paper, results obtained using the \textit{Gaia} data are referred to as the ``\textit{Gaia} only" mode, whereas those obtained using the \textit{Gaia} data alongside the CatWISE infrared photometry are identified as the ``\textit{Gaia}-IR" mode.
This paper is organised as follows. 
Section 2 provides a brief description of the data used for classification.
Section 3 describes the methodology, algorithms, and architecture.
Section 4 evaluates the classification performance in terms of completeness and purity for a labelled test set relative to the sources' sky positions and brightnesses. 
We present the full range of extragalactic classifications from DSC in GDR4 and discuss the physical properties of quasar and galaxy candidates. 
We provide recommendations of quality cuts to the \textit{Gaia} photometry and astrometry to improve the purity of the DSC extragalactic sample.
We summarise our key results in Section 5.

\begin{figure*}[htp!]
	\centering	
	\includegraphics[scale=0.27]{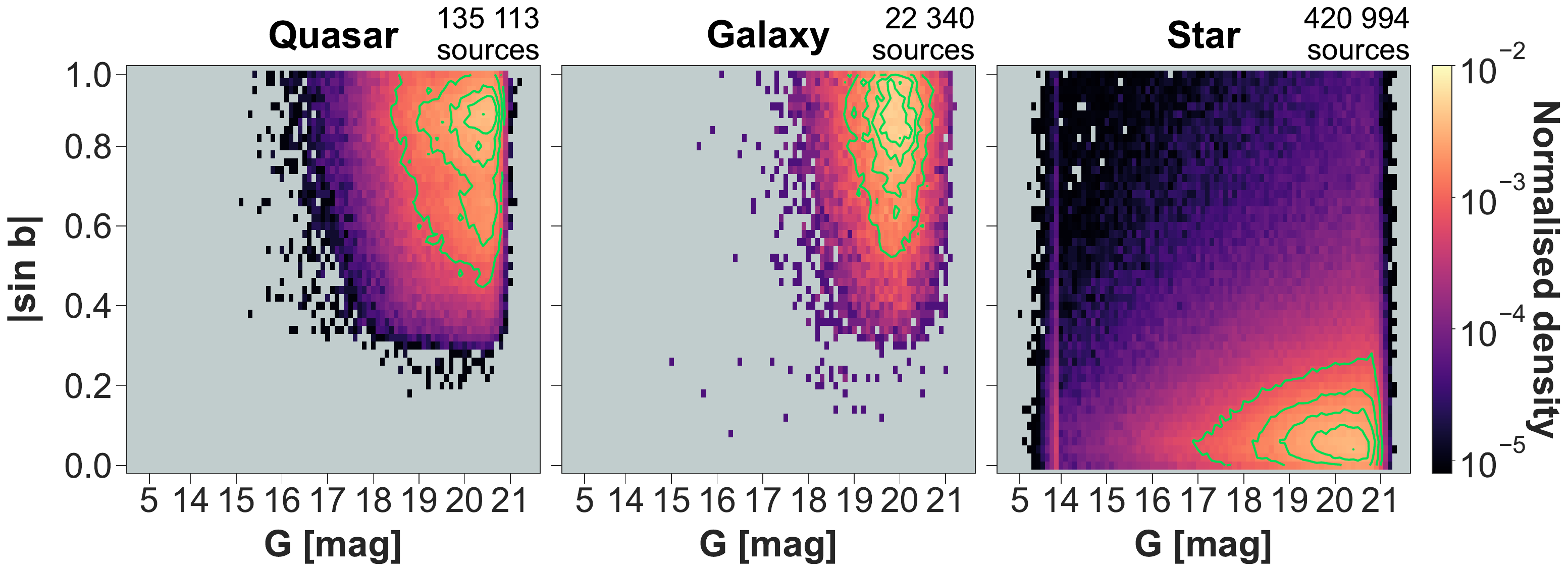}
	
	\caption{Representation in 2D of the distribution of sources as a function of Galactic latitude 
	and brightness of the training data set.  Each distribution is normalised by the total number of sources in that panel. 
	The colour scale is set such that brighter colours refer to regions of higher density compared to darker colours, 
	which have fewer sources.
	The 2D representation of the density of sources is defined on a 50$\times$91 bin grid in $|\sin b|$ and $G$. 
	Binning into magnitude bins compresses the magnitude range at $G$<14. 
	Contours (green lines) show the normalised density on a log scale of the highest-density regions.
	}
	\label{fig:2d_alldata_testset}
\end{figure*}

\section{{Data}} \label{sec:2_gaiadata}
From its launch in December 2013 to its decommissioning in March 2025, the \textit{Gaia} ESA satellite has observed over two billion sources across the entire sky. 
Examples of the scientific results stemming from \textit{Gaia} include detailed, multidimensional maps of the Milky Way \citep{vallenari_gaia_2023}
and the identification of new black holes \citep{el-badry_red_2023,panuzzo_discovery_2024}.

The upcoming \textit{Gaia} Data Release 4 (GDR4) is based on 66 months of data collected by the satellite between 2014 and 2020. 
The previous \textit{Gaia} Data Release 3 (GDR3; \citeauthor{prusti_gaia_2016}, \citeyear{prusti_gaia_2016}, \citeauthor{vallenari_gaia_2023}, \citeyear{vallenari_gaia_2023}) published data for approximately 1.9 billion sources down to ${G}$$\simeq$20~mag, while the GDR4 will publish all the processed objects, approximately 2.8 billion sources down to slightly fainter magnitudes. 
The new catalogue will contain astrometry, photometry, low-resolution BP/RP (XP) spectroscopy, and classifications, including DSC results. 
A summary of the expected content of GDR4 is available via the public ESA link\footnote{\url{https://www.cosmos.esa.int/web/gaia/dr4}}.
In the GDR4, DSC provides posterior probabilities and class labels in the main table (\texttt{gaia\_source}), the astrophysical parameter tables (\texttt{ap\_class} and \texttt{ap\_class\_ir}), as well as the extragalactic tables (\texttt{qso\_candidates} and \texttt{galaxy\_candidates}).
The \texttt{ap\_class} reports the classifications using the \textit{Gaia} data, while the \texttt{ap\_class\_ir} table provides classifications using a combination of \textit{Gaia} data and mid-infrared photometry from CatWISE.

The CatWISE2020 catalogue \citep{marocco_catwise2020_2021} has observed over 1.8 billion sources across the entire sky selected from the Wide-field Infrared Survey Explorer (WISE) and the NEOWISE survey data collected between 2010 January 7 and 2018 December 13.
The catalogue provides an extensive list of infrared sources with an improved magnitude detection limit in the $W1$ (3.4 $\mu$m) and $W2$ (4.6 $\mu$m) infrared passbands, reaching 90\% completeness depth at $W1$ = 17.7\ mag and $W2$ = 17.5\ mag.
We combine the mid-infrared photometry bands $W1$ and $W2$ of CatWISE2020 with the optical data from \textit{Gaia} to classify a source in the \textit{Gaia}-IR mode.

\section{Classification methodology} \label{sec:3_DSCmethodology}

	\subsection{An outline of DSC in GDR4}\label{subsec:DSC}
    
DSC is composed of six classifiers that compute posterior class probabilities: three baseline classifiers trained using a set of labelled sources, \texttt{Allosmod}, \texttt{Specmod}, and \texttt{Specmod-Supp}, complemented with three combined classifiers, \texttt{Combmod}, \texttt{Combmod-$\alpha$}, and \texttt{Combmod-Supp}. 
\texttt{Specmod} is trained exclusively on the low-resolution XP spectra,
\texttt{Allosmod} is trained on various astrometric and photometric features, not the XP spectra,
and \texttt{Specmod-Supp} is trained on both the XP spectra and various astrometric and photometric features.
In \textit{Gaia}-IR mode, the CatWISE mid-infrared photometry is supplemented to train the auxiliary classifiers \texttt{Allosmod(*)} and \texttt{Specmod-Supp(*)}\footnote{The symbol \textit{(*)} indicates the models using the \textit{Gaia} data alongside infrared photometry from CatWISE.}.
The list of features is summarised in Appendix \ref{sec:list_features_classifier}.
The combined classifiers are defined as follows.
\texttt{Combmod} combines the \texttt{Allosmod} and \texttt{Specmod} posterior probabilities as a weighted product, as was done in GDR3 \citep{LL:CBJ-094, ulla_gaia_2022}.
\texttt{Combmod-$\alpha$} is a parametric additive combiner that weights the contributions of \texttt{Allosmod} and \texttt{Specmod} probabilities. The optimal parameter set $\boldsymbol{\alpha}$ maximises the classification score\footnote{The score is a metric that measures the classifier's performance between 0 and 1, where 1 indicates an ideal result.} of the quasar and galaxy classes \citep{jamal_improved_2024}.
 \texttt{Combmod-Supp} selects the probabilities of \texttt{Specmod-Supp} when available, otherwise \texttt{Specmod}.
 \texttt{Combmod-Supp} is equal to \texttt{Specmod-Supp} if all inputs are available; otherwise, it is equal to \texttt{Specmod}.

Our approach of using multiple classifiers allows for greater flexibility when processing sources with missing input data. 
For example, sources lacking parallaxes and proper motions, but with valid BP and RP spectra, are classified using \texttt{Specmod}, and not processed using \texttt{Allosmod} nor \texttt{Specmod-Supp}.
We have also decided to withdraw the white dwarf and binary star classes from DSC due to their poor classification performance in GDR3 \citep{delchambre_gaia_2023}. 
Consequently, we retain a supervised classification scheme that identifies a source as a quasar, a galaxy, or a star.
In GDR3, we used a decision tree-based classifier, Extratrees \citep{geurts_extremely_2006}, for Specmod, and Gaussian Mixture Models \citep{reynolds_gaussian_2015} for Allosmod.
In GDR4, all three DSC trained models are neural networks.
To compute posterior probabilities, DSC uses a global prior that accounts for the rarity of the quasar and the galaxy classes compared to the larger star class.
The prior probabilities of any source being a quasar or a galaxy are 1/1000 and 1/5000, respectively, corresponding to the same values used in GDR3.

Our reference data set, with labelled sources used for training and testing, is built from reliable external catalogues and data releases. 
The quasar and galaxy classes correspond to a sample of spectroscopically confirmed sources from the SDSS (Sloan Digital Sky Survey) DR17 catalogue \citep{Abdurro'uf2021}.
The crossmatch to the SDSS catalogue within a one arcsecond radius yields a large number of sources (\num{563233} quasars and \num{1403392} galaxies). 
Our star class corresponds to a random sample of \num{1097000} sources from \textit{Gaia}, excluding the extragalactic selection.
We reduce this selection as we require valid XP spectra and all features to be available for training and testing (no missing data). We further clean the sample by requiring at least five transits to contribute to the mean BP and RP spectra.
Finally, we remove quasar and galaxy candidates found in the following non-\textit{Gaia} catalogues from our star class:
Milliquas v7.2-v8 \citep{flesch_half_2015, flesch_million_2023}, 
LAMOST DR7 \citep{luo_vizier_2022}, 
SDSS DR17 photometry \citep{Abdurro'uf2021}, 
 the R90 and C75 WISE AGN catalogues \citep{assef_wise_2018}
as well as a private communication of the \textit{Gaia} DPAC CU4-EO (Extended Objects) list of candidates.
The final reference set contains \num{270436} quasars, \num{44894} galaxies and \num{857977} stars.

The selected set is randomly divided into equal-sized training and testing subsets.
When assessing performance, the test set excludes the Magellanic Clouds. These regions have a higher level of stellar contamination, which masks the full range of extragalactic classifications across the entire sky. 
We remove sources within 9$^{\circ}$ and 6$^{\circ}$ radii of the centres of the LMC and the SMC, respectively (see Appendix B, \citeauthor{bailer-jones_gaia_2023} \citeyear{bailer-jones_gaia_2023}). 
Because the SDSS footprint does not cover the LMC and SMC regions, only our star class includes sources in those regions. 
We exclude these regions from the assessment of the performance on the test set, not from the training data. 
Figure \ref{fig:2d_alldata_testset} displays the latitude and magnitude distribution of the training set.
In our reference set, the relative density of the quasars and galaxies peaks at fainter magnitudes at higher latitudes, while the density of stars peaks closer to the Galactic plane and at very bright magnitudes $G$>14. 
Since the SDSS footprint shows a disparity between the northern and southern Galactic hemispheres, we randomise the Galactic latitudes $b$ by drawing from a uniform distribution over $|\sin b |$ during the training only, to emulate a uniform distribution of quasars and galaxies across the sky, and prevent the classifier from learning the SDSS footprint for these objects.
This randomisation strategy was also employed in DSC in GDR3 \citep{delchambre_gaia_2023}.
We do not use longitudes because the impact of sky position on classification is already encoded in Galactic latitudes.

	\subsection{Methods}\label{subsec:methods}
This section outlines the methods and the training strategies employed by DSC.

		\subsubsection{Multi-layer perceptron model}
Neural networks (NNs) are models in artificial intelligence that use specialised blocks to process input data and propagate information to produce an output.
In its simplest form, the multilayer perceptron consists of an input layer, an output layer, and fully connected hidden layers \citep{rumelhart_learning_1986}.
The advantage of NNs lies in their flexible design and the selection of an objective function. 
However, designing an optimal model remains an empirical process that involves a grid search over hyperparameters, such as the number of layers, activation functions, and loss functions.
For DSC, we train multiple MLP configurations and select the models that perform best on the test set (i.e., the data set never seen during training).
The training set is split 80:20 for training and validation. 
At each epoch, a stopping criterion is set to monitor the validation score, a metric that measures the classifier's performance between 0 and 1 (where 1 indicates an ideal result).
The MLP architectures and the list of hyperparameters for DSC models are provided in Appendix \ref{sec:mlp_dsc_summary}.
A comparison of the different algorithms and classification strategies explored for DSC is available in a separate technical note (\citeauthor{jamal_dsc_tn_2026} in prep.).

		\subsubsection{Combined classifiers} \label{subsubsec:combined_classifiers_definition}
The combined classifiers exploit the individual results of the three baseline classifiers, \texttt{Allosmod}, \texttt{Specmod}, and \texttt{Specmod-Supp}. 
As in GDR3, \texttt{Combmod} probabilities are computed from a product of the \texttt{Specmod} and \texttt{Allosmod} posterior probabilities, assuming that the two classifiers use the same prior and are independent (Eq. \ref{eq:eq_dsc_combmod}). We adapt the GDR3 definition to account for three classes. 
The parametric combiner \texttt{Combmod-$\alpha$}, introduced in \cite{jamal_improved_2024}, is computed from a weighted addition of the \texttt{Specmod} and \texttt{Allosmod} probabilities, such that the combination maximises the classification score (Eq. \ref{eq:eq_dsc_combmodalpha}). 
The probabilities of the combined classifiers are normalised across all classes.
\begin{subequations}
\begin{equation}
	P_{k}^{c} = P_{k}^{s} P_{k}^{a} \dfrac{1}{\pi_k}, 
	\label{eq:eq_dsc_combmod}
\end{equation}
\vspace{-0.6cm}
\begin{equation}
	P_{k}^{c_{\alpha}} = (1- \alpha_k) P_{k}^{s} + \alpha_k P_{k}^{a},
	\label{eq:eq_dsc_combmodalpha}
\end{equation}
\end{subequations}
where $k$ refers to the class,
$\pi_k$ the class prior,
$\alpha_k$ the parameter of the class $k$,
and ($P_k^c$, $P_{k}^{c_{\alpha}}$, $P_k^s$, $P_k^a$) the posterior probabilities for \texttt{Combmod}, \texttt{Combmod-$\alpha$}, \texttt{Specmod} and {\texttt{Allosmod}}, respectively.

For the parametric combiner, we perform a grid search within a limited discrete space over the parameters $\alpha_k$$\in$[0,1] and select the optimal set that achieves the highest geometric score (Eq. \ref{eq:geoscore}). 
For the quasar, galaxy and star classes, we find that the optimal result is given by the values $\boldsymbol{\alpha}$=\{0.9,0.8,0.3\}. 
These values imply a greater contribution of the \texttt{Allosmod} probabilities in defining the \texttt{Combmod-$\alpha$} probabilities.
In the case of missing data, \texttt{Combmod} and \texttt{Combmod}-$\alpha$ are equal to the probabilities of \texttt{Specmod}. The \texttt{Combmod-Supp} probabilities are set to the probabilities of \texttt{Specmod-Supp} when available; otherwise, they are set to the probabilities of \texttt{Specmod}.

\begin{table*}[htp!]
    \caption{Summary of the overall classification performance evaluated on the test data set for models trained on \textit{Gaia} data only. }               
    \label{table:tab_overall_performances_testset}  
    \centering \fontsize{8.5}{9}\selectfont 
    \renewcommand{\arraystretch}{1.1}
    \setlength{\tabcolsep}{3pt}  
     \begin{tabular}{ll *{2}{wl{5em}} c *{2}{wl{5em}} l} 
        \hline \hline 
        \multirow{3}{7em}{Magnitude bins}  
        	& \multirow{3}{6em}{Classifiers} 
    	& \multicolumn{2}{c}{\multirow{2}{8em}{\centering Overall\\ completeness}} 
    	&& \multicolumn{2}{c}{\multirow{2}{8em}{\centering Overall\\ purity}} 
	& \multicolumn{1}{c}{\multirow{3}{4em}{\centering Geometric score}} \\
   	&&&&&&&\\
         \cline{3-4} \cline{6-7} 
        & & quasar & galaxy && quasar & galaxy & \\
        \hline \hline 
        \multirow{5}{9em}{\centering All magnitudes\\ (\textit{q}:\num{135323}, \textit{g}:\num{22447})} 
    	 & \texttt{Specmod} 
	 	& \cellcolor{gray!40}0.702 $\rm^{\color{blue!70}(+29\,pp)}$ & \cellcolor{gray!40}0.722 $\rm^{\color{red!90}(-13\,pp)}$ 
		&&\cellcolor{gray!40}0.691 $\rm^{\color{blue!70}(+25\,pp)}$ & \cellcolor{gray!40}0.618 $\rm^{\color{blue!70}(+26\,pp)}$ 
		& \cellcolor{gray!40}0.682 $\rm^{\color{blue!70}(+19\,pp)}$ \\ 	 
        	 & \texttt{Allosmod} 
	 	& 0.869 $\rm^{\color{blue!70}(+3\,pp)}$ & 0.773 $\rm^{\color{red!90}(-18\,pp)}$ 
		&& \cellcolor{gray!40}0.737 $\rm^{\color{blue!70}(+7\,pp)}$ & \cellcolor{gray!40}0.703 $\rm^{\color{blue!70}(+27\,pp)}$ 
		& 0.768 $\rm^{\color{blue!70}(+7\,pp)}$ \\ 	 
        	 & \texttt{Specmod-Supp} 
	 	& 0.862 & \cellcolor{gray!40}0.723 
	 	&& \cellcolor{gray!40}0.749 & 0.778 
		& 0.776 \\	 	
    	 & \texttt{Combmod} 
	 	& 0.972 $\rm^{\color{blue!70}(+5\,pp)}$ & 0.941 $\rm^{\color{red!90}(-2\,pp)}$ 
	 	&& \cellcolor{gray!40}0.608 $\rm^{\color{blue!70}(+20\,pp)}$ & \cellcolor{red!20}0.373 $\rm^{\color{blue!70}(+20\,pp)}$ 
		& \cellcolor{gray!40}0.675 $\rm^{\color{blue!70}(+14\,pp)}$ \\ 	 
        	 & \texttt{Combmod-$\alpha$} 
	 	& 0.843 $\rm^{\color{blue!70}(+2\,pp)}$ & 0.762 $\rm^{\color{red!90}(-17\,pp)}$ 
		&& 0.762 $\rm^{\color{blue!70}(+6\,pp)}$ & \cellcolor{gray!40}0.741 $\rm^{\color{blue!70}(+23\,pp)}$ 
		& 0.776 $\rm^{\color{blue!70}(+5\,pp)}$ \\ 
 	\hline	
    \end{tabular}
    \tablefoot{
   	 The number of quasars (\textit{q}) and galaxies (\textit{g}) in the test set is indicated. 
    	The performance of the \texttt{Combmod-Supp} classifier is equivalent to \texttt{Specmod-Supp} in the test set, given the availability of all required inputs.
    	Results highlighted in grey, light red, and dark red correspond to values below 0.75, 0.50, and 0.25, respectively.
    	The numbers in pp (percentage points) show the change in performance relative to that achieved in GDR3 as evaluated on the same test set. 
	The DSC stellar predictions from GDR3 are merged into one star class for this purpose.
    }
 \end{table*}
 
 \begin{figure*}[htp!]
	 \centering
	 \includegraphics[scale=0.34]{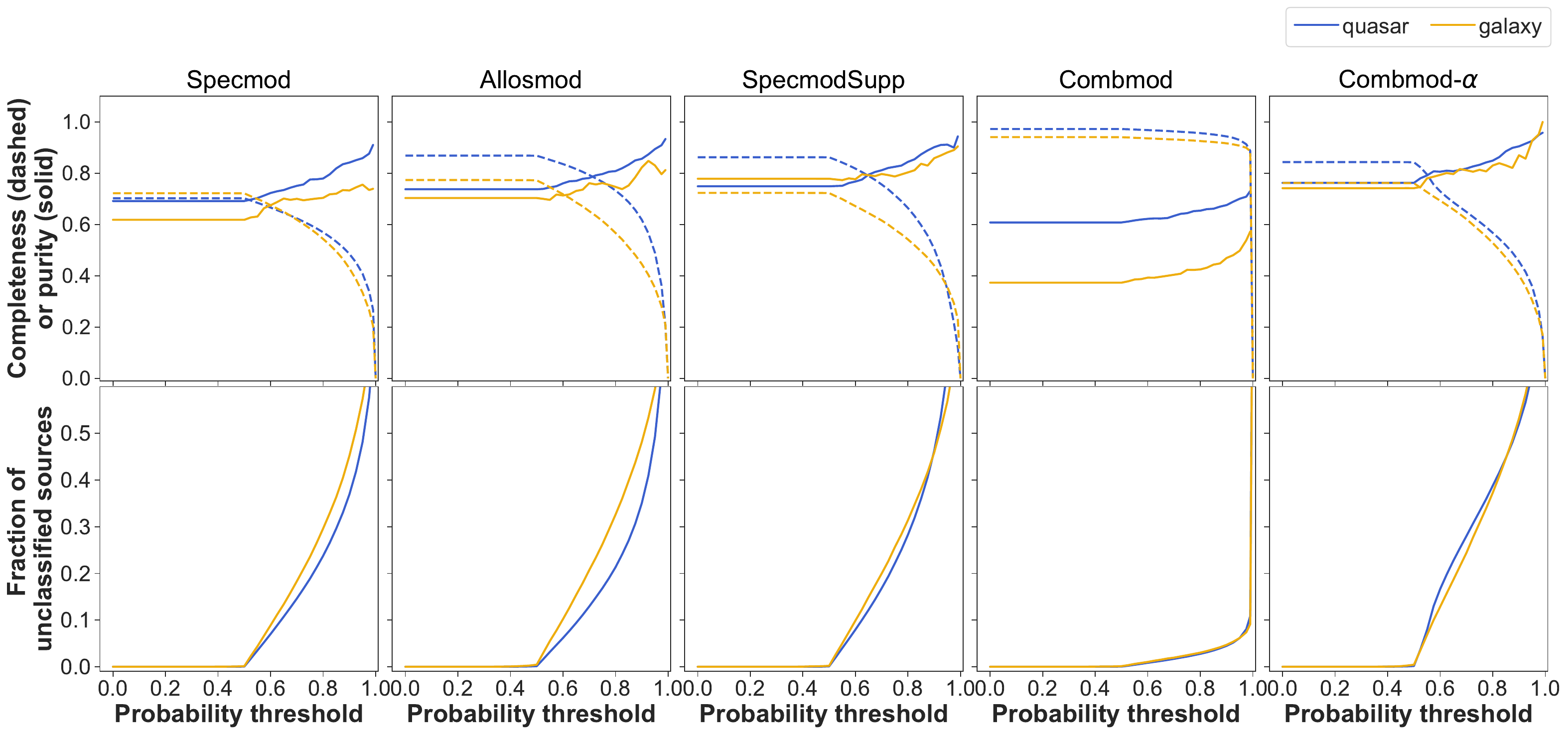}
	\caption{
		\textit{Top}: Variation in the completeness and the purity of each class in the test set as a function of the posterior probability threshold used to assign a class.
		\textit{Bottom}: Variation in the proportion of the unclassified sources with probabilities below the threshold, relative to the total number of sources in each class. 
		The $y$-axis is limited to 0.5 for visualisation purposes only.
	}
	\label{fig:perfo_metrics_testset_thresholds}
\end{figure*}

\section{{DSC performance in GDR4}} \label{sec:4_dsc_performance}

	\subsection{Evaluation on the test set} 
	
		\subsubsection{Performance metrics}	
To evaluate the performance of a classifier, we compute two metrics from the confusion matrix, namely the completeness and the purity. 
The completeness is the proportion of correct predictions in each class relative to the total number of true members of that class. 
The purity is the proportion of correct predictions in each class relative to the total number of predicted members of that class.
The geometric score is defined as the geometric mean of the completeness and the purity of the quasar and galaxy classes.
The metrics are defined as follows.
\begin{subequations}
\begin{equation}
	\rm completeness_{\,k} = \frac{TP_k}{TP_k+FN_k},  
	\qquad
	purity_{\,k} = \frac{TP_k}{TP_k+FP_k},   
	\label{eq:metrics}
\end{equation}
\begin{equation}
	\begin{split}
		\rm score_{\rm geometric} 
	 	&= \Big( \rm completeness_{\rm quasar}  \times\,  \rm completeness_{\rm galaxy}  \\
	 	&   \times\, \rm purity_{\rm quasar} \times\, \rm purity_{\rm galaxy} \Big)^{1/4},
		\end{split}	
	\label{eq:geoscore}
\end{equation}
\end{subequations}
where $k$ refers to the class, and TP, FN, and FP refer to the true positives, the false negatives, and the false positives, respectively. 
Both high completeness and high purity are desirable, but there is an unavoidable trade-off between them.
Low completeness but high purity indicates a classifier that rejects false positives (i.e., sources from other classes incorrectly classified as the target class) but overlooks true objects.
High completeness but low purity, in contrast, reflects a classifier that can identify the true classes but also introduces false detections.
In DSC, a source is assigned to the class that achieves the highest classification probability above a fixed threshold. 
If this maximum probability is below the threshold, the source is marked as unclassified. 
A high threshold rejects more false positives but reduces completeness by also rejecting true positives with probabilities below the threshold.

In classification tasks, class imbalance affects performance on minority classes because the loss function is overweighted by the dominant class during training. 
In our case, a classifier can easily overlook the rare quasar and galaxy classes in favour of the much larger star class.
We address class imbalance using an approach similar to that developed in \cite{bailer-jones_quasar_2019} and adopted in \cite{delchambre_gaia_2023, jamal_improved_2024}. First, we apply a prior function to the probabilities to reflect the expected fraction of each class in the data.
Second, we adjust the confusion matrix to align the test set with the overall class distributions expected in the \textit{Gaia} survey.
This adjustment affects only the purities, not the completeness.
Throughout this work, we report adjusted metrics. Ignoring these adjustments (as is often the case in the literature) would give the false appearance of better performance.
We do not report the performance for the star class, since its completeness and purity are always close to unity due to the dominance of stars. 
Classifying everything as a star would also produce purity and completeness close to unity.
\texttt{Combmod-Supp} is equal to \texttt{Specmod-Supp} if all inputs are available; otherwise, it is equal to \texttt{Specmod}.

\begin{figure*}[htp!]
	 \begin{subfigure}{1\textwidth} \centering
	 \includegraphics[scale=0.40]{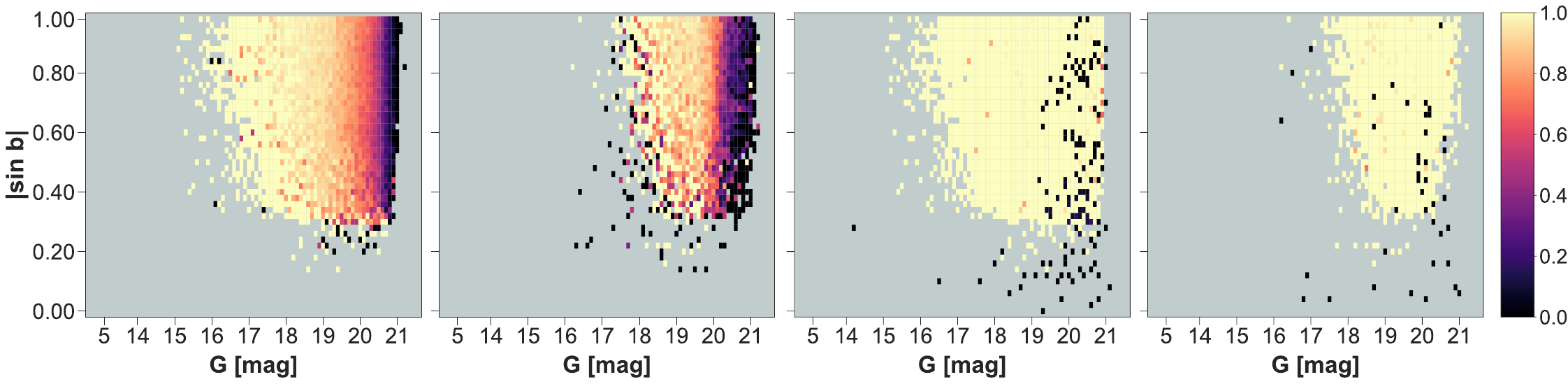}
	 \vspace{-0.15cm} \caption{}
	\label{fig:perfo_metrics_testset_specmod}
    	\end{subfigure}
	
    	 \begin{subfigure}{1\textwidth} \centering
	 \includegraphics[scale=0.40]{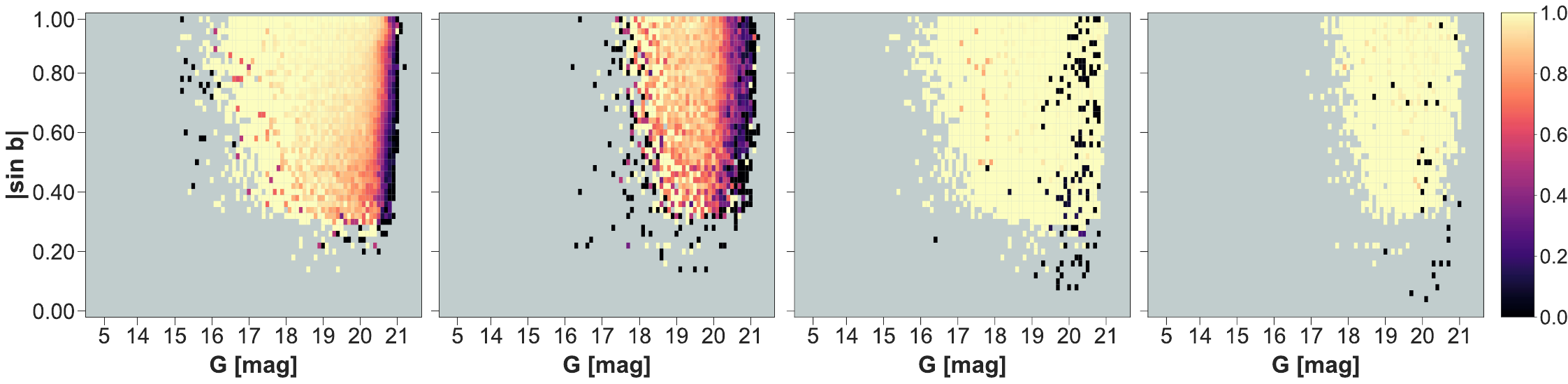}
	\vspace{-0.15cm}  \caption{}
	\label{fig:perfo_metrics_testset_allosmod}
    	\end{subfigure}
	
	\begin{subfigure}{1\textwidth}
         \centering
	\includegraphics[scale=0.40]{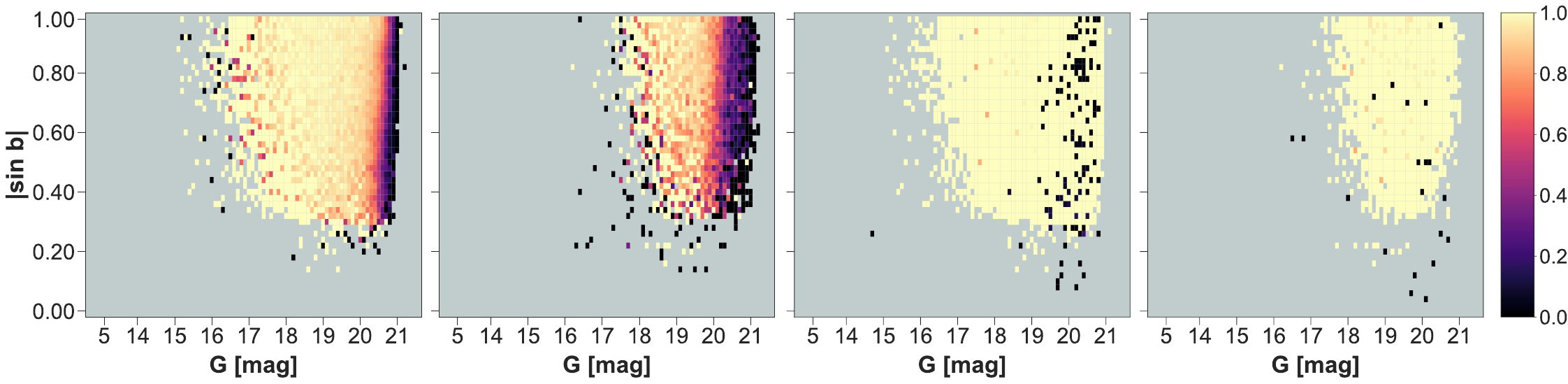}
	\vspace{-0.15cm} \caption{}
	\label{fig:perfo_metrics_testset_specmodsupp}
    	\end{subfigure}
	
	\begin{subfigure}{1\textwidth}
         \centering
	 \includegraphics[scale=0.40]{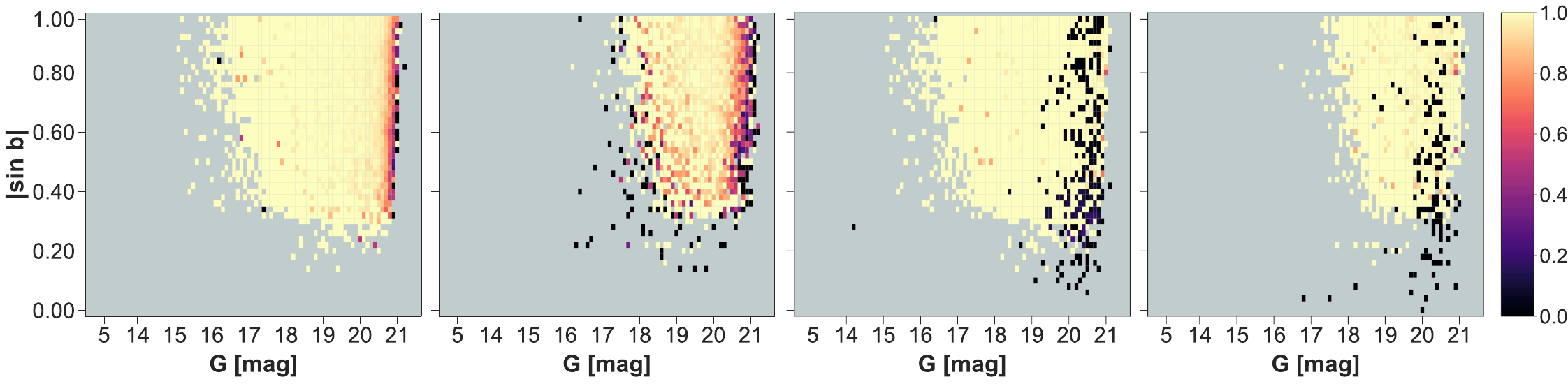}
	 \vspace{-0.15cm}\caption{}
	\label{fig:perfo_metrics_testset_combmod}
    	\end{subfigure}
    
	 \begin{subfigure}{1\textwidth} \centering
	 \includegraphics[scale=0.40]{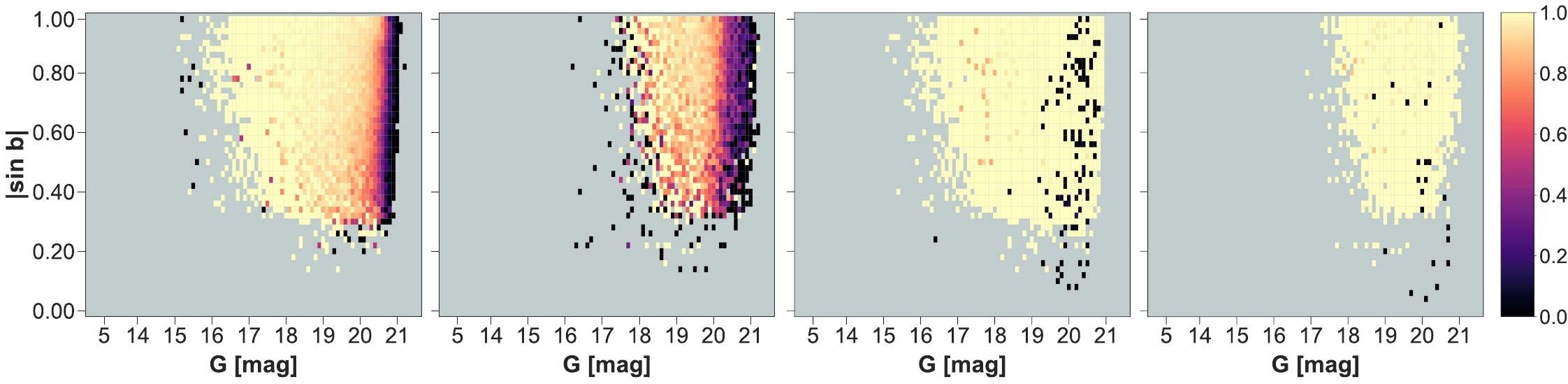}
	\vspace{-0.15cm} \caption{}
	\label{fig:perfo_metrics_testset_combmodalpha}
    	\end{subfigure}
	
	\vspace{-0.2cm} 
	\caption{
		DSC classification performance on the test data set as a function of source Galactic latitude and G-magnitude. 
		Predicted labels are obtained by selecting the class with the highest posterior probability, using the global prior.
		Left to right, the completenesses and the purities of the quasar and galaxy classes.
		Rows $(a)$, $(b)$, $(c)$, $(d)$ and $(e)$ refer to \texttt{Specmod}, \texttt{Allosmod}, \texttt{Specmod-Supp}, \texttt{Combmod}, and \texttt{Combmod-$\alpha$}, respectively.
        }
	\label{fig:perfo_metrics_testset_2d}
\end{figure*} 
  
\begin{figure*}[htp!]
	 \centering
	  \begin{subfigure}{1\textwidth} \centering
	 \includegraphics[scale=0.70]{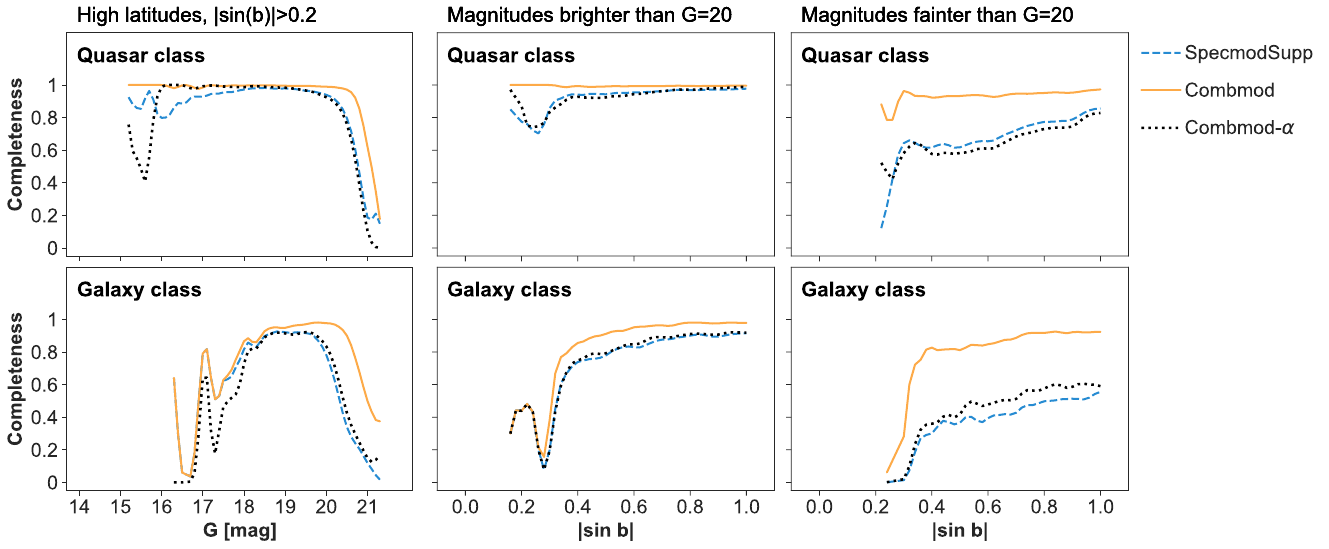}
	\caption{}
	\label{fig:perfo_metrics_testset_combmodalpha}
	 \end{subfigure}
	 
	 \vspace{.3cm}
         \begin{subfigure}{1\textwidth}
	  \centering
	 \includegraphics[scale=0.70]{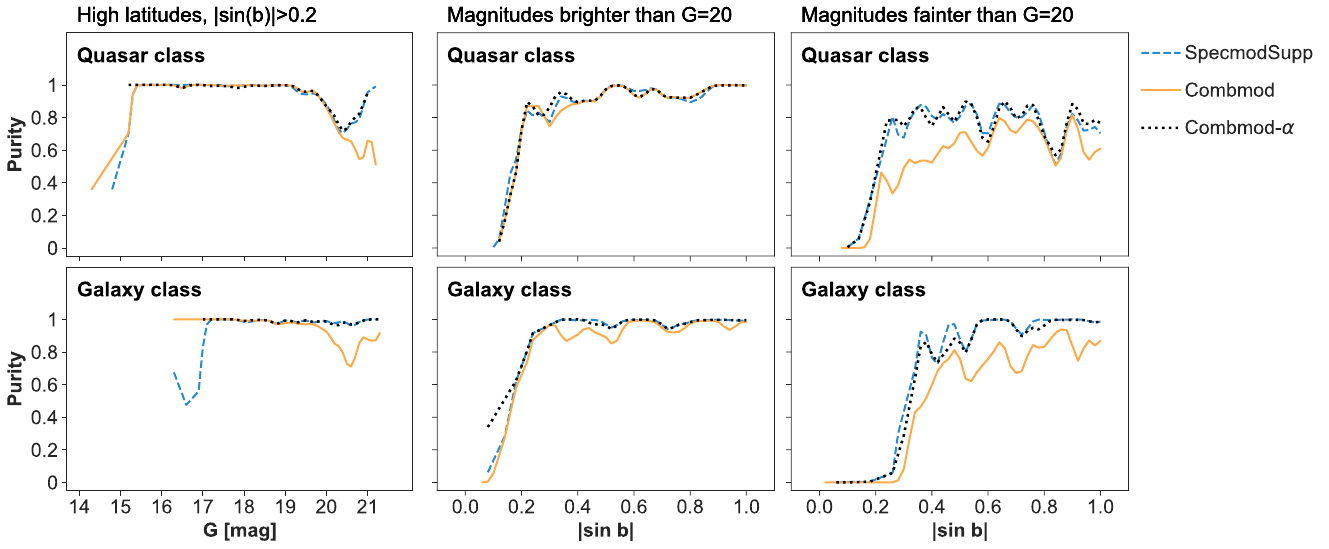}
	 \caption{}
	\label{fig:perfo_metrics_testset_combmodalpha}
	\end{subfigure}
	 
	\caption{
		Variation in performance as a function of magnitude for sources at high latitudes as a function of absolute Galactic latitude for two brightness ranges 
		(middle and right) for each class (rows). 
		The one-dimensional plots are a marginalisation of the two-dimensional representations in Figure \ref{fig:perfo_metrics_testset_2d}.
		\textit{Top}: completeness. \textit{Bottom}: purity.
		The left panels show completeness and purity as functions of magnitude, averaged over latitudes with |$\sin{b}$| > 0.2. 
		The middle and right panels show completeness and purity as functions of latitude, averaged over magnitudes brighter than $G$=20 and 
		fainter than $G$=20, respectively.
        }
	\label{fig:perfo_metrics_testset_2d_1davg}
\end{figure*} 

\begin{table*}[htp!]
    \caption{Summary of the 2D classification performance evaluated on the test data set at different magnitude limits for models trained on \textit{Gaia} data only.}               
    \label{table:tab_overall_performances_testset_magnitudes_avg2d}  
    \centering  \fontsize{8.5}{9}\selectfont
    \renewcommand{\arraystretch}{1.1}
    \setlength{\tabcolsep}{3pt}  
    \begin{tabular}{ll *{2}{wc{3.5em}} c *{3}{wc{3.5em}}} 
        \hline \hline 
        \multirow{3}{7em}{Magnitude bins}  
        	& \multirow{3}{6em}{Classifiers} 
    	& \multicolumn{2}{c}{\multirow{2}{8em}{\centering Weighted average\\ 2D completeness}} 
    	&& \multicolumn{2}{c}{\multirow{2}{8em}{\centering Weighted average\\ 2D purity}} 
	& \multicolumn{1}{c}{\multirow{3}{4em}{\centering Geometric score}} \\
   	&&&&&&&\\
        \cline{3-4} \cline{6-7}  
        & & \multicolumn{1}{c}{\centering quasar} & \multicolumn{1}{c}{\centering galaxy} && \multicolumn{1}{c}{\centering quasar} & \multicolumn{1}{c}{\centering galaxy} &\\
        \hline \hline 
        	\multirow{6}{9em}{$G$<19\\ (\textit{q}:\num{22557}, \textit{g}:\num{2217})}
	        	& \texttt{Specmod} & 0.959 & 0.890 && 0.998 & 0.983 & 0.957	\\
        & \texttt{Allosmod} & 0.981 & 0.849 && 0.996 & 0.990 & 0.952	\\
        & \texttt{Specmod-Supp} & 0.975 & 0.894 && 0.998 & 0.986 & 0.962	\\
        & \texttt{Combmod} & 0.996 & 0.920 && 0.998 & 0.978 & 0.972	\\
        & \texttt{Combmod-$\alpha$} & 0.987 & 0.878 && 0.996 & 0.988 & 0.961	\\
        \hline    	   	   
        	\multirow{6}{9em}{19$\leq$$G$<20\\ (\textit{q}:\num{47868}, \textit{g}:\num{11415})}  
        & \texttt{Specmod} & 0.841 & 0.884 && 0.953 & 0.986 & 0.914	\\
        & \texttt{Allosmod} & 0.956 & 0.880 && 0.955 & 0.986 & 0.943	\\
        & \texttt{Specmod-Supp} & 0.961 & 0.878 && 0.943 & 0.988 & 0.942	\\
        & \texttt{Combmod} & 0.993 & 0.974 && 0.947 & 0.961 & 0.969	\\
        & \texttt{Combmod-$\alpha$} & 0.957 & 0.895 && 0.949 & 0.988 & 0.947	\\
        \hline  	   	   
          \multirow{6}{9em}{$G$<20\\ (\textit{q}:\num{70425}, \textit{g}:\num{13632})} 
         & \texttt{Specmod} & 0.879 & 0.885 && 0.969 & 0.986 & 0.929	\\
        & \texttt{Allosmod} & 0.964 & 0.875 && 0.968 & 0.987 & 0.947	\\
        & \texttt{Specmod-Supp} & 0.965 & 0.881 && 0.961 & 0.987 & 0.948	\\
        & \texttt{Combmod} & 0.994 & 0.965 && 0.963 & 0.963 & 0.971	\\
        & \texttt{Combmod-$\alpha$} & 0.966 & 0.893 && 0.964 & 0.988 & 0.952	\\
       \hline  \hline 
       \multirow{6}{9em}{20$\leq$$G$<20.5\\ (\textit{q}:\num{38027}, \textit{g}:\num{6824})}  
        & \texttt{Specmod} & \cellcolor{gray!40}0.633 & \cellcolor{gray!40}0.550 && 0.811 & 0.957 & \cellcolor{gray!40}0.721	\\
        & \texttt{Allosmod} & 0.885 & \cellcolor{gray!40}0.701 && 0.750 & 0.964 & 0.818	\\
        & \texttt{Specmod-Supp} & 0.878 & \cellcolor{gray!40}0.557 && \cellcolor{gray!40}0.739 & 0.982 & 0.772	\\
        & \texttt{Combmod} & 0.982 & 0.958 && \cellcolor{gray!40}0.711 & 0.832 & 0.864	\\
        & \texttt{Combmod-$\alpha$} & 0.855 & \cellcolor{gray!40}0.646 && 0.762 & 0.972 & 0.800	\\
       \hline 
        \multirow{6}{9em}{$G$$\geq$20\\ (\textit{q}:\num{64898}, \textit{g}:\num{8815})} 
        & \texttt{Specmod} & \cellcolor{gray!40}0.510 & \cellcolor{red!40}0.469 && 0.820 & 0.955 & \cellcolor{gray!40}0.658	\\
        & \texttt{Allosmod} & 0.765 & \cellcolor{gray!40}0.617 && 0.750 & 0.964 & 0.764	\\
        & \texttt{Specmod-Supp} & 0.750 & \cellcolor{red!40}0.479 && 0.753 & 0.980 & \cellcolor{gray!40}0.718	\\
        & \texttt{Combmod} & 0.949 & 0.902 && \cellcolor{gray!40}0.670 & 0.821 & 0.828	\\
        & \texttt{Combmod-$\alpha$} & \cellcolor{gray!40}0.710 & \cellcolor{gray!40}0.560 && 0.772 & 0.971 & \cellcolor{gray!40}0.739	\\
        \hline
     \multirow{6}{9em}{$G$$\geq$20.5\\ (\textit{q}:\num{26871}, \textit{g}:\num{1991})} 
         & \texttt{Specmod} & \cellcolor{red!40}0.337 & \cellcolor{red!20}0.190 && 0.845 & 0.936 & \cellcolor{red!40}0.474	\\
        & \texttt{Allosmod} & \cellcolor{gray!40}0.596 & \cellcolor{red!40}0.326 && 0.750 & 0.959 & \cellcolor{gray!40}0.611	\\
        & \texttt{Specmod-Supp} & \cellcolor{gray!40}0.569 & \cellcolor{red!20}0.215 && 0.785 & 0.961 & \cellcolor{gray!40}0.551	\\
        & \texttt{Combmod} & 0.902 & \cellcolor{gray!40}0.713 && \cellcolor{gray!40}0.607 & 0.770 & \cellcolor{gray!40}0.740	\\
        & \texttt{Combmod-$\alpha$} & \cellcolor{gray!40}0.504 & \cellcolor{red!40}0.267 && 0.799 & 0.965 & \cellcolor{gray!40}0.568	\\
        \hline
    \end{tabular}
   \tablefoot{Same as Table \ref{table:tab_overall_performances_testset}.
   }
 \end{table*}

 \begin{table*}[htp!]
    \caption{
    	Summary of the 2D classification performance evaluated on the test data set at different magnitude limits 
    	for models trained on \textit{Gaia} data and mid-infrared photometry from CatWISE.
    }    
    \label{table:tab_overall_performances_testset_magnitudes_avg2d_IR}  
    \centering  \fontsize{8.5}{9}\selectfont 
    \renewcommand{\arraystretch}{1.1}
    \setlength{\tabcolsep}{5pt}  
   \begin{tabular}{ll *{2}{wl{5em}} c *{2}{wl{5em}} l} 
        \hline \hline 
        \multirow{3}{7em}{\centering Magnitude bins} 
        	& \multirow{3}{6em}{\centering Classifiers} 
	& \multicolumn{2}{c}{\centering \multirow{2}{8em}{\centering Weighted average\\ 2D completeness}} 
    	&& \multicolumn{2}{c}{\centering \multirow{2}{8em}{\centering Weighted average\\ 2D purity}} 
	& \multicolumn{1}{c}{\multirow{3}{4em}{\centering Geometric score}} \\
   	&&&&&&&\\
        \cline{3-4} \cline{6-7}  
        & & \multicolumn{1}{c}{\centering quasar} & \multicolumn{1}{c}{\centering galaxy} && \multicolumn{1}{c}{\centering quasar} & \multicolumn{1}{c}{\centering galaxy} \\			
        \hline \hline		
            \multirow{6}{9em}{\centering $G$<19\\ (\textit{q}:\num{20529}, \textit{g}:\num{1743})}			
     & \texttt{Specmod} & 0.959 $^{(-)}$  & 0.956 $^{(-)}$  && 0.999 $^{(-)}$  & 0.987 $^{(-)}$  & 0.975 $^{(-)}$ \\	
         & \texttt{Allosmod(*)} & 0.985 $^{(-)}$  & 0.963 $\rm^{\color{blue!70}(+5\,pp)}$  && 0.998 $^{(-)}$  & 0.987 $^{(-)}$  & 0.983 $\rm^{\color{blue!70}(+1\,pp)}$ \\	
         & \texttt{SpecmodSupp(*)} & 0.983 $\rm^{\color{blue!70}(+1\,pp)}$  & 0.950 $\rm^{\color{red!90}(-1\,pp)}$  && 0.998 $^{(-)}$  & 0.994 $\rm^{\color{blue!70}(+1\,pp)}$  & 0.981 $^{(-)}$ \\	
         & \texttt{Combmod(*)} & 0.997 $^{(-)}$  & 0.977 $\rm^{\color{blue!70}(+1\,pp)}$  && 0.998 $^{(-)}$  & 0.984 $^{(-)}$  & 0.989 $^{(-)}$ \\	
         & \texttt{Combmod-$\alpha$(*)} & 0.991 $^{(-)}$  & 0.964 $\rm^{\color{blue!70}(+2\,pp)}$  && 0.999 $^{(-)}$  & 0.989 $^{(-)}$  & 0.986 $\rm^{\color{blue!70}(+1\,pp)}$ \\	
        \hline			
            \multirow{6}{9em}{\centering 19$\leq$$G$<20\\ (\textit{q}:\num{42290}, \textit{g}:\num{8857})}			
        & \texttt{Specmod} & 0.843 $^{(-)}$  & 0.900 $^{(-)}$  && 0.965 $^{(-)}$  & 0.994 $^{(-)}$  & 0.924 $^{(-)}$ \\	
         & \texttt{Allosmod(*)} & 0.979 $\rm^{\color{blue!70}(+2\,pp)}$  & 0.924 $\rm^{\color{blue!70}(+3\,pp)}$  && 0.952 $\rm^{\color{red!90}(-2\,pp)}$  & 0.991 $^{(-)}$  & 0.961 $\rm^{\color{blue!70}(+1\,pp)}$ \\	
         & \texttt{SpecmodSupp(*)} & 0.969 $\rm^{\color{blue!70}(+1\,pp)}$  & 0.951 $\rm^{\color{blue!70}(+6\,pp)}$  && 0.961 $^{(-)}$  & 0.991 $^{(-)}$  & 0.968 $\rm^{\color{blue!70}(+2\,pp)}$ \\	
         & \texttt{Combmod(*)} & 0.996 $^{(-)}$  & 0.989 $^{(-)}$  && 0.960 $^{(-)}$  & 0.974 $^{(-)}$  & 0.980 $^{(-)}$ \\	
         & \texttt{Combmod-$\alpha$(*)} & 0.980 $\rm^{\color{blue!70}(+2\,pp)}$  & 0.923 $\rm^{\color{blue!70}(+1\,pp)}$  && 0.961 $^{(-)}$  & 0.991 $^{(-)}$  & 0.963 $\rm^{\color{blue!70}(+1\,pp)}$ \\	
        \hline			
            \multirow{6}{9em}{\centering $G$<20\\ (\textit{q}:\num{62819}, \textit{g}:\num{10600})}			
        & \texttt{Specmod} & 0.881 $^{(-)}$  & 0.910 $^{(-)}$ & & 0.977 $^{(-)}$  & 0.992 $^{(-)}$  & 0.939 $^{(-)}$ \\	
         & \texttt{Allosmod(*)} & 0.981 $\rm^{\color{blue!70}(+2\,pp)}$  & 0.930 $\rm^{\color{blue!70}(+4\,pp)}$  && 0.967 $\rm^{\color{red!90}(-1\,pp)}$  & 0.991 $^{(-)}$  & 0.967 $\rm^{\color{blue!70}(+1\,pp)}$ \\	
         & \texttt{SpecmodSupp(*)} & 0.974 $\rm^{\color{blue!70}(+1\,pp)}$  & 0.951 $\rm^{\color{blue!70}(+5\,pp)}$  && 0.973 $^{(-)}$  & 0.992 $^{(-)}$  & 0.972 $\rm^{\color{blue!70}(+1\,pp)}$ \\	
         & \texttt{Combmod(*)} & 0.996 $^{(-)}$  & 0.987 $^{(-)}$  && 0.973 $^{(-)}$  & 0.975 $^{(-)}$  & 0.983 $^{(-)}$ \\	
         & \texttt{Combmod-$\alpha$(*)} & 0.984 $\rm^{\color{blue!70}(+2\,pp)}$  & 0.929 $\rm^{\color{blue!70}(+1\,pp)}$  && 0.973 $^{(-)}$  & 0.991 $^{(-)}$  & 0.969 $\rm^{\color{blue!70}(+1\,pp)}$ \\	
         \hline\hline			
            \multirow{6}{9em}{\centering 20$\leq$$G$<20.5\\ (\textit{q}:\num{32893}, \textit{g}:\num{4990})}			
        & \texttt{Specmod} &\cellcolor{gray!40}0.636 $^{(-)}$  & \cellcolor{gray!40}0.544 $^{(-)}$  && 0.862 $^{(-)}$  & 0.972 $^{(-)}$  & \cellcolor{gray!40}0.734 $^{(-)}$ \\	
         & \texttt{Allosmod(*)} & 0.959 $\rm^{\color{blue!70}(+7\,pp)}$  & \cellcolor{gray!40}0.717 $\rm^{\color{blue!70}(+2\,pp)}$  && 0.837 $\rm^{\color{blue!70}(+1\,pp)}$  & 0.983 $^{(-)}$  & 0.867 $\rm^{\color{blue!70}(+2\,pp)}$ \\	
         & \texttt{SpecmodSupp(*)} & 0.952 $\rm^{\color{blue!70}(+7\,pp)}$  & 0.800 $\rm^{\color{blue!70}(+25\,pp)}$  && 0.851 $\rm^{\color{blue!70}(+2\,pp)}$  & 0.959 $\rm^{\color{red!90}(-3\,pp)}$  & 0.888 $\rm^{\color{blue!70}(+10\,pp)}$ \\	
         & \texttt{Combmod(*)} & 0.992 $\rm^{\color{blue!70}(+1\,pp)}$  & 0.970 $^{(-)}$  && 0.804 $^{(-)}$  & 0.857 $\rm^{\color{red!90}(-1\,pp)}$  & 0.902 $^{(-)}$ \\	
         & \texttt{Combmod-$\alpha$(*)} & 0.949 $\rm^{\color{blue!70}(+9\,pp)}$  & \cellcolor{gray!40}0.650 $\rm^{\color{blue!70}(+1\,pp)}$  && 0.826 $\rm^{\color{red!90}(-1\,pp)}$  & 0.993 $^{(-)}$  & 0.843 $\rm^{\color{blue!70}(+2\,pp)}$ \\	
        \hline			
        \multirow{6}{9em}{\centering $G$$\geq$20\\ (\textit{q}:\num{55687}, \textit{g}:\num{6231})}			
        & \texttt{Specmod} & \cellcolor{gray!40}0.514 $^{(-)}$  & \cellcolor{red!20}0.471 $^{(-)}$  && 0.866 $^{(-)}$  & 0.968 $^{(-)}$  & \cellcolor{gray!40}0.671 $^{(-)}$ \\	
         & \texttt{Allosmod(*)} & 0.919 $\rm^{\color{blue!70}(+15\,pp)}$  & \cellcolor{gray!40}0.628 $\rm^{\color{blue!70}(+1\,pp)}$  && 0.797 $\rm^{\color{red!90}(-2\,pp)}$  & 0.981 $^{(-)}$  & 0.820 $\rm^{\color{blue!70}(+3\,pp)}$ \\	
         & \texttt{SpecmodSupp(*)} & 0.915 $\rm^{\color{blue!70}(+16\,pp)}$  & \cellcolor{gray!40}0.692 $\rm^{\color{blue!70}(+22\,pp)}$ & & 0.809 $\rm^{\color{red!90}(-2\,pp)}$  & 0.959 $\rm^{\color{red!90}(-3\,pp)}$  & 0.837 $\rm^{\color{blue!70}(+10\%}$ \\	
         & \texttt{Combmod(*)} & 0.976 $\rm^{\color{blue!70}(+3\,pp)}$  & 0.916 $^{(-)}$  && \cellcolor{gray!40}0.746 $\rm^{\color{red!90}(-2\,pp)}$  & 0.845 $\rm^{\color{red!90}(-2\,pp)}$  & 0.866 $^{(-)}$ \\	
         & \texttt{Combmod-$\alpha$(*)} & 0.884 $\rm^{\color{blue!70}(+17\,pp)}$  & \cellcolor{gray!40}0.563 $^{(-)}$  && 0.797 $\rm^{\color{red!90}(-4\,pp)}$  & 0.990 $^{(-)}$  & 0.792 $\rm^{\color{blue!70}(+3\,pp)}$ \\	
	  \hline			
        \multirow{6}{9em}{\centering $G$$\geq$20.5\\ (\textit{q}:\num{22794}, \textit{g}:\num{1241})}			
        & \texttt{Specmod} & \cellcolor{red!20}0.338 $^{(-)}$  & \cellcolor{red!40}0.177 $^{(-)}$  && 0.876 $^{(-)}$  & 0.931 $^{(-)}$  & \cellcolor{red!20}0.470 $^{(-)}$ \\	
         & \texttt{Allosmod(*)} & 0.862 $\rm^{\color{blue!70}(+27\,pp)}$  & \cellcolor{red!20}0.268 $\rm^{\color{red!90}(-1\,pp)}$  && \cellcolor{gray!40}0.733 $\rm^{\color{red!90}(-7\,pp)}$  & 0.959 $^{(-)}$  & \cellcolor{gray!40}0.635 $\rm^{\color{blue!70}(+4\,pp)}$ \\	
         & \texttt{SpecmodSupp(*)} & 0.862 $\rm^{\color{blue!70}(+29\,pp)}$  & \cellcolor{red!20}0.259 $\rm^{\color{blue!70}(+8\,pp)}$  && \cellcolor{gray!40}0.743 $\rm^{\color{red!90}(-9\,pp)}$  & 0.962 $^{(-)}$  & \cellcolor{gray!40}0.632 $\rm^{\color{blue!70}(+10\,pp)}$ \\	
         & \texttt{Combmod(*)} & 0.954 $\rm^{\color{blue!70}(+5\,pp)}$  & \cellcolor{gray!40}0.699 $^{(-)}$  && \cellcolor{gray!40}0.658 $\rm^{\color{red!90}(-5\,pp)}$  & 0.786 $\rm^{\color{red!90}(-2\,pp)}$  & 0.766 $\rm^{\color{red!90}(-1\,pp)}$ \\	
         & \texttt{Combmod-$\alpha$(*)} & 0.789 $\rm^{\color{blue!70}(+29\,pp)}$  & \cellcolor{red!40}0.213 $\rm^{\color{red!90}(-2\,pp)}$  && \cellcolor{gray!40}0.747 $\rm^{\color{red!90}(-9\,pp)}$  & 0.959 $^{(-)}$  & \cellcolor{gray!40}0.589 $\rm^{\color{blue!70}(+4\,pp)}$ \\	
          \hline
    \end{tabular}
    \tablefoot{
    \textit{(*)} indicates the models trained with \textit{Gaia} data and infrared photometry from CatWISE.
    The numbers in pp are equal to the difference in performance between models trained with \textit{Gaia} data and CatWISE, and models trained with \textit{Gaia} data only, evaluated on the exact test baseline of sources with valid infrared data.
    The symbol \textit{(-)} indicates no change. 
    Colour code is similar to Table \ref{table:tab_overall_performances_testset}.\\
    }
 \end{table*}

		\subsubsection{Overall performance}
We assign objects to classes with maximum probabilities above a threshold.
Figure \ref{fig:perfo_metrics_testset_thresholds} shows the variation in the overall performance evaluated on the test set as a function of the posterior probability threshold.
As the threshold increases, completeness decreases while purity increases. 
The results of the different classifiers show \texttt{Combmod} as the best-performing classifier in terms of completeness of the extragalactic classes. 
The fractions of unclassified sources (i.e. sources with probabilities below the threshold) in each class are comparable in the other classifiers, which indicates a smoother probability distribution than that of Combmod, where probabilities are close to unity.
Not applying a threshold to the maximum probability is equivalent to setting the threshold to zero.

In this study, we assess the performance of DSC without applying a probability threshold, thereby maintaining the highest possible completeness of the extragalactic classes.
The overall performance, as evaluated on the test set, is summarised in Table \ref{table:tab_overall_performances_testset}. 
\texttt{Allosmod} achieves high completenesses of 87\% and 77\% and purities of 73\% and 70\% for quasars and galaxies, respectively. 
\texttt{Specmod} performs less well, with completenesses of 16 percentage points (pp)\footnote{Percentage points, here noted as pp, are equal to the arithmetic difference between two percentages.} and 5 pp lower, and purities of 5 pp and 9 pp lower, for quasars and galaxies, respectively.
\texttt{Specmod-Supp} has a similar overall performance to Allosmod, the largest differences being a 5 pp decrease in completeness and a 7 pp increase in purity for the galaxy class.
Combmod achieves the highest completeness among the models, at 97\% for quasars and 94\% for galaxies, but at the cost of very low purity (37\%) for galaxies. 
\texttt{Combmod-$\alpha$} and \texttt{Specmod-Supp} show the best compromise in terms of the geometric score defined by equation~\ref{eq:geoscore}.

In GDR3, \texttt{Specmod} achieved low completeness for the quasars and low purity for the extragalactic classes, whereas \texttt{Allosmod} and \texttt{Combmod} achieved high completeness but low purity \citep{delchambre_gaia_2023}.
Compared to these results, we now achieve an increase in quasar completeness in all classifiers but a decrease in the galaxy completeness. 
However, purity has significantly improved in all classifiers for both classes, which is one of the main goals of this work.
The percentage points in Table \ref{table:tab_overall_performances_testset} compare these results to those in GDR3 on the same test set.
For this purpose, we mapped the three GDR3 stellar classes into one star class.
The highest increase in completeness for quasars is achieved by \texttt{Specmod} as a result of changing the algorithm, from Extratrees to an MLP. The highest increase in purity is 25 pp for quasars by \texttt{Specmod} and 27 pp for galaxies by \texttt{Allosmod}.
Despite the reduced completeness of the galaxy class in Combmod-$\alpha$, the performance for the combined classifier shows a large increase in purity for both classes.

The analysis of the feature importance for DSC, described in Appendix \ref{appendix_feature_importance}, shows that the most informative features are proper motions, astrometric excess noise, the colour \texttt{bp\_g}, relative variability in the $G$ band, and the colour excess factor. 
Moreover, the analysis of each BP/RP pixel input to \texttt{Specmod} shows that feature importance is distributed across a range of wavelengths, reflecting the model's reliance on diverse spectral signatures to identify the source type.

		\subsubsection{Discussion on calibration}
The high completeness and low purity of \texttt{Combmod} using the same thresholds as the other classifiers may suggest that its outputs do not behave like well-calibrated probabilities, and should perhaps instead be interpreted more like confidence-like scores.
Interestingly, the \texttt{Combmod}-$\alpha$ output is a weighted sum of two probabilities, so it cannot be strictly interpreted as a probability, and yet this does not exhibit the same significant drop in purity seen in \texttt{Combmod}.
We note that some of our models take as inputs both the BP/RP spectra and the colours formed from the integrated BP and RP fluxes, or they combine probabilities from other models that use these inputs. Those models, therefore, have inputs that are not strictly independent. This may lead to overconfident outputs that are not strictly posterior probabilities.
We emphasise that the combined classifiers are primarily intended as an aggregation tool to optimise performance, rather than as a rigorous probabilistic merger.
We nonetheless choose to retain the term ‘probabilities’ for their outputs to remain consistent with the mathematical framework of the DSC pipeline.
As we do not implement any calibration on the DSC outputs, a comprehensive examination of calibration is beyond the scope of this study, although we will discuss it in a technical note that will accompany the data release. 
The thresholding approach for DSC outputs selects the class with the highest probability above a defined limit. We recommend that users consider these values carefully when conducting studies that require precise probability calibration.

\begin{table*}[htp!]
    \renewcommand{\arraystretch}{1.15}     
    \caption{ Quasar and galaxy candidates classified by DSC across all magnitudes and latitudes but excluding the LMC and SMC. }   
    \label{tab:ref_counts_predictions_fullruns_maglims}  
    \centering  \fontsize{8.5}{9}\selectfont   
    \renewcommand{\arraystretch}{1.1}
    \setlength{\tabcolsep}{2pt}                   
    \begin{tabular}{r | l | r *{4}{wr{4em}} | r *{4}{wr{4em}}  }        
        \hline\hline                  
        		\multicolumn{2}{c|}{\multirow{2}{5em}{\bf \centering Classifier}} 
		& \multicolumn{5}{c|}{\bf quasar candidates} & \multicolumn{5}{c}{\bf galaxy candidates} \\
		 \cline{3-12}
		 \multicolumn{2}{c|}{}& \multicolumn{1}{r}{total}&  \multicolumn{1}{r}{G<20} & \multicolumn{1}{r}{20$\leq$$G$<20.5}
		 	& \multicolumn{1}{r}{20.5$\leq$$G$<21} & \multicolumn{1}{r|}{$G$$\geq$21}
		 & \multicolumn{1}{r}{total}&  \multicolumn{1}{r}{G<20} & \multicolumn{1}{r}{20$\leq$$G$<20.5}
		 	& \multicolumn{1}{r}{20.5$\leq$$G$<21} & \multicolumn{1}{r}{$G$$\geq$21}  \\
	\hline 
        \multirow{6}{*}{(a)}  
         & \texttt{Specmod} & \num{2108425} & 40 (32) & 22 (18) & 18 (12) & 20 (1) & \num{1885485} & 22 (17) & 20 (18) & 21 (12) & 38 (11) \\
         & \texttt{Allosmod} & \num{1692662} & 46 (43) & 33 (32) & 20 (19) & 0.4 (0.1) & \num{623087} & 52 (45) & 35 (30) & 10.6 (7) & 3 (1) \\
         & \texttt{Specmod-Supp} & \num{1651764} & 51 (45) & 32 (30) & 17 (16) & 0.1 (0.03) & \num{551551} & 59 (54) & 33 (31) & 6 (5) & 2 (0.8) \\
         & \texttt{Combmod} &\num{3010203} & 29 (26) & 23 (22) & 33 (27) & 15 (1) & \num{2302881} & 19 (16) & 29 (26) & 27 (19) & 25 (4) \\
         & \texttt{Combmod-$\alpha$} & \num{2021603} & 39 (36) & 26 (24) & 16 (11) & 19 (0.1) & \num{1326305} & 27 (22) & 19 (16) & 14 (4) & 40 (4) \\
         & \texttt{Combmod-Supp} &\num{2156151} & 40 (34) & 25 (23) & 17 (13) & 18 (0.1) & \num{1236203} & 29 (24) & 16 (14) & 13 (3) & 42 (4) \\
        \hline  \multirow{6}{*}{(b)}  
         & \texttt{Specmod} & \num{1084516} & 53 (51) & 28 (27) & 16 (15) & 3 (1) & \num{763628} & 30 (29) & 28 (27) & 19 (18) & 23 (18) \\
         & \texttt{Allosmod(*)} & \num{1726112} & 39 (37) & 30 (29) & 30 (28) & 0.8 (0.4) & \num{386739} & 56 (54) & 37 (35) & 6 (6) & 0.8 (0.6) \\
         & \texttt{Specmod-Supp(*)} & \num{1755423} & 38 (37) & 30 (28) & 31 (29) & 1 (0.6) & \num{413421} & 52 (49) & 39 (38) & 8 (8) & 1 (1) \\
         & \texttt{Combmod} & \num{1995096} & 35 (33) & 28 (26) & 35 (32) & 3 (1) & \num{1196593} & 22 (21) & 38 (37) & 32 (30) & 8 (5) \\
         & \texttt{Combmod-$\alpha$} & \num{1600652} & 41 (40) & 31 (30) & 26 (25) & 2 (0.3) & \num{440195} & 50 (48) & 33 (31) & 6 (5) & 12 (6) \\
         & \texttt{Combmod-Supp} & \num{1782062} & 38 (36) & 29 (28) & 31 (28) & 2 (1) & \num{468333} & 46 (44) & 35 (33) & 8 (7) & 12 (6) \\
        \hline  \multirow{4}{*}{(c)}    
         & \texttt{Allosmod(*)} & \num{420567} & 10 (9) & 20 (18) & 67 (60) & 3 (1) & \num{79638} & 31 (29) & 55 (51) & 12 (10) & 3 (2) \\
         & \texttt{Specmod-Supp(*)} & \num{540010} & 10 (8) & 23 (20) & 64 (56) & 4 (2) & \num{141003} & 18 (15) & 60 (58) & 18 (17) & 4 (3) \\
         & \texttt{Combmod} & \num{150774} & 6 (5) & 12 (10) & 71 (58) & 11 (5) & \num{125606} & 3 (2) & 14 (13) & 68 (63) & 15 (14) \\
         & \texttt{Combmod-$\alpha$} & \num{366773} & 10 (9) & 24 (22) & 63 (59) & 2 (1) & \num{44283} & 26 (23) & 58 (54) & 13 (12) & 4 (3) \\
    \hline
    \end{tabular}
    \tablefoot{
    	The symbol \textit{(*)} indicates models trained using \textit{Gaia} data and infrared photometry from CatWISE.
     	 Rows $(a)$, $(b)$, and $(c)$ refer to candidates identified in \textit{Gaia} only mode, \textit{Gaia}-IR mode or exclusive to \textit{Gaia}-IR mode, respectively.
    	 Detections exclusive to \textit{Gaia}-IR mode refer to the candidates identified only in this mode and not in the \textit{Gaia} only mode.
   	 The percentages of sources in each magnitude bin are shown in the columns. 
	 The numbers in parentheses refer to the percentage of sources in the magnitude bin at higher galactic 
	 latitudes |$\sin b$|>0.2 after the application of the photometric quality cut \texttt{phot\_[bp;rp]\_n\_obs}$\geq$10.
	The number of sources in each magnitude bin is equal to the percentage of sources in that bin multiplied by the total number of sources.
	\texttt{Combmod-Supp(*)} predictions identified exclusively in the \textit{Gaia}-IR mode are equal to the candidates 
     	 from \texttt{Specmod-Supp(*)}, as the data required for the classifier is available.
         }
\end{table*}

		\subsubsection{Magnitude and latitude-based assessment of the performance} \label{subsection415}	
As noted also in \cite{hughes_quasar_2022} and \cite{jamal_improved_2024}, the summary performance evaluated on the entire test set does not fully reflect a classifier's strengths and weaknesses.
We therefore evaluate the classification as a function of the sources' brightness and Galactic latitude. 
We bin the test data on a 2D grid of 4550 bins in |$\sin{b}$|$\in$[0,1] and $G$$\in$[5,21], with coarser binning in the bright magnitude regime of $G$<14.
In each bin, we adjust the confusion matrix to account for class imbalance and compute completeness and purity.

Figure \ref{fig:perfo_metrics_testset_2d} shows this 2D completeness and purity for the quasar and galaxy classes.
We use grey for undefined values, light colours for high values, and black for very small to zero values.
Undefined and zero values can be explained by Eq. \ref{eq:metrics}.
Zero completeness occurs when TP is zero, but TN is not, meaning that all objects in the class are misclassified.
Conversely, zero purity refers to zero TP but non-zero FP, as the classifier only predicts false detections.
An undefined completeness occurs when TP and TN are both null, meaning that there are no objects from the class to predict.
An undefined purity occurs when TP and FP are both null, meaning that the classifier does not introduce any contamination (FP null) nor any correct prediction (TP null).
The performance is unaffected when the metrics are undefined, unlike zero values, which indicate very poor classifier performance.

The performance decreases towards fainter magnitudes in Figure~\ref{fig:perfo_metrics_testset_2d}, as we would expect.
For \texttt{Specmod}, the quasar class completeness begins to decrease at magnitudes fainter than $G$=19, whereas \texttt{Allosmod} and \texttt{Specmod-Supp} show a decrease only from $G$$\geq$20. 
This result indicates a limitation of \texttt{Specmod}, which uses only the XP spectra to predict a label, and these spectra will generally be noisier than the features used by Allosmod.
The galaxy completeness remains comparable over the three baseline classifiers, with a continuous decrease from $G$$\geq$20.
The purities are high overall, with contamination mostly localised at magnitudes fainter than $G$=20.
Across all magnitudes, \texttt{Combmod} achieves the highest completenesses amongst the classifiers, but the lowest purities, particularly for the galaxy class. 
\texttt{Combmod-$\alpha$} achieves lower completeness but a higher gain in purity, compared to \texttt{Combmod}.

Figure \ref{fig:perfo_metrics_testset_2d_1davg} illustrates how these metrics change with latitude across two distinct magnitude ranges.
We compute these one-dimensional representations by averaging the two-dimensional representations along one axis.
In Figure \ref{fig:perfo_metrics_testset_2d_1davg}, the completeness at $G<20$\,mag is high for all classifiers when far from the Galactic plane. 
At magnitudes fainter than $G$=20\,mag, however, the completeness and purities decrease. 
Purity improves at higher latitudes where there is less dust extinction, crowding and stellar density, with the best purities reached by \texttt{Allosmod}, \texttt{Combmod-$\alpha$} and \texttt{Specmod-Supp}.
Overall, the completeness of the quasar class remains higher compared to that of the galaxy class.
As previously stated, \texttt{Combmod} achieves the highest completeness amongst the classifiers, but the lowest purity for the galaxy class. 

To summarise the performance across magnitude and latitude, we compute the average metrics, weighted by the denominators in Eq. \ref{eq:metrics}, i.e., the number of true sources per bin for completeness and the number of predictions per bin for purity. 
We would like to emphasise that, since the adjustment only affects purity, the average value does not correspond to the global metric, unlike the average completeness. Therefore, the adjustment made to the overall metrics cannot be compared with the per-bin adjustments applied to the 2D metrics.
The average metrics for different magnitude ranges are reported in Table \ref{table:tab_overall_performances_testset_magnitudes_avg2d}.
The classifiers perform best at magnitudes brighter than $G$=20, achieving an average 2D completeness of 87-99\% and an average 2D purity of 96-98\% for the extragalactic classes.
At magnitudes of 20<$G$$\leq$20.5, purity decreases rapidly, demonstrating the classifiers' inability to reject the false positives.
Also, while \texttt{Allosmod} and the combined classifiers maintain a 2D completeness of at least 75\% for quasars, it drops considerably for the galaxies.%
At magnitudes fainter than $G$=20.5, both completeness and purity are significantly reduced. 
The results also show that \texttt{Specmod} performance is more limited at faint magnitudes than that of the other classifiers, indicating a limited use of XP spectra for reliable classifications at such magnitudes. 
The results also show that, unlike \texttt{Combmod}, which prioritises completeness, \texttt{Combmod-$\alpha$} and \texttt{Specmod-Supp} improve purity, particularly at $G$$\geq$20.
At magnitudes of 20<$G$$\leq$20.5, the latter two classifiers increase the purity by 5 and 15 pp, but completeness is reduced by 13 and 50 pp, for quasars and galaxies, respectively. 
At magnitudes fainter than $G$=20.5, they largely reduce stellar contamination by 18 to 20 pp at the cost of missing out true classes, especially galaxies.

	\subsection{Augmented DSC predictions with IR data} 
We now train separate classifiers using mid-infrared catWISE photometry and \textit{Gaia} data.
We supplement the inputs with the mid-infrared photometry bands $W1$ and $W2$ to train the \texttt{Allosmod}(*) and \texttt{Specmod-Supp}(*) classifiers (see Section \ref{sec:3_DSCmethodology}). The other \textit{Gaia} inputs remain the same as before (see Appendix \ref{sec:list_features_classifier}). 
The \texttt{Specmod} classifier is the same as before (no IR inputs). We then combine the posterior probabilities of these classifiers, as done before, to compute new combined \textit{Gaia}-IR classifiers. 
As shown in Tables \ref{table:tab_overall_performances_testset_magnitudes_avg2d} and \ref{table:tab_overall_performances_testset_magnitudes_avg2d_IR}, models in \textit{Gaia}-IR mode use a different training and test data set composed of sources with valid CatWISE photometry than models in \textit{Gaia} only mode.

The performance on the infrared-augmented test set at different magnitude limits is provided in Table \ref{table:tab_overall_performances_testset_magnitudes_avg2d_IR}. 
The percentage points provide a direct comparison between the \textit{Gaia}-IR and \textit{Gaia}-only modes, on the same test set that requires valid infrared photometry.
The classification results demonstrate that combining the \textit{Gaia} optical data with mid-IR data from CatWISE improves the completeness of quasars and galaxies at the expense of a slight reduction in purity.
The improvement in completeness at bright magnitudes $G$<20 reaches a maximum increase of 5 pp. 
This effect is more significant at fainter magnitudes, where the increase in completeness is substantial at the slight expense of a reduction in purity by 1 to 9 pp. 
Here, the maximum increase in completeness reaches 25 pp for galaxies at magnitudes 20$\leq$$G$<20.5 by \texttt{Specmod-Supp}, and 29 pp for quasars at $G$$\geq$20.5 by \texttt{Specmod-Supp} and \texttt{Combmod-$\alpha$}. 

The analysis of feature importance for \texttt{Allosmod}(*) in Appendix \ref{appendix_feature_importance} confirms that the infrared colour $W1$-$W2$ is a key feature for quasars completeness, alongside the \textit{Gaia} colours, proper motions and the astrometric excess noise.

To summarise, adding the IR photometry to the \textit{Gaia} data significantly improves DSC completeness, primarily for faint quasars at magnitudes $G$$\geq$20.

\begin{figure*}[htp!]
	\begin{subfigure}{0.3333\textwidth}
         \centering
	 \includegraphics[scale=0.30]{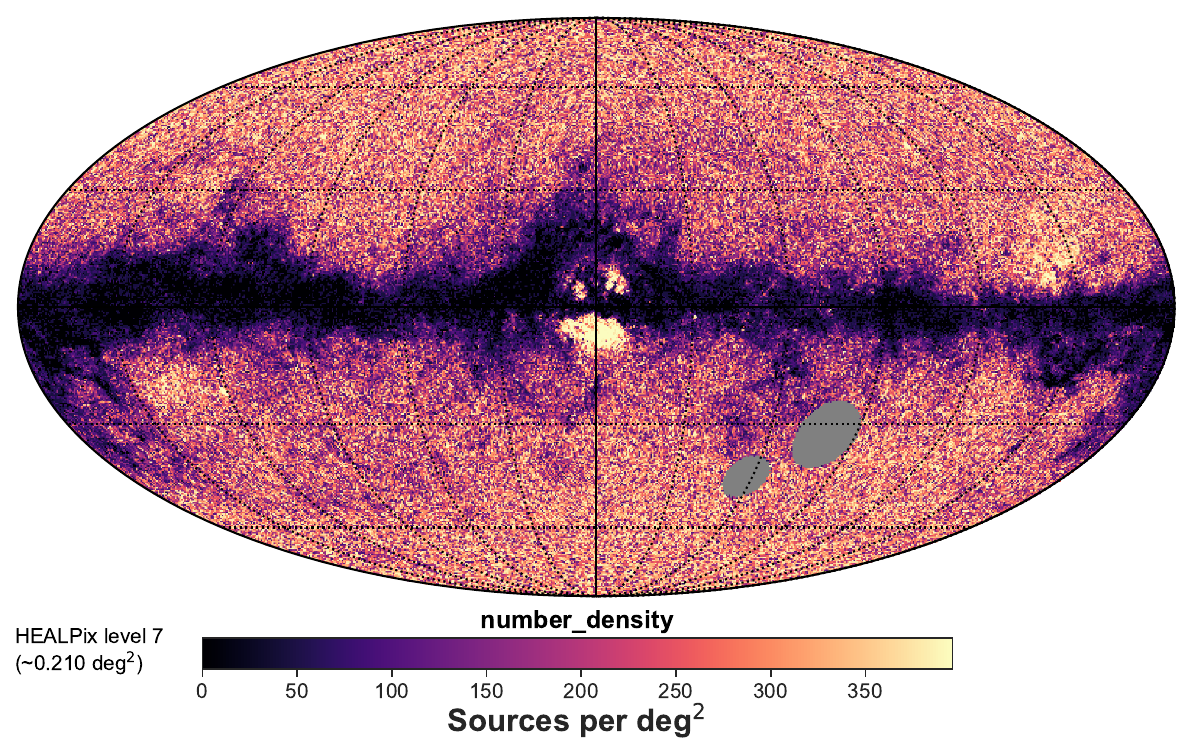}
	\caption{}
    	\end{subfigure}
	\hspace{0.06cm}
	\begin{subfigure}{0.3333\textwidth}
         \centering
	 \includegraphics[scale=0.30]{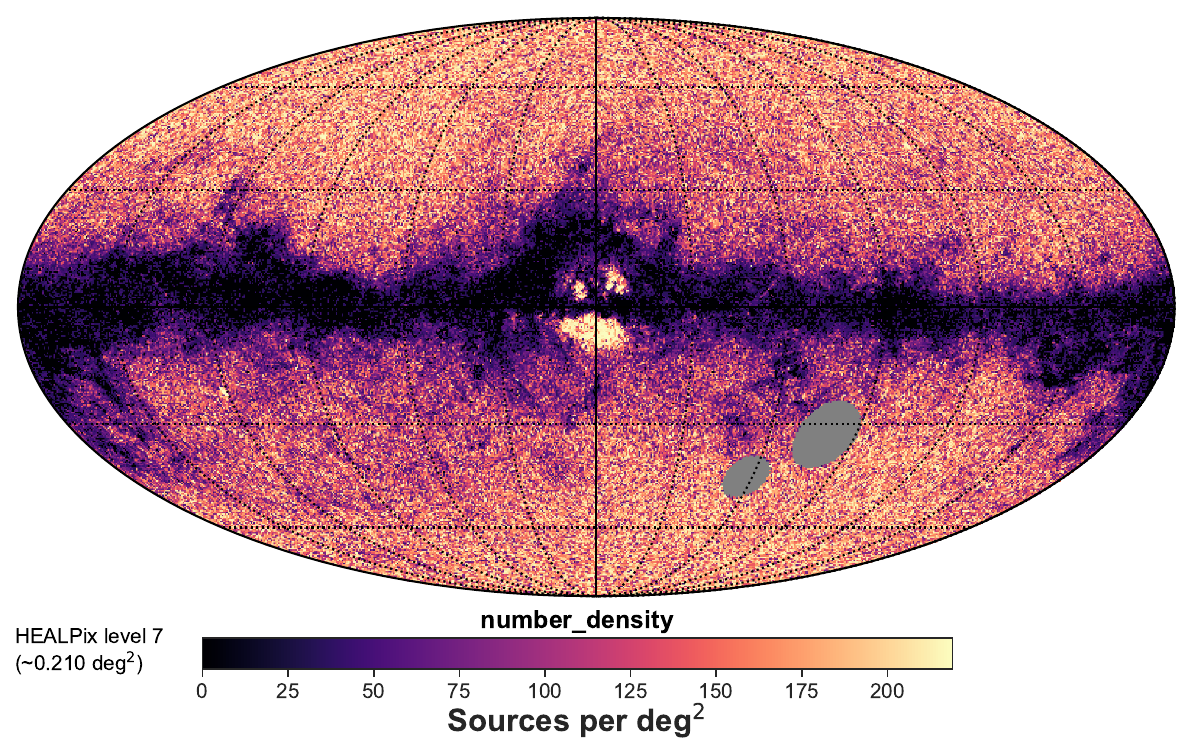}
	\caption{}
	\end{subfigure}
	\hspace{0.06cm}
	\begin{subfigure}{0.3333\textwidth}
         \centering
	 \includegraphics[scale=0.30]{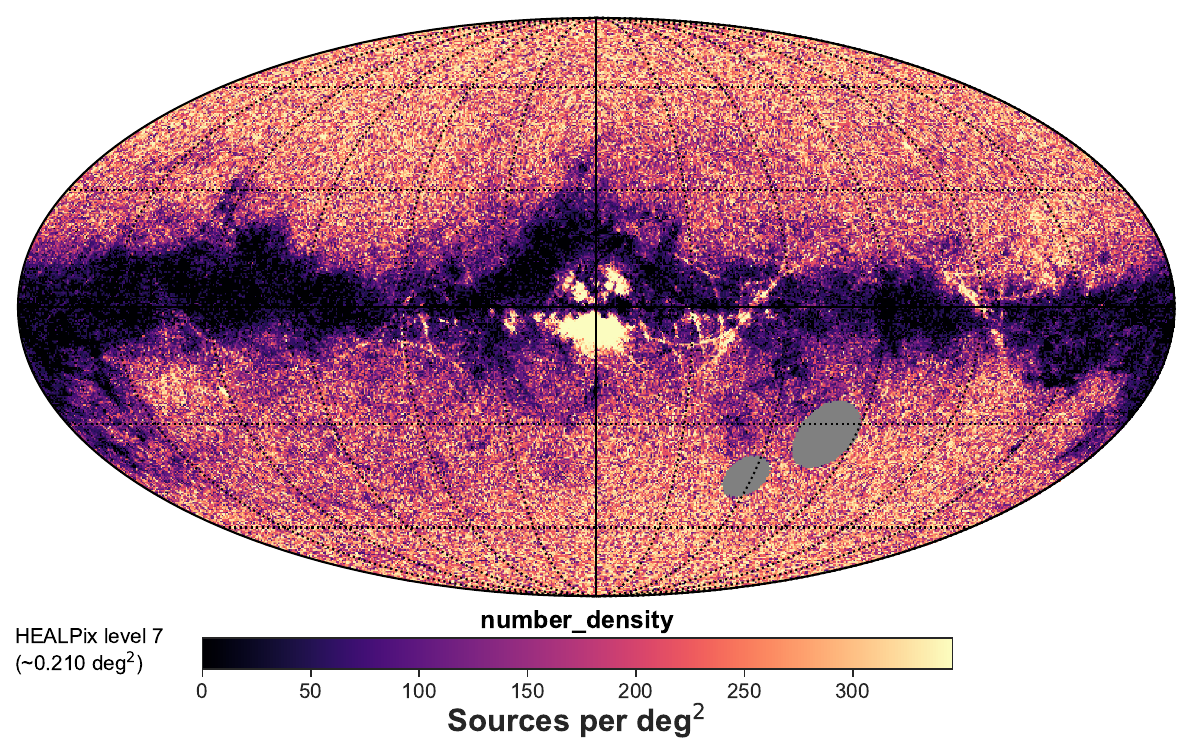}
	\caption{}
	\end{subfigure}

	\begin{subfigure}{0.3333\textwidth}
         \centering
	\includegraphics[scale=0.30]{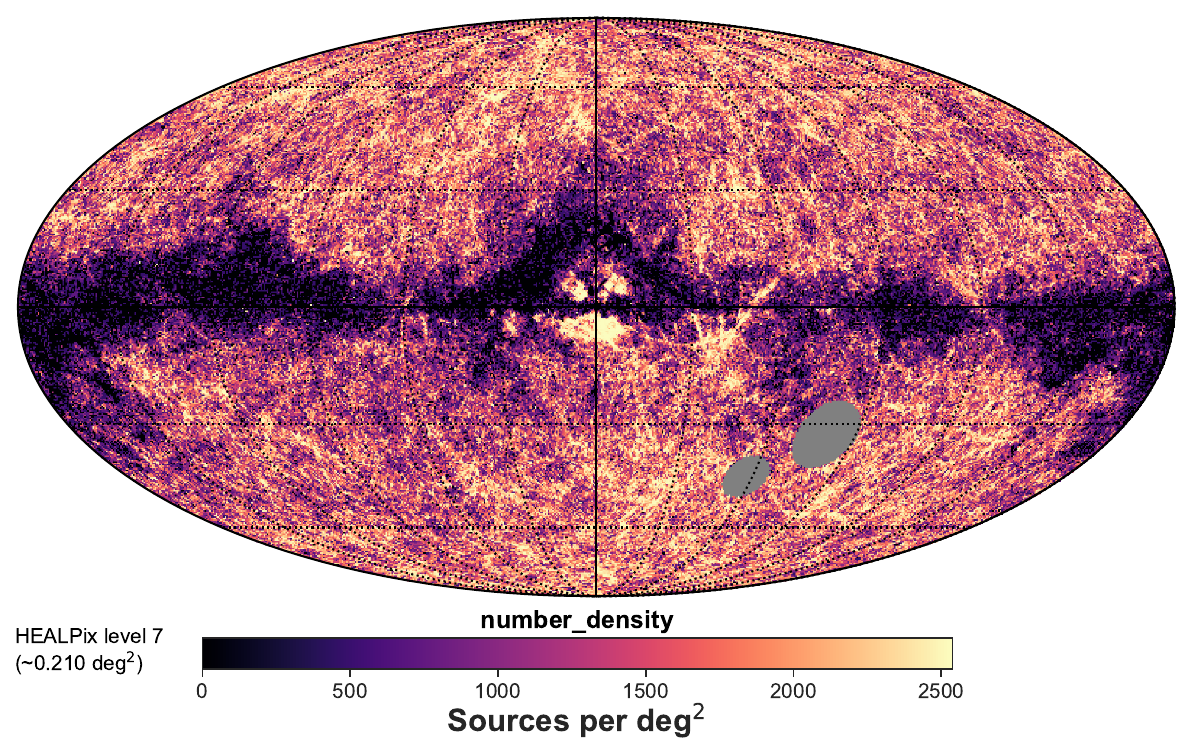}
	\caption{}
	\end{subfigure}
	\hspace{0.06cm}
	\begin{subfigure}{0.3333\textwidth}
         \centering
	\includegraphics[scale=0.30]{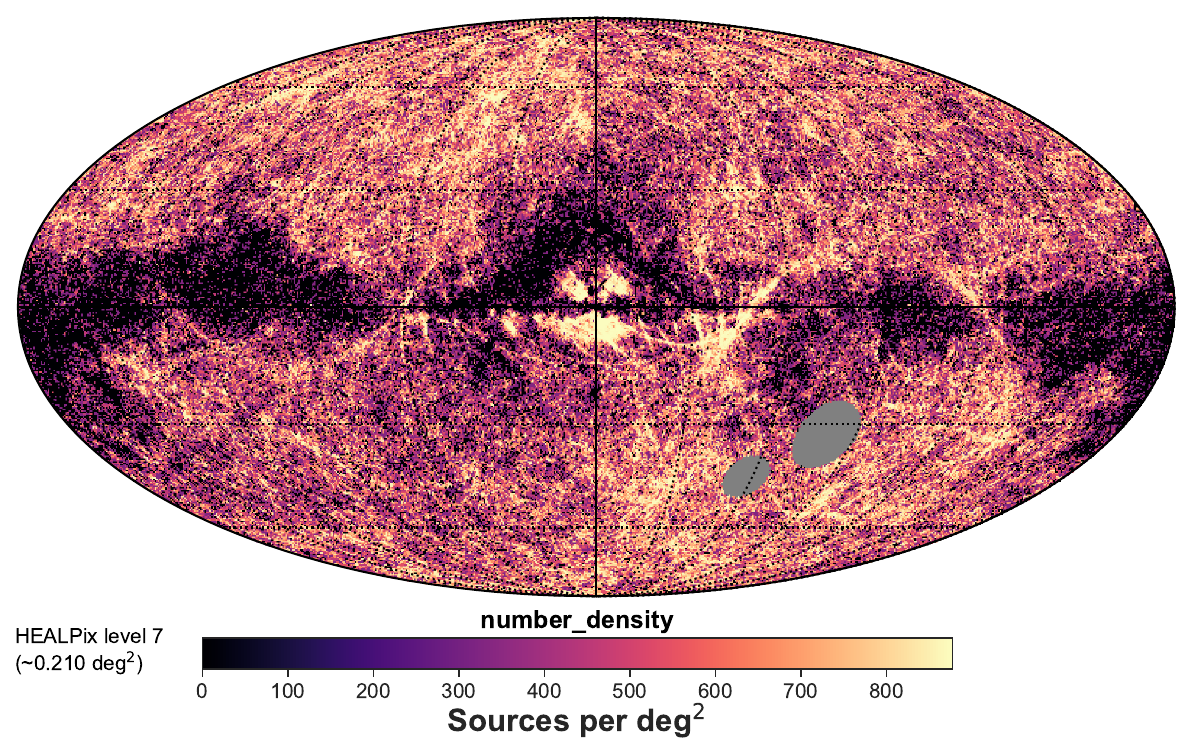}
	\caption{}
	\end{subfigure}
	\hspace{0.06cm}
	\begin{subfigure}{0.3333\textwidth}
         \centering
	\includegraphics[scale=0.30]{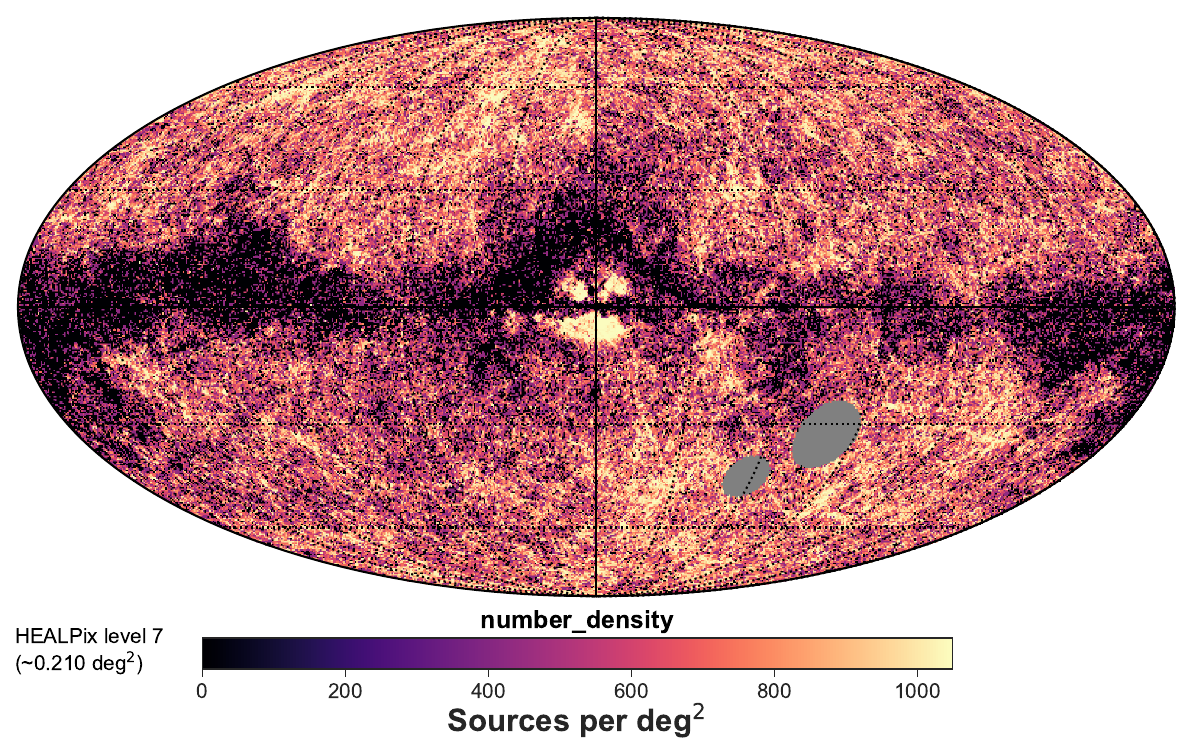}
	\caption{}
	\end{subfigure}
	
\caption{Galactic sky distribution of sources at magnitudes $G$<20.5 classified from maximum probabilities 
	by DSC combined classifiers using the global prior at HEALpixel level 7 in Mollweide projection. 
	\textit{Top}: quasars. \textit{Bottom}: galaxies.
	The colour map uses bright colours for high-density regions, while darker colours refer to regions with fewer observations.
	The LMC and SMC regions are masked in grey.
	   \texttt{Combmod} identifies \num{1568861} quasars $(a)$ and \num{1114782} galaxies $(d)$. 
	   \texttt{Combmod-$\alpha$} identifies \num{1313150} quasars $(b)$ and \num{608781} galaxies $(e)$.
	   \texttt{Combmod-Supp} identifies \num{1400549} quasars $(c)$ and \num{550413} galaxies $(f)$. 
	}
\label{fig:perfo_metrics_fullrun_205mag}
\end{figure*} 

\begin{figure*}[htp!]
	\begin{subfigure}{0.3333\textwidth}
         \centering
	 \includegraphics[scale=0.30]{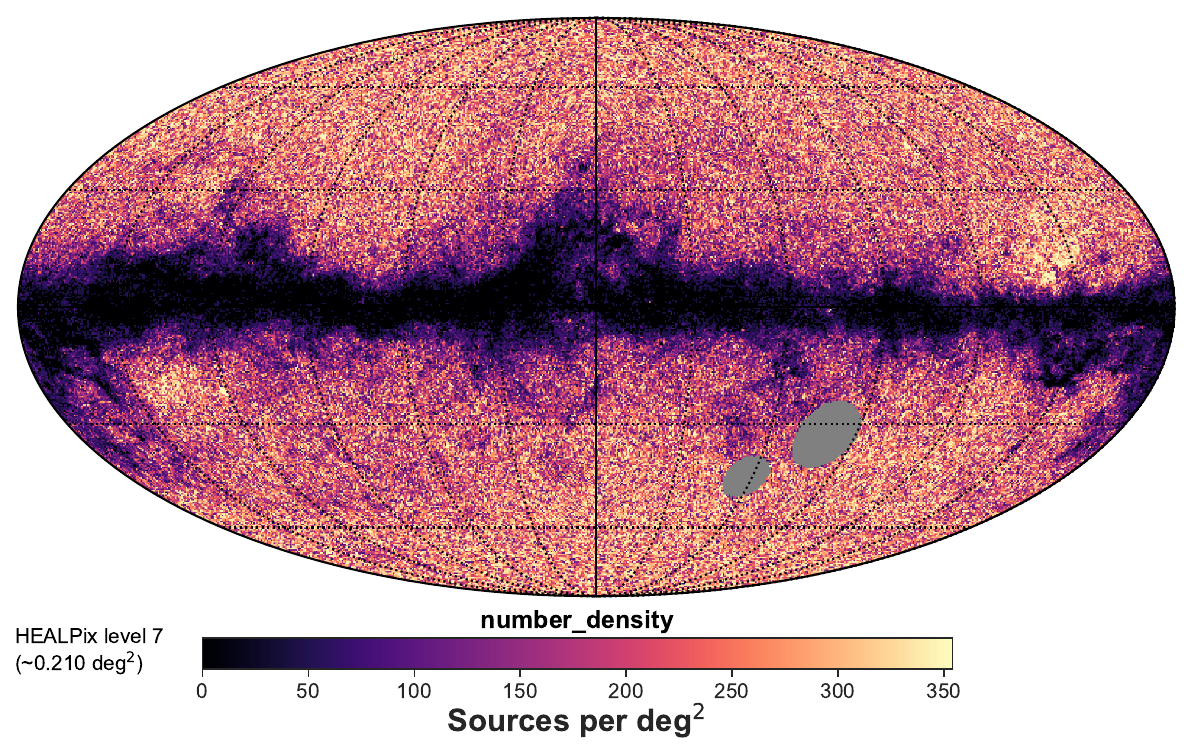}
	\caption{}
    	\end{subfigure}
	\hspace{0.06cm}
	\begin{subfigure}{0.3333\textwidth}
         \centering
	 \includegraphics[scale=0.30]{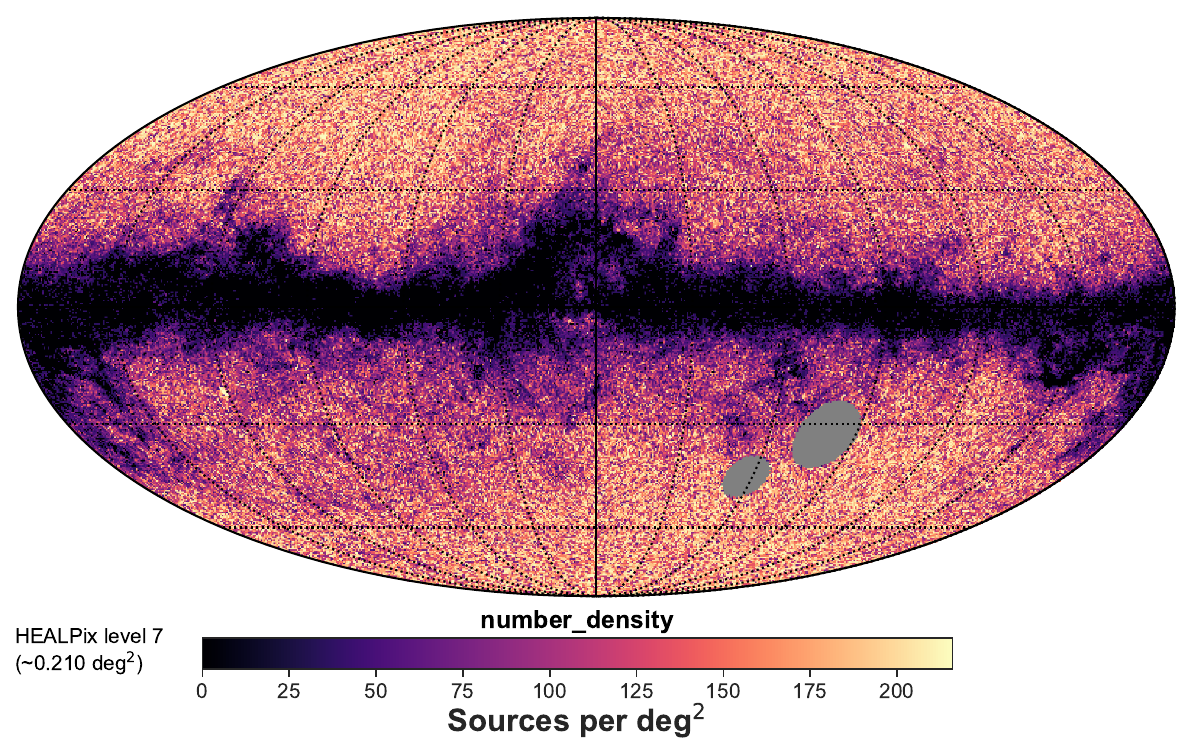}
	\caption{}
    	\end{subfigure}
	\hspace{0.06cm}
	\begin{subfigure}{0.3333\textwidth}
         \centering
	 \includegraphics[scale=0.30]{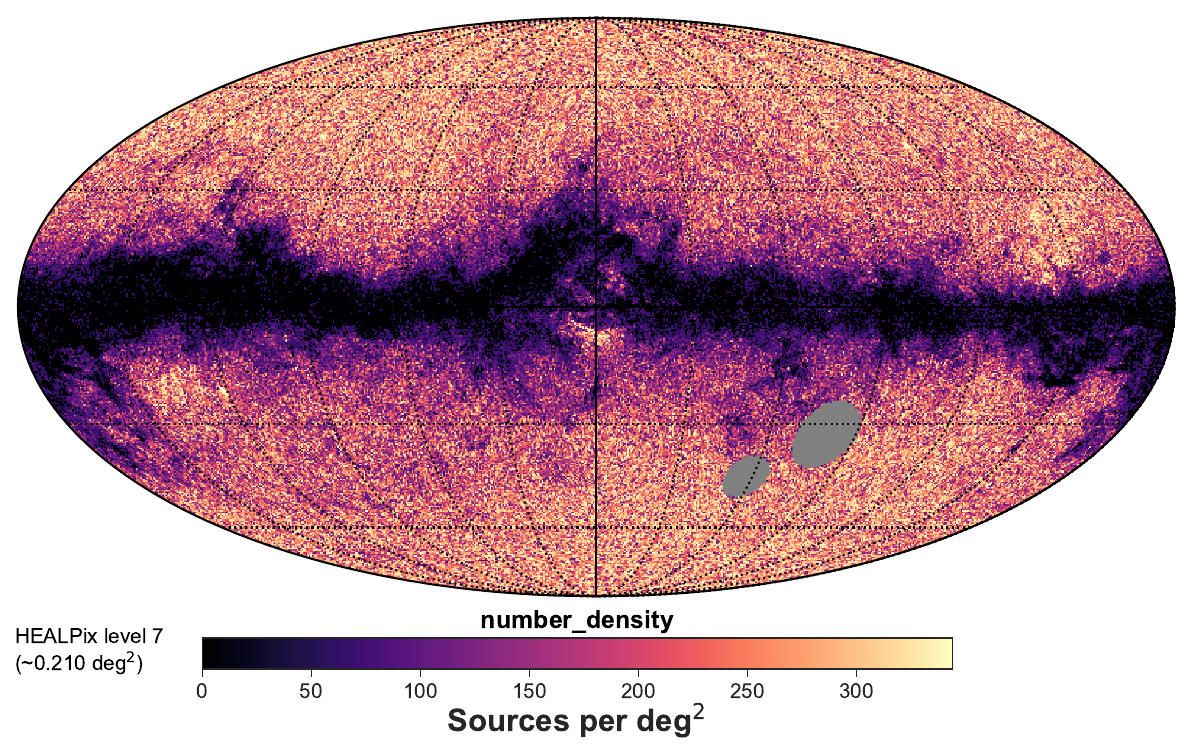}
	\caption{}
    	\end{subfigure}
	
	\begin{subfigure}{0.3333\textwidth}
         \centering
	\includegraphics[scale=0.30]{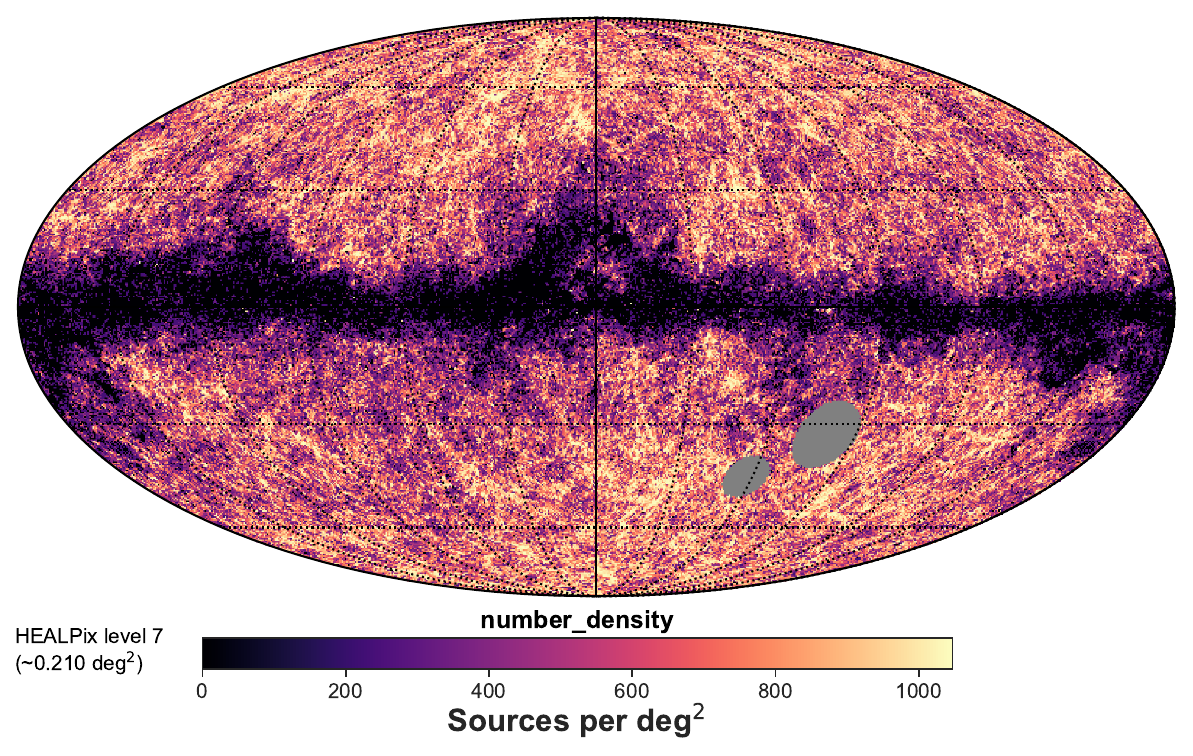}
	\caption{}
    	\end{subfigure}
	\hspace{0.06cm}
	\begin{subfigure}{0.3333\textwidth}
         \centering
	 \includegraphics[scale=0.30]{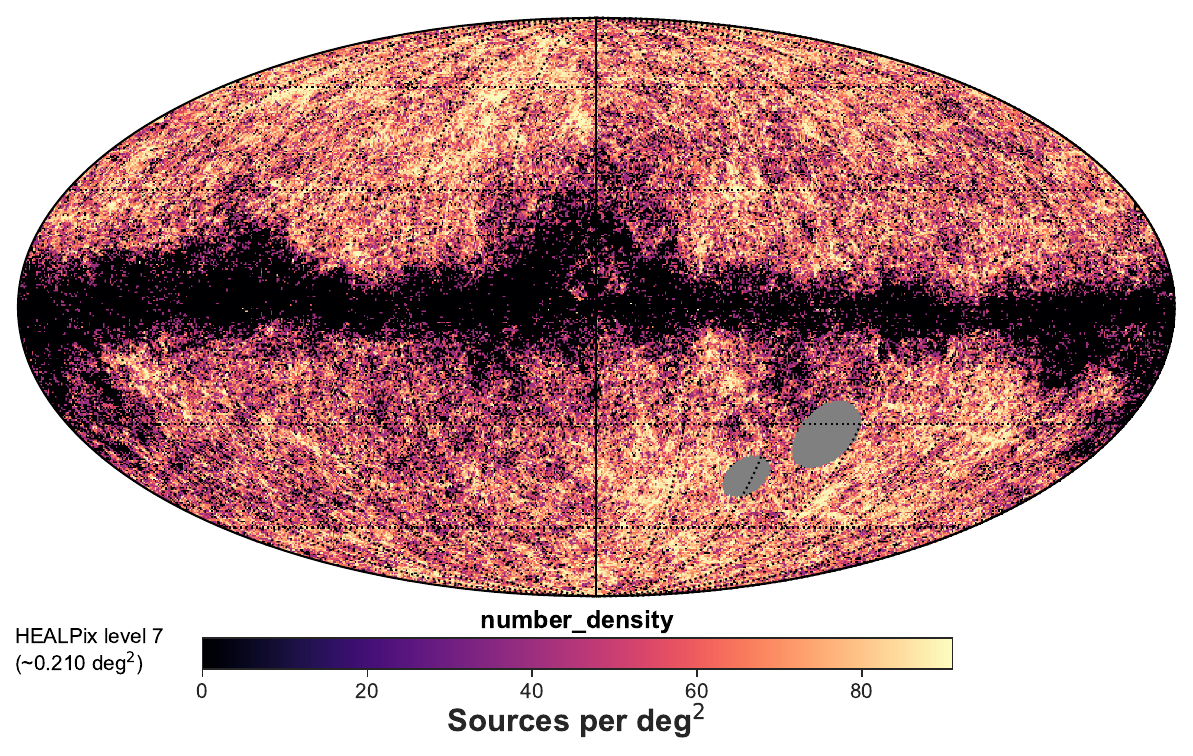}
	\caption{}
    	\end{subfigure}
	\hspace{0.06cm}
	\begin{subfigure}{0.3333\textwidth}
         \centering
	\includegraphics[scale=0.30]{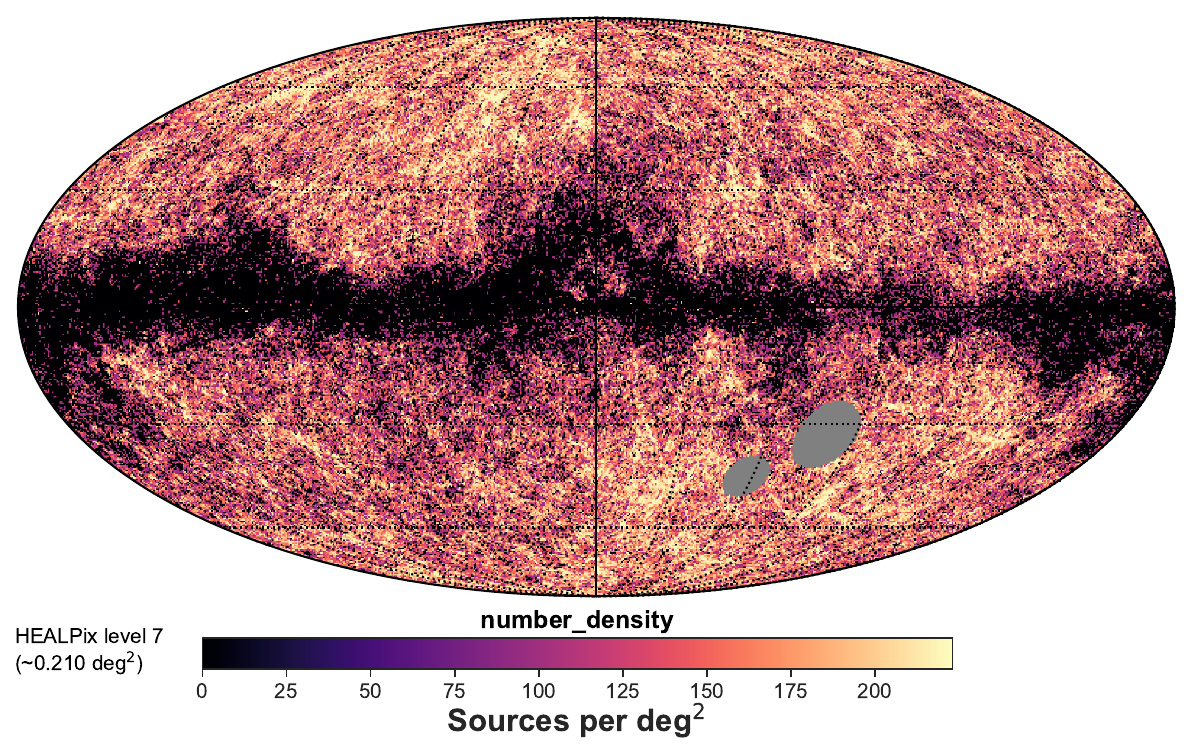}
	\caption{}
	\end{subfigure}
		
\caption{
	Same as Figure \ref{fig:perfo_metrics_fullrun_205mag}, but with a quality cut imposed 
	on the number of observations contributing to the integrated photometry in BP and RP 
	\texttt{phot\_[bp,rp]\_n\_obs}$\geq$10.
	\textit{Top}: quasars. \textit{Bottom}: galaxies.
	\texttt{Combmod} identifies \num{1513811} quasars $(a)$ and \num{1031405} galaxies $(d)$.
	\texttt{Combmod-$\alpha$} identifies \num{1268369} quasars $(b)$ and \num{526707} galaxies $(e)$.
	\texttt{Combmod-Supp} identifies \num{1294542} quasars $(c)$ and \num{494717} galaxies $(f)$. 
	}
\label{fig:perfo_metrics_fullrun_205mag_qualitycut10}
\end{figure*} 

\begin{figure*}[htp!]
	 \begin{subfigure}{1\textwidth} \centering
         \includegraphics[scale=0.33]{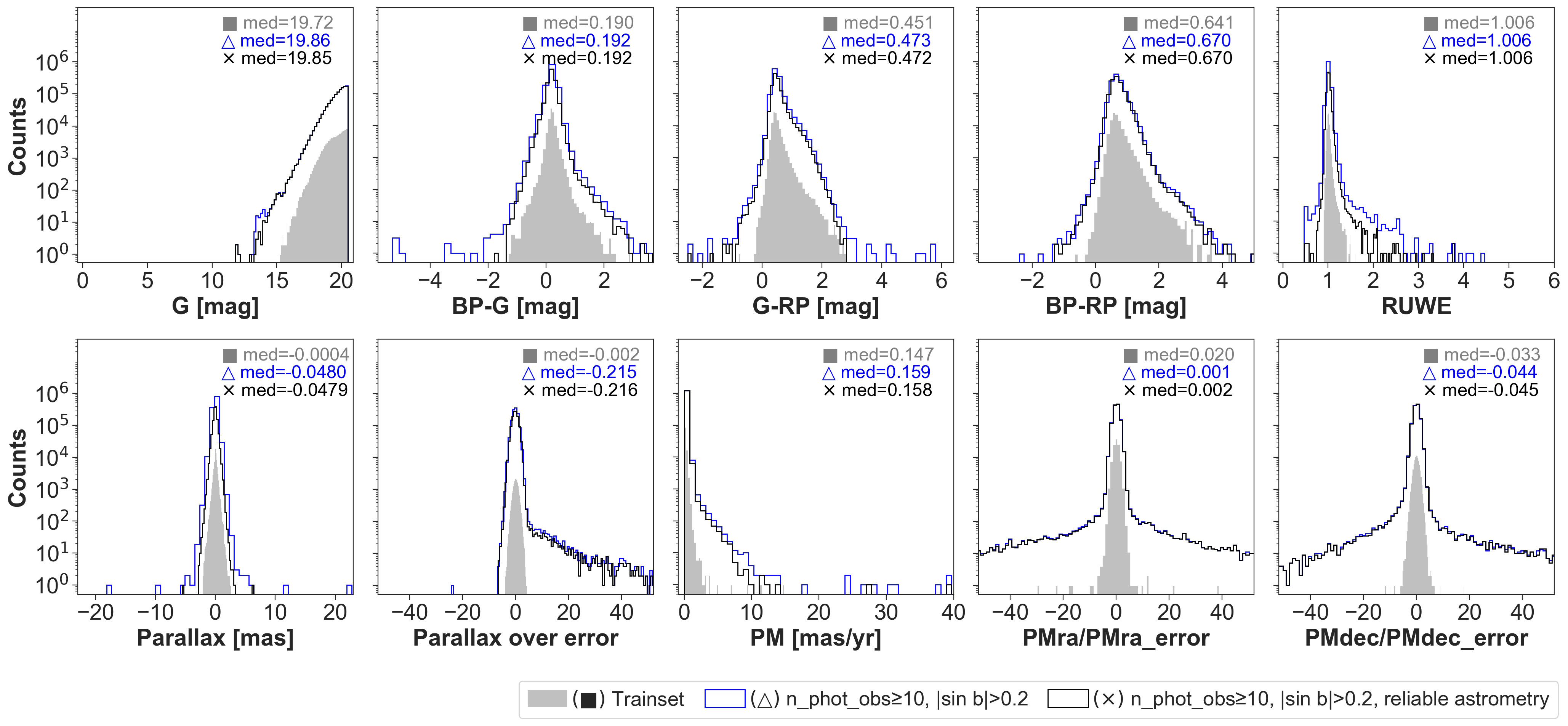}
         \caption{}
    	\end{subfigure}
	
	\begin{subfigure}{1\textwidth}
         \centering
         \includegraphics[scale=0.33]{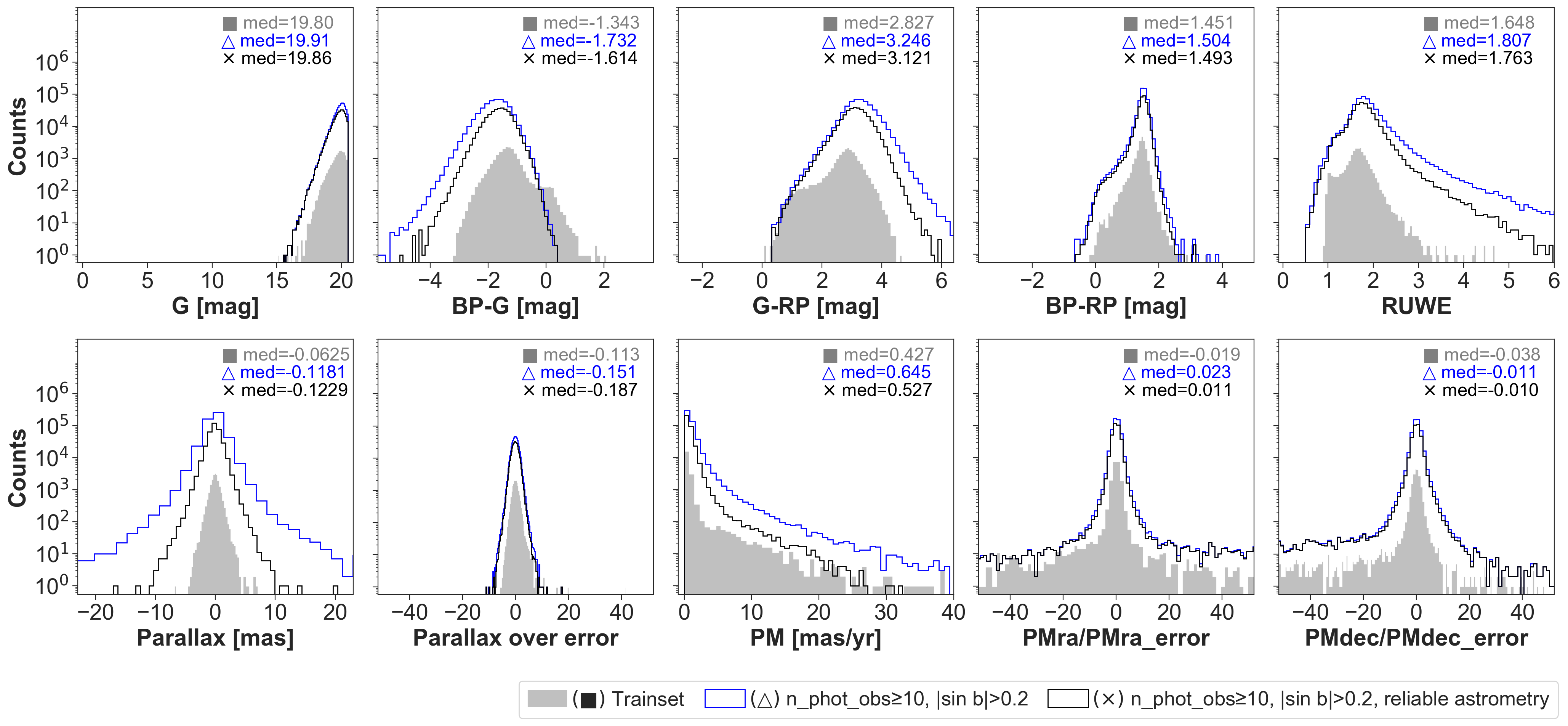}
         \caption{}
    	\end{subfigure}
	\caption{
		Feature distributions of extragalactic candidates identified in the \textit{Gaia} only mode by \texttt{Combmod-$\alpha$} 
		using the global prior, at magnitudes $G$<20.5 and higher latitudes $|\sin b|$>0.20.
		Sources located in the LMC and SMC are excluded.
		$(a)$ Quasar candidates. $(b)$ Galaxy candidates.  
		The quality cut is applied to \textit{Gaia} photometry and astrometry.
		The photometric quality cut requires a minimum of ten photometric observations in the BP- and RP- bands (\texttt{phot\_[bp,rp]\_n\_obs}$\geq$10). 
		The astrometric quality cut uses a Boolean flag, \texttt{unreliable\_astrometry}, in the GDR4 table \texttt{all\_source\_astrometry}. 
		The flag is set to \texttt{'false'} to indicate reliable astrometric measurements.
		The training dataset of the \textit{Gaia} only mode is shown in grey and is associated with the square symbol $\blacksquare$.
		The list of candidates after the photometric quality cut is displayed in blue and is associated with the triangle symbol $\bigtriangleup$.
		Lastly, the list of candidates after the photometric and astrometric quality cut is displayed in black and is associated with the cross symbol $\times$.
		The median value for each dataset is indicated.
		The RUWE stands for Renormalised Unit Weight Error.
		PMra and PMdec refer to the proper motions in the right ascension and declination, respectively.
		In $(a)$, the classifier identifies as quasars, \num{1215830} sources ($\bigtriangleup$) and \num{1207285} sources ($\times$).
		In $(b)$, the classifier identifies as galaxies, \num{502209} sources ($\bigtriangleup$) and \num{344184} sources ($\times$). 
	}
	\label{fig:perfo_fullrun_distribution_features_combmodalpha_0-G-20.5_Pmax}
\end{figure*}

	\subsection{Application of DSC to all 2.8 billion sources in GDR4} 
 
		\subsubsection{Quasar and galaxy candidates}\label{subsubsec:qso_gal_gdr4}
We apply DSC to the entire \textit{Gaia} data and report the anticipated classification statistics for GDR4.
Table \ref{tab:ref_counts_predictions_fullruns_maglims} provides the total number of quasar and galaxy candidates, excluding sources in the Magellanic Clouds. 
Recall that we assign the class that achieves the highest probability, no matter how large or small that probability is.
Candidates located in the LMC and SMC comprise about 2\% to 6\% of the total extragalactic predictions, except for \texttt{Specmod} and \texttt{Combmod}, which account for 15\% and 9\% of extragalactic predictions, respectively.
We expect most classifications in dense regions to be false detections.
Therefore, we have excluded the LMC and SMC from our assessment.

Figure \ref{fig:perfo_metrics_fullrun_205mag} shows the Galactic sky distributions of the quasar and galaxy candidates of the combined classifiers, namely \texttt{Combmod}, \texttt{Combmod-$\alpha$} and \texttt{Combmod-Supp}.
Due to stellar contamination, we expect to see Galactic features in the extragalactic maps.
If the contamination rate were constant across the sky, the pattern of the Galaxy would be visible in the number of contaminants (i.e. stars), even when added to a flat distribution of true detections.
The sky distributions show a pattern in latitude that follows regions of high dust extinction and high stellar density, as observed by \textit{Gaia}.
In particular, we observe a depletion of candidates along the Galactic plane far from the bulge but a higher density in the bulge due to stellar contamination.
False extragalactic detections result from a limitation of the classifier or from a reduced quality of the astrometry and photometry.
Far from the Galactic plane, the overall sky distribution of quasars appears smoother, with no peculiar aggregates in distinct regions. 
For the galaxies, however, a clustering pattern is noticeable across the whole sky. 
This pattern is not new and can be observed in the GDR3 results \citep{delchambre_gaia_2023, jamal_improved_2024}.

There are regions of the sky that have been observed only a few times by \textit{Gaia}, resulting in few photometric observations or transits and noisy measurements.
To filter out peculiar patterns, we impose a quality cut on the number of observations contributing to the integrated photometry in the BP and RP bands.
Setting a minimum number of photometric observations to \texttt{phot\_[bp,rp]\_n\_obs} =10 significantly reduces the densities at the bulge and the scanning law patterns, as shown in Figure \ref{fig:perfo_metrics_fullrun_205mag_qualitycut10}.
The Appendix Figures \ref{fig:perfo_skyplots_fullrun_magbins_combmodalpha_quasars} to \ref{fig:perfo_skyplots_fullrun_magbins_combmodalpha_galaxies} show the Galactic sky distributions of the quasar and galaxy candidates at different magnitude limits for \texttt{Combmod-$\alpha$}.
The pattern from the scanning law, which is detectable at all magnitude limits, is reduced by the quality cut. 
A large number of sources are filtered at fainter magnitudes $G$$\geq$20.5. 
The density at the bulge is also filtered out when a higher threshold on the number of photometric observations is imposed.
At magnitudes fainter than $G$=21, the quality cut rejects most of the noise, but the predictions are still affected by the scanning law and crowding near the Galactic plane.
The DSC results in the GDR4 astrophysical parameters tables (\texttt{ap\_class} and \texttt{ap\_class\_ir}) are not subject to the quality cut, enabling users to apply custom filters when querying the \textit{Gaia} catalogue. 
However, as a conservative approach, a quality cut of \texttt{phot\_[bp,rp]\_n\_obs}$\geq$6 is applied to the DSC \texttt{Combmod} results reported in the GDR4 main source catalogue (\texttt{gaia\_source}).

\begin{figure*}[htp!]
	\vspace{-.5cm}
	 \begin{subfigure}{1\textwidth} \centering
        \includegraphics[scale=0.746]{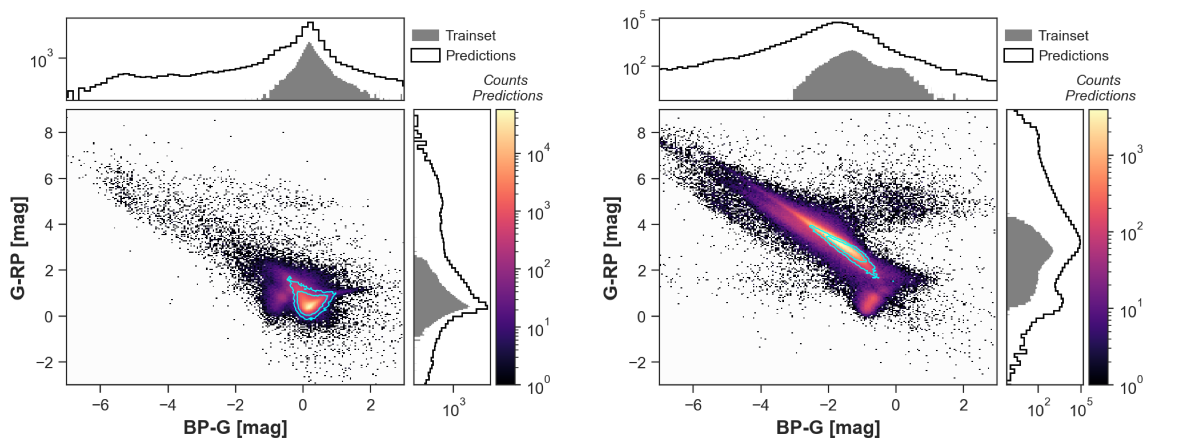}
	\vspace{-0.5cm}\caption{}
    	\end{subfigure}
	
	 \begin{subfigure}{1\textwidth} \centering
         \includegraphics[scale=0.746]{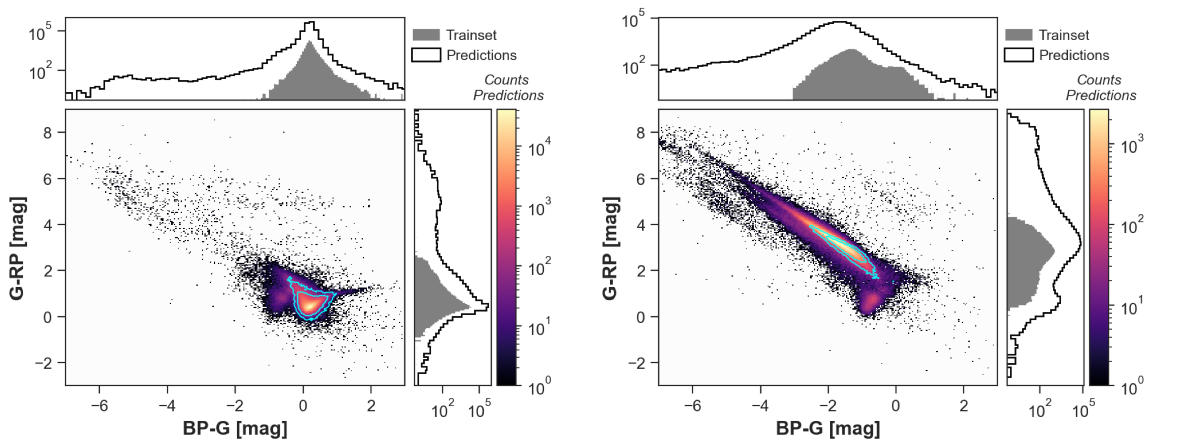}
	\vspace{-0.5cm}\caption{}
    	\end{subfigure}
	
	 \begin{subfigure}{1\textwidth} \centering
         \includegraphics[scale=0.746]{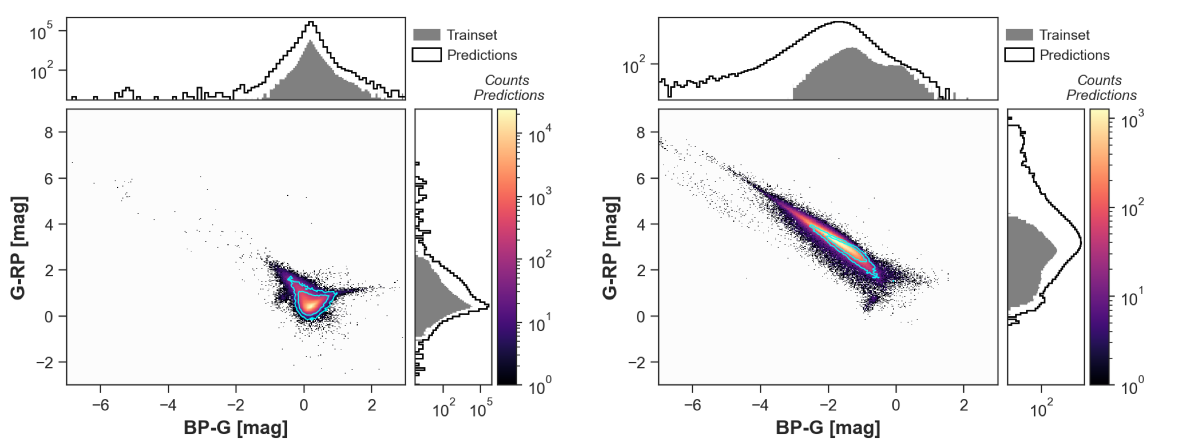}
	\vspace{-0.5cm}\caption{}
    	\end{subfigure}
	
	 \begin{subfigure}{1\textwidth} \centering
         \includegraphics[scale=0.746]{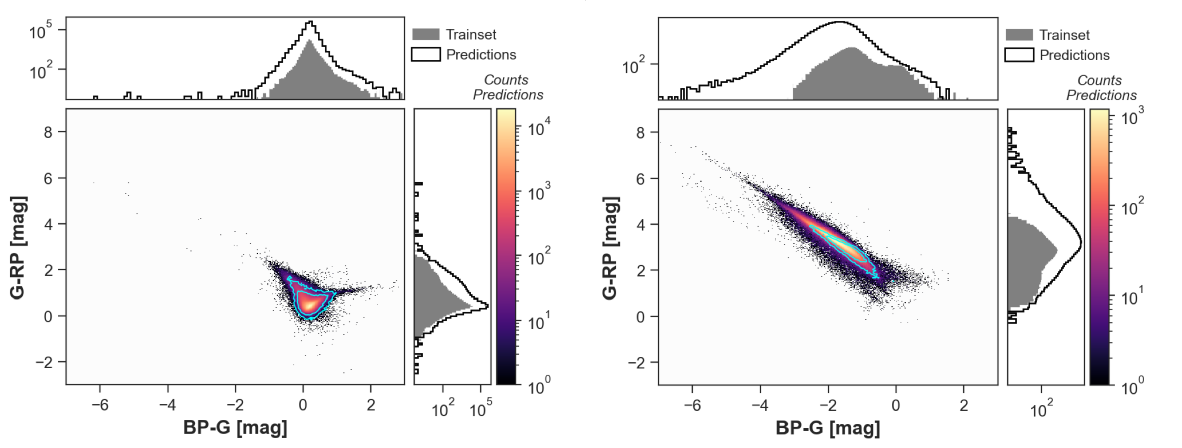}
	\vspace{-0.5cm}\caption{}
    	\end{subfigure}
	\caption{
		Colour-colour distributions of sources classified by  \texttt{Combmod-$\alpha$} 
		using the global prior at magnitudes of $G$<20.5. 
		\textit{Left}: quasars. \textit{Right}: galaxies.
		Sources located in the LMC and SMC are excluded.
		Contours (cyan lines) show the normalised density on a log scale of the highest-density regions for the trainset.
		(a) \num{1313150} quasars and \num{608781} galaxies identified across the whole sky.
		(b) \num{1234753} quasars and \num{531634} galaxies identified at higher latitudes $|\sin b|$>0.20.
		(c) \num{1218707} quasars and \num{504906} galaxies identified at higher latitudes $|\sin b|$>0.20 and \texttt{phot\_[bp,rp]\_n\_obs}$\geq$6.
		(d) \num{1215830} quasars and \num{502209} galaxies identified at higher latitudes $|\sin b|$>0.20 and \texttt{phot\_[bp,rp]\_n\_obs}$\geq$10.
		}
	\label{fig:perfo_fullrun_distribution_combmodalpha}
\end{figure*}

Figures \ref{fig:perfo_fullrun_distribution_features_combmodalpha_0-G-20.5_Pmax} and \ref{fig:perfo_fullrun_distribution_combmodalpha} show the feature distributions of the quasar and galaxy candidates identified by the combined classifier \texttt{Combmod-$\alpha$}, at magnitudes $G$<20.5. 
The feature distribution is overlaid with our training data. 
We require a minimal number of photometric observations in the BP- and RP- bands and also use the Boolean flag \texttt{unreliable\_astrometry} provided in GDR4, set to \texttt{'false'} to indicate reliable AGIS (Astrometric Global Iterative Solution) processing. 
The filtered set of candidates exhibits properties similar to those of the spectroscopically confirmed quasars and galaxies in our SDSS DR17 training set. 
However, we observe deviations from the training set in the signal-to-noise ratios of the proper motions and parallaxes (proper motions over errors and parallaxes over errors). 
Compared to the DSC trainset, the excess of negative values in $G_{BP}$-$G$ and positive values in $G$-$G_{RP}$ indicate the presence of extended sources DSC identifies as galaxies.
A deviation from our training data can be attributed to either a broader range of sources beyond our controlled training sample or contamination.
As illustrated in Appendix Figure \ref{fig:perfo_fullrun_distribution_combmodalpha_pmax}, applying a higher threshold to the probabilities produces cleaner samples.
The filtering is evident in Appendix Figure \ref{fig:perfo_fullrun_distribution_features_combmodalpha_0-G-20.5_Pmax0.9} showing reduced scatter in the astrometry, especially for quasars.
We note, however, that even after implementing a quality cut and applying a higher threshold to the probabilities, the galaxy candidates still exhibit an excess of negative values in $G_{BP}$-$G$ and positive values in $G$-$G_{RP}$ compared to our training sample of galaxies, which indicates these extended sources as reliable candidates. 
The difference with the DSC trainset may be due to a selection bias in the galaxies identified in SDSS.

	\subsubsection{Extragalactic candidates identified using \textit{Gaia} and CatWISE} 
In \textit{Gaia}-IR mode, DSC provides classifications for \textit{Gaia} sources crossmatched to the CatWISE2020 catalogue. 
Table \ref{tab:ref_counts_predictions_fullruns_maglims} reports the total number of quasar and galaxy candidates identified by DSC, excluding sources in the LMC and SMC.
As explained in Section \ref{subsubsec:qso_gal_gdr4}, contamination is reduced by implementing quality cuts and excluding sources located at the Galactic plane.
Appendix Figure \ref{fig:perfo_fullrun_distribution_ir_combmodalpha} shows the candidates identified exclusively in the \textit{Gaia}-IR.
At magnitudes brighter than $G$=21, the colour-colour distribution of the identified quasars and galaxies is similar to that of the extragalactic selection of the training set.
However, at magnitudes fainter than $G$=21, the colour-colour distribution reveals redder sources that are most likely galaxies misidentified as quasars. 
This result may be due to misclassifications between AGN-type quasars and redder galaxies.
Nevertheless, this confusion only affects a few sources to such a magnitude.
The level of stellar contamination introduced by the addition of infrared photometry for classification is low and can be evident in the deviation from the training set and along the stellar locus in the colour-colour plane at magnitudes $G$<20.5 (panel $(a)$ in Appendix Figure \ref{fig:perfo_fullrun_distribution_ir_combmodalpha}).
This result can also be seen in the feature distributions in Appendix Figure \ref{fig:perfo_fullrun_distribution_features_ir_combmodalpha_0-G-20.5}, where a fraction of sources likely erroneously identified as quasar candidates displays bright magnitudes ($G$<12) and higher proper motions.
       
	\subsection{Recommended selections and caveats}
In this section, we summarise quality cuts recommended to improve the selection of quasar and galaxy candidates selected by DSC.
To reject peculiar patterns due to the scanning law and crowding close to the Galactic bulge, we recommend setting the minimum number of photometric observations to \texttt{phot\_[bp,rp]\_n\_obs}$\geq$10 (or 6 for a less restrictive cut).
Further rejection of stellar contaminants in the Galactic plane can be achieved by selecting only sources at higher Galactic latitudes $|\sin b|$>0.20.
In this work, we have been allocating classes based simply on the maximum probability. A user could instead allocate a class only if the maximum probability is above some threshold. This approach removes contaminants at the expense of missing true detections. 
We suggest a probability threshold of $P$>0.50 for a less restrictive cut, while a threshold of $P$>0.90 provides a stricter cut.
When available, we also recommend using any reliability flags provided in the GDR4 tables\footnote{\url{https://www.cosmos.esa.int/web/gaia/dr4}} on the \textit{Gaia} astrometry and the photometry (\texttt{all\_source\_astrometry} and \texttt{all\_source\_photometry}). The astrometric reliability flag rejects sources with spurious astrometric measurements that show large negative parallaxes. 
A technical note on DSC will show how the astrometric flag correction improves the classifications (\citeauthor{jamal_dsc_tn_2026} in prep.).

Finally, DSC processes all sources with valid BP and RP spectra across all magnitudes.
However, classifications at magnitudes fainter than $G$=21 are highly contaminated by stellar objects due to increased noise in the data.
In the bright regime of magnitudes $G$<21, stellar contamination is expected, particularly for DSC models trained on the \textit{Gaia} XP spectra, \texttt{Specmod} and \texttt{Specmod-Supp}, as well as the combined classifiers, \texttt{Combmod} and \texttt{Combmod-Supp}.
We recommend setting the magnitude range to exclude fainter magnitudes $G$>21.

\section{Conclusions} \label{sec:5_conclusions}
We have improved purity at the expense of a moderate loss of completeness for the extragalactic classes identified by DSC in GDR4 compared to GDR3.
The best improvement in completeness has been achieved for quasars by the MLP \texttt{Specmod} classifier with an increase of 29 percentage points (pp) compared to GDR3 Extratrees \texttt{Specmod}, while the other classifiers have shown a moderate gain of only 2 to 5 pp. 
However, the completeness of galaxies has decreased, indicating greater difficulty in identifying extended sources.
Nevertheless, the largest increase in purity has reached $\sim$25 pp for both quasars and galaxies.
When re-evaluated as a function of magnitude and sky position, the best DSC models have achieved an average completeness of 87-99\% and a purity of 96-98\% for the extragalactic classes, at magnitudes brighter than $G$=20. At magnitudes fainter than $G$=20, the performance drops, with an even greater reduction at magnitudes fainter than $G$=20.5.
When comparing all DSC models, the \texttt{Combmod} classifier prioritises completeness, unlike \texttt{Combmod-$\alpha$} and \texttt{Specmod-Supp}, which improve purity, particularly at $G$$\geq$20.
We emphasise that DSC outputs are not calibrated probabilities, and the underlying assumptions of the combined models may lead to overconfidence. But we have shown that thresholds on the DSC outputs can be used effectively to identify classes.

For GDR4, we have also trained independent models that combine \textit{Gaia} optical data with mid-infrared photometry from the CatWISE2020 catalogue for classification. 
This combination has been shown to improve the completeness of extragalactic data, particularly for quasars, albeit with a slight reduction in purity, primarily at fainter magnitudes.
Compared to models trained on \textit{Gaia} data alone, including infrared data has increased quasar completeness by up to 29 percentage points at magnitudes fainter than $G$=20.5.

Overall, the quasar and galaxy candidates identified in GDR4 by DSC exhibit physical properties close to those in our controlled training set. 
Any deviations from the training set are either due to noisy measurements or to classifier limitations that lead to false stellar detections.
DSC candidate selection has improved when selecting sources at higher Galactic latitudes, far from the Galactic plane. 
Further improvements have been proposed through applying quality cuts to \textit{Gaia} photometry and astrometry. 
These quality cuts have been shown to effectively reduce stellar contamination and correct for the \textit{Gaia} scanning law pattern.
In the \textit{Gaia} IR mode, we have assessed the detections of extragalactic candidates, exclusively identified by DSC when combining \textit{Gaia} data with infrared photometry from CatWISE, as valid candidates despite the moderate increase of stellar contamination.

\begin{acknowledgements} 
This work has used data from the European Space Agency (ESA) mission {\it Gaia} (\url{https://www.cosmos.esa.int/gaia}), processed by the {\it Gaia} Data Processing and Analysis Consortium (DPAC, \url{https://www.cosmos.esa.int/web/gaia/dpac/consortium}). 
Funding for the DPAC has been provided by national institutions, in particular the institutions participating in the {\it Gaia} Multilateral Agreement.
This publication also uses data products from the Wide-field Infrared Survey Explorer (WISE), a joint project of the University of California, Los Angeles, and the Jet Propulsion Laboratory/California Institute of Technology, funded by the National Aeronautics and Space Administration.
The authors would like to thank the referee for their insightful comments and suggestions, which helped improve the quality of this manuscript.
The authors would also like to thank Morgan Fouesneau and Rene Andrae for useful discussions, and Antonella Vallenari, Anthony Brown, David Teyssier, and Johannes Sahlmann for reviewing the manuscript, as well as the technical staff and collaborators at DPAC, CNES (Centre National d'Études Spatiales) and ESA.
SJ contributed to the development and implementation of the models, technical and scientific validation, documentation, analysis of the results and wrote the draft of the manuscript.
CBJ contributed to the project definition, method design, results analysis, and manuscript editing.
OC contributed to the overall coordination of the DPAC CU8 Apsis data model, documentation, and interface with the DPCC (CNES Data Processing Centre), as well as to the technical and scientific validation and revision of the manuscript.
RC contributed to the revision of the manuscript and the analysis of the results in relation to the presence of different extragalactic populations.
This work was funded in part by the DLR (German space agency) via grant 50 QG 2102.

\end{acknowledgements}

\bibliographystyle{aa}  
\bibliography{bibliography_dsc_results.bib}  

\begin{appendix} 

\begin{minipage}{0.95\textwidth}

\section{List of features from the \textit{Gaia} astrometry and photometry used to train  DSC}
\label{sec:list_features_classifier}
This section provides a summary of the features used to train \texttt{Allosmod} and \texttt{Specmod-Supp}. 
We provide the field names as they will appear in the GDR4.
Infrared photometry features are only used in \textit{Gaia}-IR  mode.\\
\begin{multicols}{2}{
\small
\noindent  Astrometric features 
\begin{itemize} \setlength{\parskip}{0pt}
    \item Sine of the Galactic latitude, $\sin {b}$
    \item parallax, \texttt{parallax}, and its error, \texttt{parallax\_error}
    \item total proper motion, \texttt{pm}
    \item proper motion in the right ascension direction, \texttt{pmra}, \\ and its error, \texttt{pmra\_error}
    \item proper motion in the declination direction, \texttt{pmdec}, \\ and its error, \texttt{pmdec\_error}
    \item renormalised unit weight error, \texttt{ruwe}
    \item unrescaled excess noise of the source, \\ \texttt{agis\_astrometric\_excess\_noise}, and its significance measuring the fit to a normal distribution $\mathcal{N}(0,1)$.
    \\
\end{itemize}

\noindent IPD (Image Parameter Determination) parameters 
\begin{itemize} \setlength{\parskip}{0pt}
    \item fraction of successful IPD windows with one peak and high GoF (Goodness of Fit), \texttt{ipd\_frac\_high\_gof}
    \item fraction of successful IPD windows with multiple peaks, \texttt{ipd\_frac\_multi\_peak}
    \item fraction of transits with truncated windows or multiple gates, \texttt{ipd\_gof\_frac\_odd\_win}
    \item amplitude of the IPD Gof versus the position angle of scan, \texttt{ipd\_gof\_harmonic\_amplitude}
    \item phase of the IPD Gof versus the position angle of scan, \texttt{ipd\_gof\_harmonic\_phase}
    \\
\end{itemize}

\noindent Photometric features 
\begin{itemize} \setlength{\parskip}{0pt}
    \item $G$ \ band magnitude, \texttt{phot\_g\_mean\_mag}
    \item colour $G_{BP}-G$, \texttt{bp\_g}
    \item colour $G-G_{RP}$, \texttt{g\_rp}
    \item colour $G_{BP}-G_{RP}$, \texttt{bp\_rp} 
    \item relative variability in the G-band (relvarg), \\
        $\sqrt{ \texttt{phot\_g\_n\_obs} / \texttt{phot\_g\_mean\_flux\_over\_error}}$
    \item error on the G-band mean flux,\\ \texttt{phot\_g\_mean\_flux\_error}
    \item error on the BP-band mean flux,\\ \texttt{phot\_bp\_mean\_flux\_error}
    \item error on the RP-band mean flux,\\ \texttt{phot\_rp\_mean\_flux\_error}
    \item BP/RP flux excess factor with the colour dependency removed. 
    The colour excess is computed from a ratio of the mean fluxes in the BP-, RP- and G-bands,
    $(\texttt{phot\_bp\_mean\_flux}+\texttt{phot\_rp\_mean\_flux})/$\\ $\texttt{phot\_g\_mean\_flux}$
    \\
\end{itemize}

\noindent Features combining mid-infrared photometry from CatWISE2020 and \textit{Gaia} data
\begin{itemize} \setlength{\parskip}{0pt}
    \item colour $G-W1$
    \item colour $W1-W2$
    \\
\end{itemize}
}\end{multicols}
\end{minipage}

\begin{minipage}{0.95\textwidth}
\section{Summary of the MLP architectures in DSC}\label{sec:mlp_dsc_summary}
This section summarises the hyperparameters used for the MLP models in DSC. 
Several models are trained with various hyperparameter configurations (number of layers, activation function, learning rate, and batch size), including different preprocessing and normalisation strategies.
When normalised, XP fluxes are divided by the total sum of the fluxes in BP and RP, similar to GDR3. When preprocessed, features from astrometry and photometry are standardised.
The best-performing models achieve the highest completeness and purity of the extragalactic classes.

\end{minipage}

\begin{table}[htp!]
\centering
\begin{minipage}{0.95\textwidth}
    \caption{MLP configuration of hyperparameters}               
    \label{table:tab_nn_keras_summary}  
    \centering \fontsize{8.5}{9}\selectfont
    \renewcommand{\arraystretch}{1.1}
    \begin{tabular}{lllll} 
        \hline \hline 
        Classifier & MLP architecture & $n$ & Selected & Common \\
         && & hyperparameters & hyperparameters \\
        
        \hline
        \texttt{Specmod}
            	& 1 \textit{Gaia} input layer (size: 220) & \num{38723} & Learning rate 10$^{-4}$& Optimizer: \texttt{Adam} \citep{kingma_adam_2017}  \\
           	 & 3 dense layers (units: 128, 64, 32)  && Batch size: 32 & Loss: categorical crossentropy\\
            	& 1 softmax layer (units: 3) && Activations: \texttt{SELU} & Maximum training epochs: 200\\

        \cline{1-4}
        \texttt{Allosmod}
            	& 1 \textit{Gaia} input layer (size: 23)  & \num{11523} & Learning rate 10$^{-4}$ & Validation data fraction : 20\% \\
            	& 2 dense layers (units: 128, 64) && Batch size: 32 & Early stopping on the validation loss\\
            	& 1 softmax layer (units: 3) && Activations: \texttt{RELU} & Random seed : 42 \\
            
        \cline{1-4}
        \texttt{Specmod-Supp}
            	& 1 \textit{Gaia} input layer (size: 220 + 23) & \num{41667} & Learning rate 10$^{-4}$ & \\
            	& 3 dense layers (units: 128, 64, 32)  && Batch size: 32 & \\
            	& 1 softmax layer (units: 3) && Activations: \texttt{SELU} &\\
            
        \cline{1-4}
        \texttt{Allosmod(*)}
        		& 1 \textit{Gaia} input layer (size: 23) & \num{11529} & Learning rate 10$^{-4}$ &\\
            	& 2 dense layers (units: 128, 64) && Batch size: 512 &\\
            	& 1 IR input layer (size: 2) && Activations: \texttt{SELU} &\\
            	& 1 concatenate layer,  &&&\\
            	& 1 softmax layer (units: 3) &&&\\

        \cline{1-4}
        \texttt{Specmod-Supp(*)}
        		& 1 \textit{Gaia} input layer (size: 220 + 8) & \num{39753} & Learning rate 10$^{-3}$&\\
            	& 3 dense layers (units: 128, 64, 32) && Batch size: 256 &\\
            	& 1 IR layer (size: 2)  && Activations: \texttt{SELU} &\\
            	& 1 concatenate layer  &&&\\
            	& 1 softmax layer (units: 3) &&&\\
        \hline
    \end{tabular}
    \tablefoot{
    	$n$ refers to the total number of hyperparameters in the model.
    	The symbol \textit{(*)} indicates the models trained with \textit{Gaia} data and infrared photometry from CatWISE.
    	\texttt{Specmod-Supp} is trained using the XP spectra (220 pixels) and the list of features (either 23 features or 8 features).
    	Open-source libraries include the \texttt{python} \texttt{keras} library \citep{chollet2015keras} and 
	the \texttt{java} \texttt{Eclipse} \texttt{deeplearning4j} library (\url{https://deeplearning4j.konduit.ai/}).
  	The best-performing models identified in the \textit{Gaia}-IR mode use preprocessed inputs, whereas those in the \textit{Gaia} only mode use the raw inputs.
    }
\end{minipage}
\vspace{-1cm}
\end{table}

\begin{figure}[htp!]
 \begin{minipage}{0.95\textwidth}
	\begin{subfigure}{1\textwidth}\centering
         \includegraphics[scale=0.41]{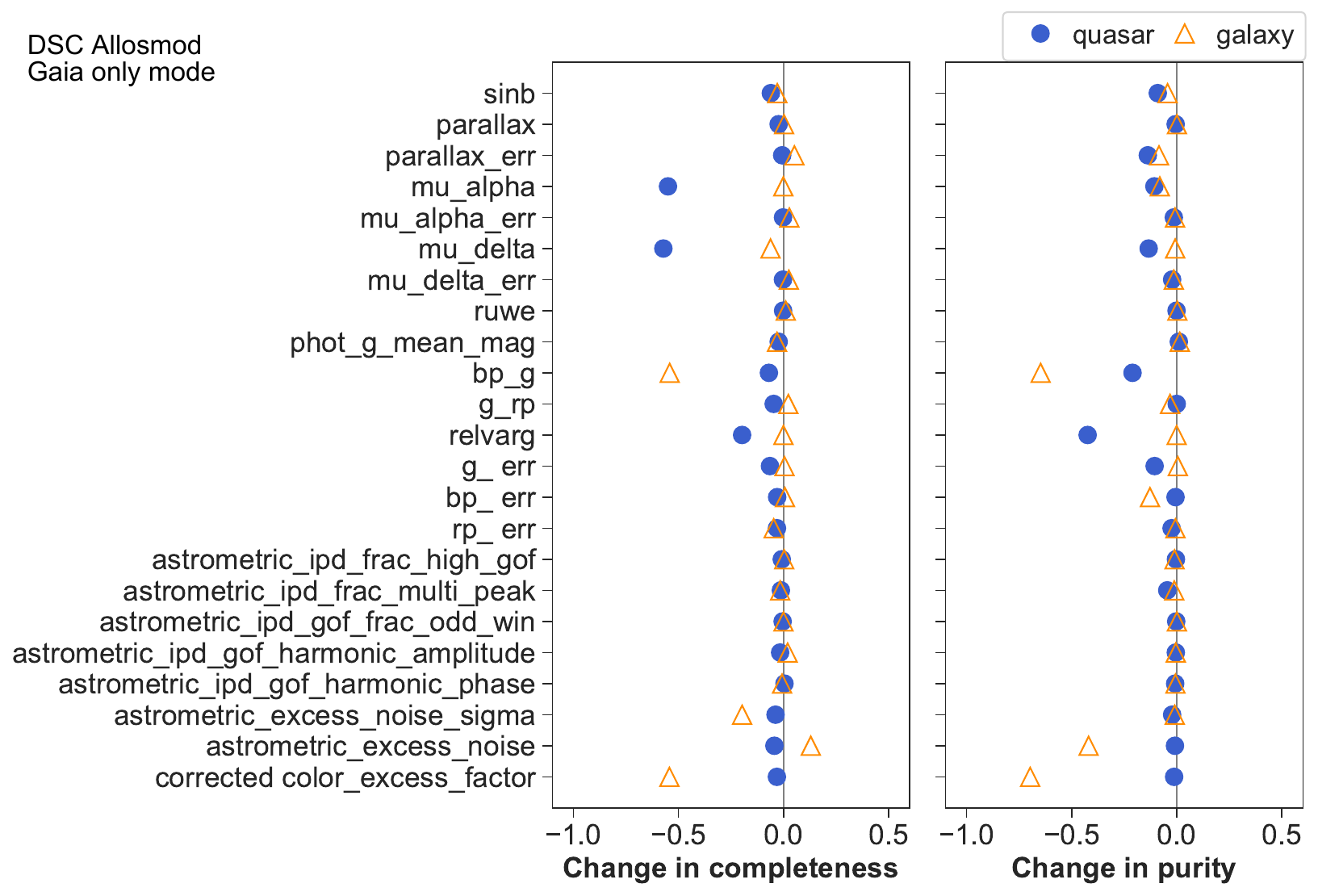}
	\vspace{-0.28cm} \caption{}
	\label{fig:features_importance_allosmod_gaiaonly}
	\end{subfigure}
	
	\begin{subfigure}{1\textwidth}
         \centering
         \includegraphics[scale=0.41]{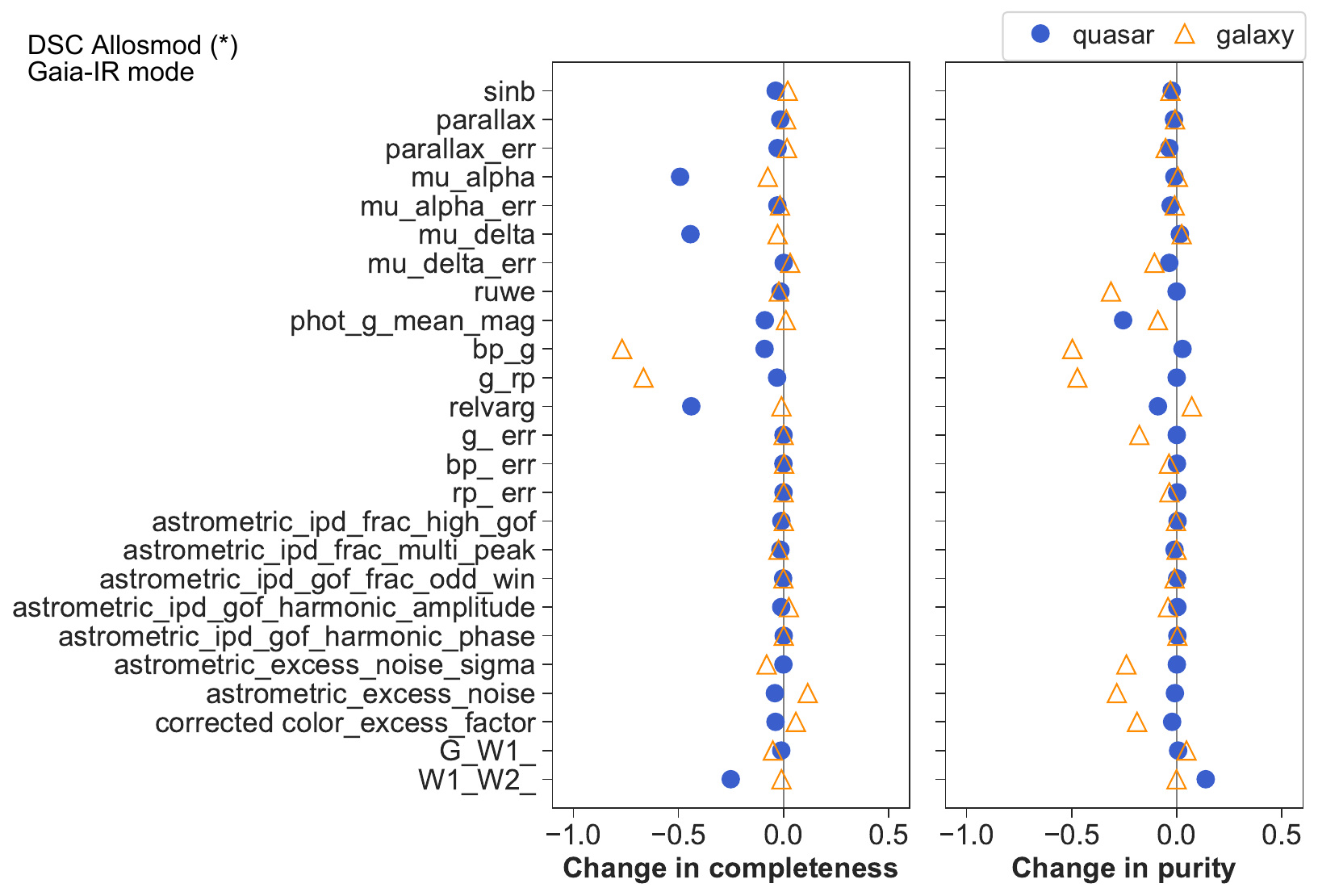}
	\vspace{-0.28cm} \caption{}
	\label{fig:features_importance_allosmod_gaiair}
	\end{subfigure}
	
	\caption{Feature importance for the DSC models trained using \textit{Gaia} astrometry and photometry. 
		The $x$-axis shows the variation of the completeness (left panel) and purity (right panel) of the extragalactic classes.
		A negative value indicates a loss in performance, identifying the feature as informative.
		The $y$-axis lists the input of the trained models.
		\textit{(a)} and \textit{(b)} show the change in performance for the \texttt{Allosmod} in the \textit{Gaia} only mode and \texttt{Allosmod(*)} in the \textit{Gaia}-IR mode, respectively.
		}
\end{minipage}
\vspace{-1cm}
\end{figure}

\newpage~\newpage
\begin{minipage}{0.95\textwidth}
\section{Features importance analysis of DSC \texttt{Allosmod} models} \label{appendix_feature_importance}

This section shows how the input features contribute to the classification performance of the trained DSC \texttt{Allosmod} model. 
We evaluate the importance of each input feature to each trained model by randomly permuting each input in turn. 
The change in the performance of the model on the test set gives a measure of the importance of the perturbed input without changing the model architecture or introducing unusual values of the input. By resampling each input across all the classes, we eliminate the information that a feature may carry about a specific class.\\
For every model, we run ten random permutations of each input in the test set and compute the completeness and purity of the extragalactic classes in the modified test set. The results, presented in Figures \ref{fig:features_importance_allosmod_gaiaonly}-\ref{fig:features_importance_allosmod_gaiair} for \texttt{Allosmod} and \texttt{Allosmod(*)}, show the change in performance, defined as the difference between the modified results (averaged across all permutations) and our initial results (refer to Table \ref{table:tab_overall_performances_testset}). 
A negative change indicates a loss in performance, and the larger the change, the larger the amount of information that feature appears to carry for that class. 
We see that the most informative features are proper motions, astrometric excess noise, the colour \texttt{bp\_g}, relative variability in the $G$ band, and the colour excess factor. 
In the \textit{Gaia}-IR mode, the infrared colour $W1$-$W2$ is also a key feature in quasars completeness.
The analysis for each BP/RP pixel input to Specmod (figure not shown) indicates that feature importance is distributed across a range of wavelengths, suggesting that it uses spectral lines at specific redshifts. \\
We explored various MLP configurations and selected the best-performing models with the highest completeness and purity for the extragalactic classes. The performance of models varies with the neural network architecture and its training. Therefore, the results discussed here are specific to the models under presentation. \\[-2em]
\end{minipage}
%

\newpage~\newpage
\begin{minipage}{0.95\textwidth}
\section{Quasar and galaxy candidates identified by DSC \texttt{Combmod-$\alpha$} with probabilities over a fixed threshold using \textit{Gaia} data}
This section shows how the quasar and galaxy candidates identified by the \texttt{Combmod-$\alpha$} classifier are distributed in colour-colour space, as a function of the threshold applied to the maximum posterior probability used for class assignment.
\end{minipage}

\begin{figure}[htp!]
\centering
 \begin{minipage}{1\textwidth}
	 \begin{subfigure}{1\textwidth}
         \vspace{-.2cm}
	 \centering
          \includegraphics[scale=0.80]{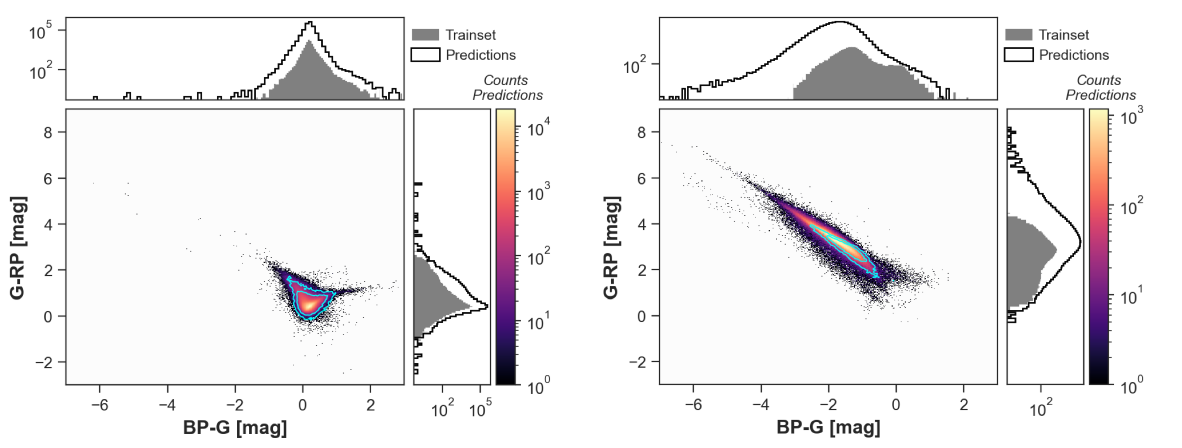}
	 \vspace{-0.15cm} \caption{}
   	 \end{subfigure}
	
	\begin{subfigure}{1\textwidth}
         \centering
         \includegraphics[scale=0.80]{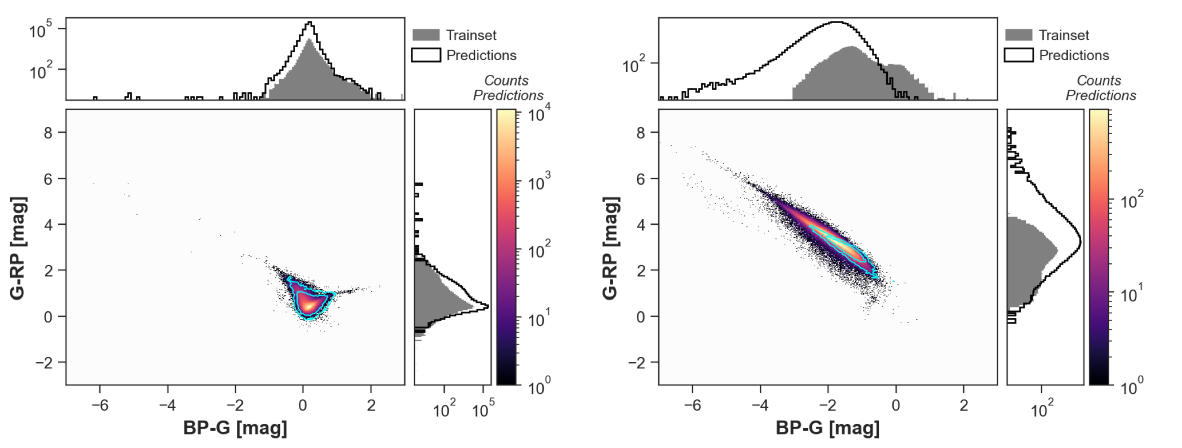}
	\vspace{-0.15cm} \caption{}
    	\end{subfigure}
	
	\begin{subfigure}{1\textwidth}
         \centering
         \includegraphics[scale=0.80]{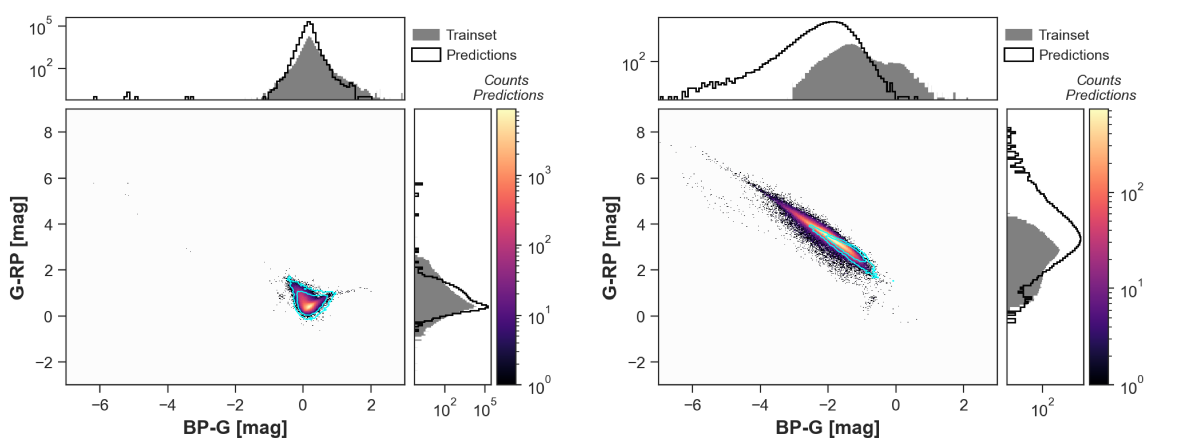}
	\vspace{-0.15cm} \caption{}
   	\end{subfigure}

	\caption{
		Same as Figure \ref{fig:perfo_fullrun_distribution_combmodalpha} showing quasar and galaxy candidates at magnitudes $G$<20.5.
		\textit{Left}: quasars. \textit{Right}: galaxies.
		Candidates are identified from the class with the highest probability above a fixed threshold. 
		The quality cut is applied for sources at higher latitudes $|\sin b|$>0.20 and \texttt{phot\_[bp,rp]\_n\_obs}$\geq$10.
		Contours (cyan lines) show the normalised density on a log scale of the highest-density regions for the trainset.
		$(a)$ \num{1212536} quasars and \num{501082} galaxies classified with a maximum probability of $P$>0.5.
		$(b)$ \num{881217} quasars and \num{396346} galaxies classified with a maximum probability of $P$>0.7.
		$(c)$ \num{557949} quasars and \num{268554} galaxies classified with a maximum probability of $P$>0.9.
		}
	\label{fig:perfo_fullrun_distribution_combmodalpha_pmax}
	\end{minipage} 	
\end{figure} 

\newpage~\newpage
\begin{figure}[htp!]
\begin{minipage}[t]{1\textwidth}
	 \begin{subfigure}{1\textwidth} \centering
         \includegraphics[scale=0.33]{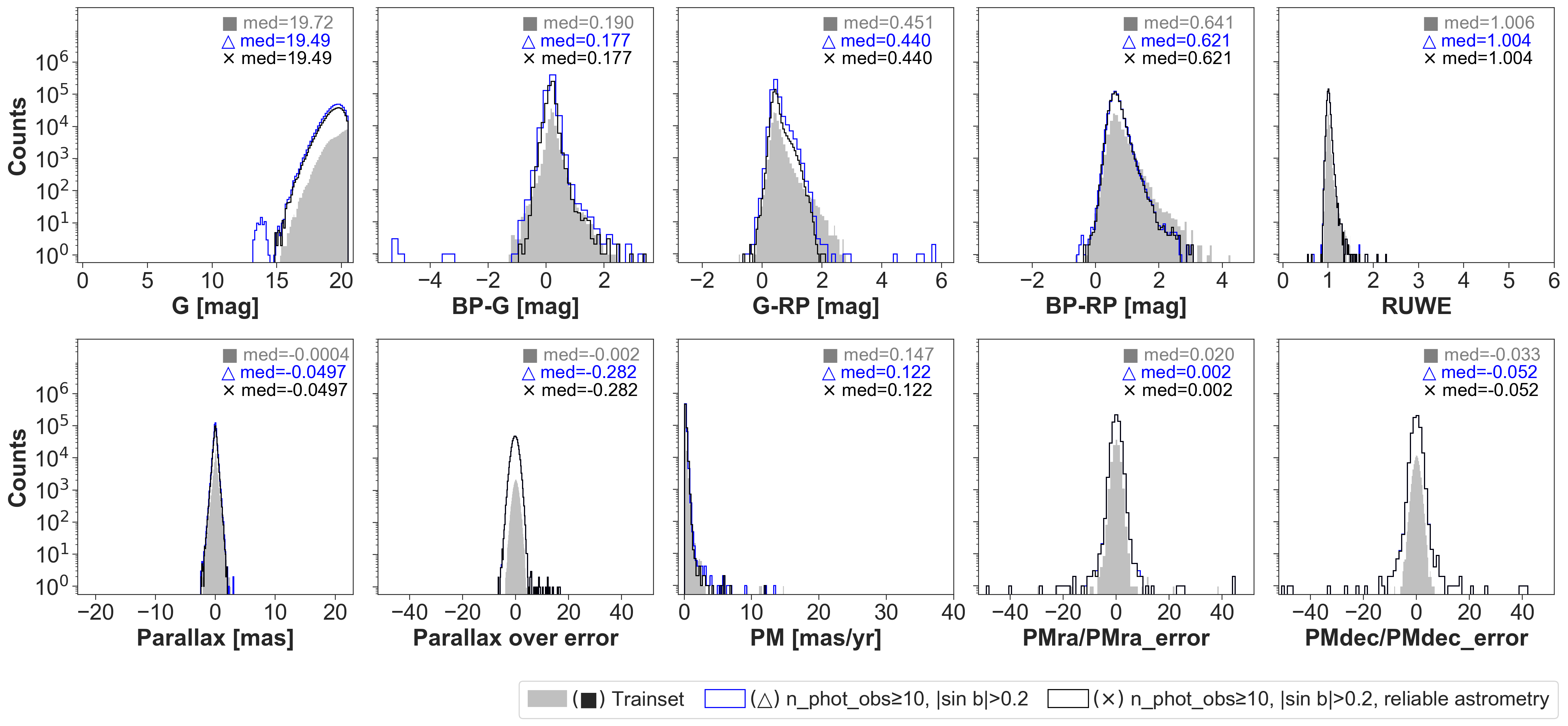}
         \caption{}
    	\end{subfigure}
	
	\begin{subfigure}{1\textwidth}
         \centering
         \includegraphics[scale=0.33]{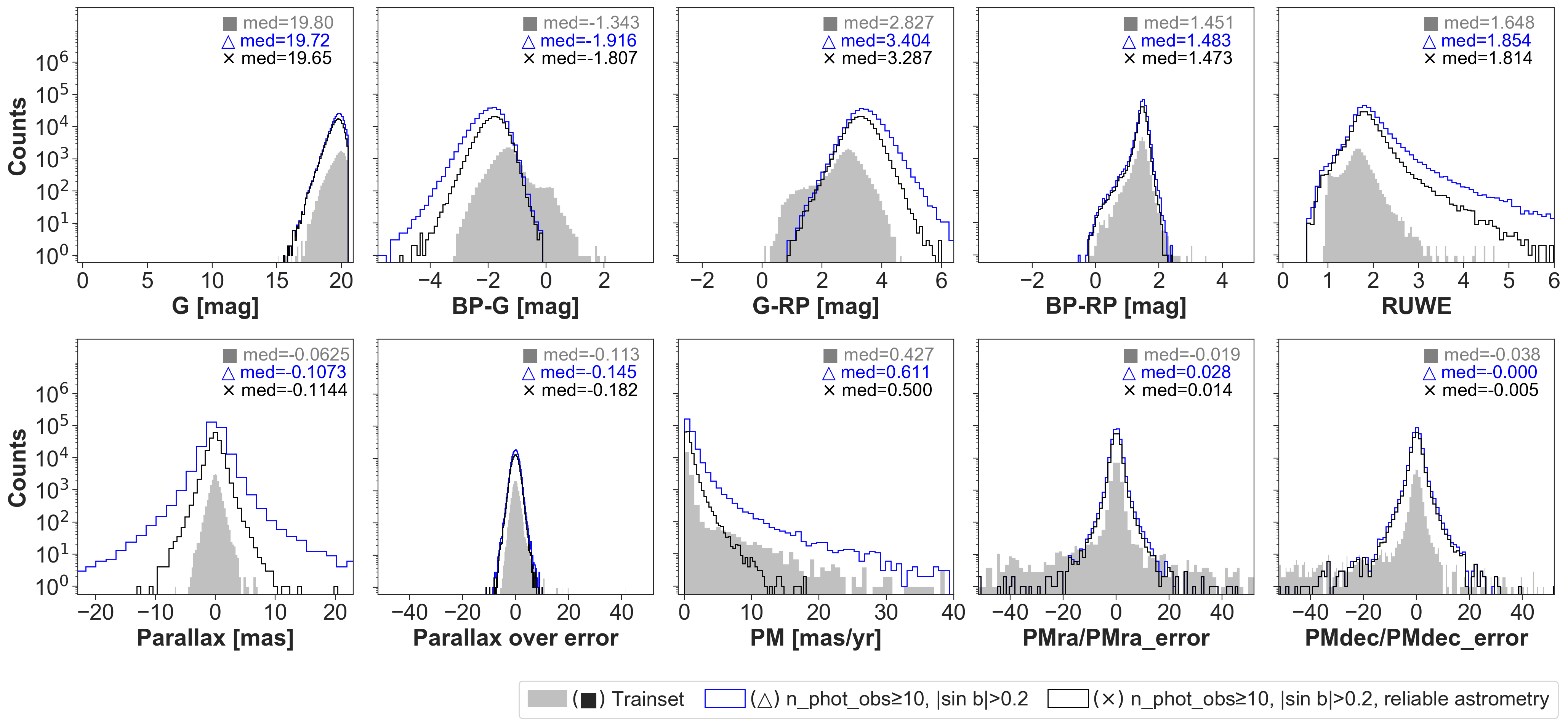}
         \caption{}
    	\end{subfigure}
	\caption{ 
		Same as Fig. \ref{fig:perfo_fullrun_distribution_features_combmodalpha_0-G-20.5_Pmax} but for the extragalactic candidates identified 
		in the \textit{Gaia} only mode by \texttt{Combmod-$\alpha$} using the global prior, at magnitudes $G$<20.5,  with a maximum probability of $P$>0.9.
		$(a)$ Quasar candidates. $(b)$ Galaxy candidates.  
		In $(a)$, the classifier identifies as quasars, \num{557949} sources ($\bigtriangleup$) and \num{557163} sources ($\times$). 
        		In $(b)$, the classifier identifies as galaxies, \num{268554} sources ($\bigtriangleup$) and \num{187764} sources ($\times$).
		}
	\label{fig:perfo_fullrun_distribution_features_combmodalpha_0-G-20.5_Pmax0.9}
	\end{minipage}
\vspace{1cm}
\end{figure}

\begin{figure}[htp!]
\begin{minipage}[t]{1\textwidth}
	\vspace{-.25cm}  
	 \textbf{\texttt{Combmod-$\alpha$} classifier predictions of quasar candidates}\\
	\begin{subfigure}{0.3333\textwidth}
         \centering
	 \includegraphics[scale=0.30]{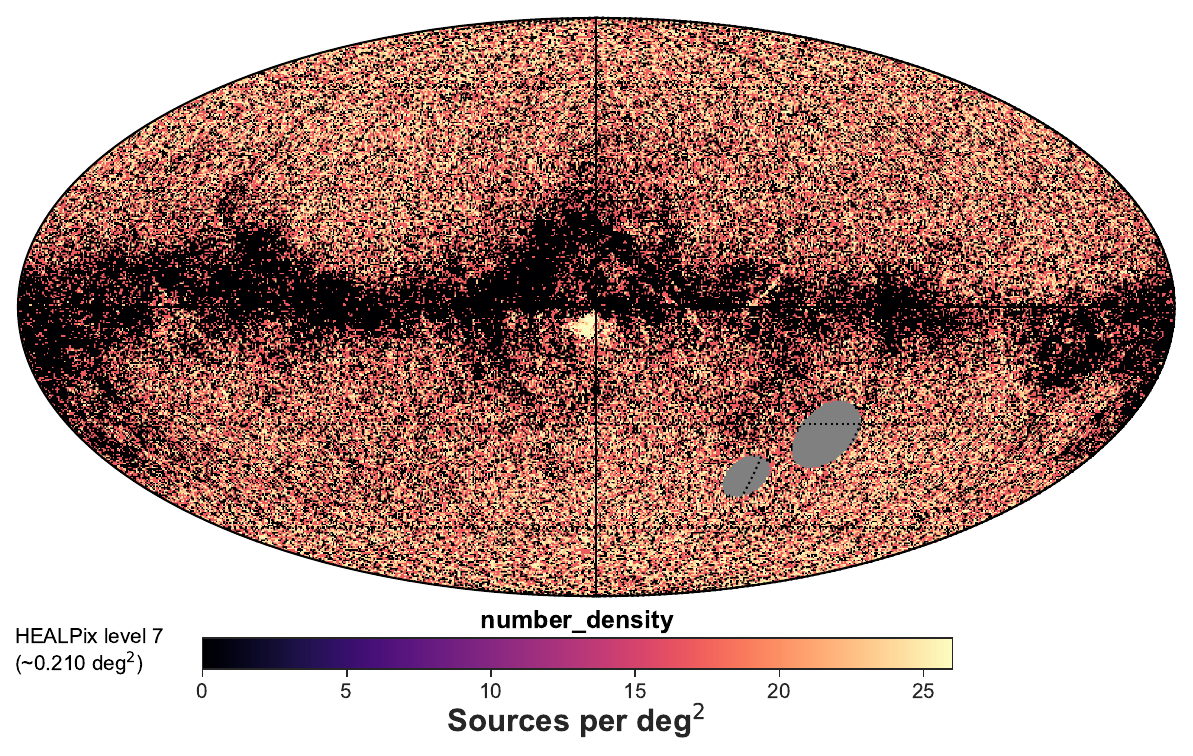}
	  \caption{\num{201792} quasars, 14.5$\leq$$G$<19\\\,}
	\end{subfigure}
	\hspace{0.1cm}
	\begin{subfigure}{0.3333\textwidth}
         \centering
	 \includegraphics[scale=0.30]{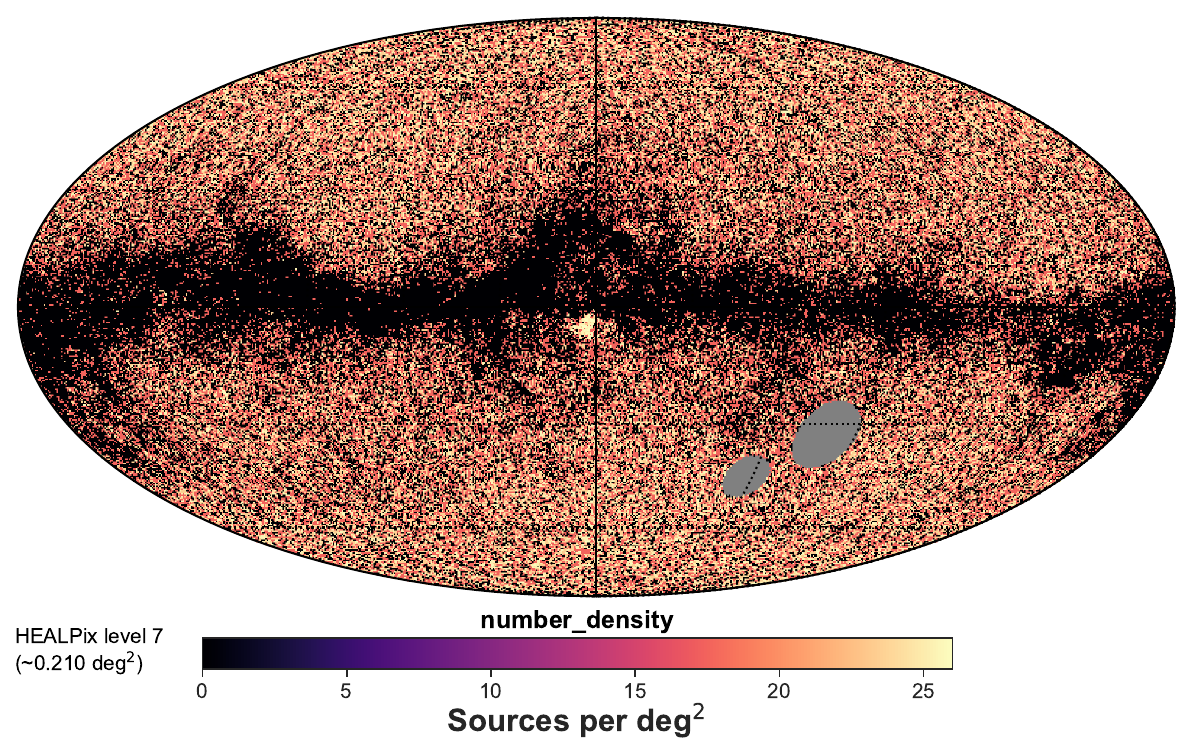}
	  \caption{\num{192153} quasars, 14.5$\leq$$G$<19 and \texttt{phot\_[bp,rp]\_n\_obs}$\geq$6$\,$}
	\end{subfigure}
	\hspace{0.1cm}
	\begin{subfigure}{0.3333\textwidth}
         \centering
	 \includegraphics[scale=0.30]{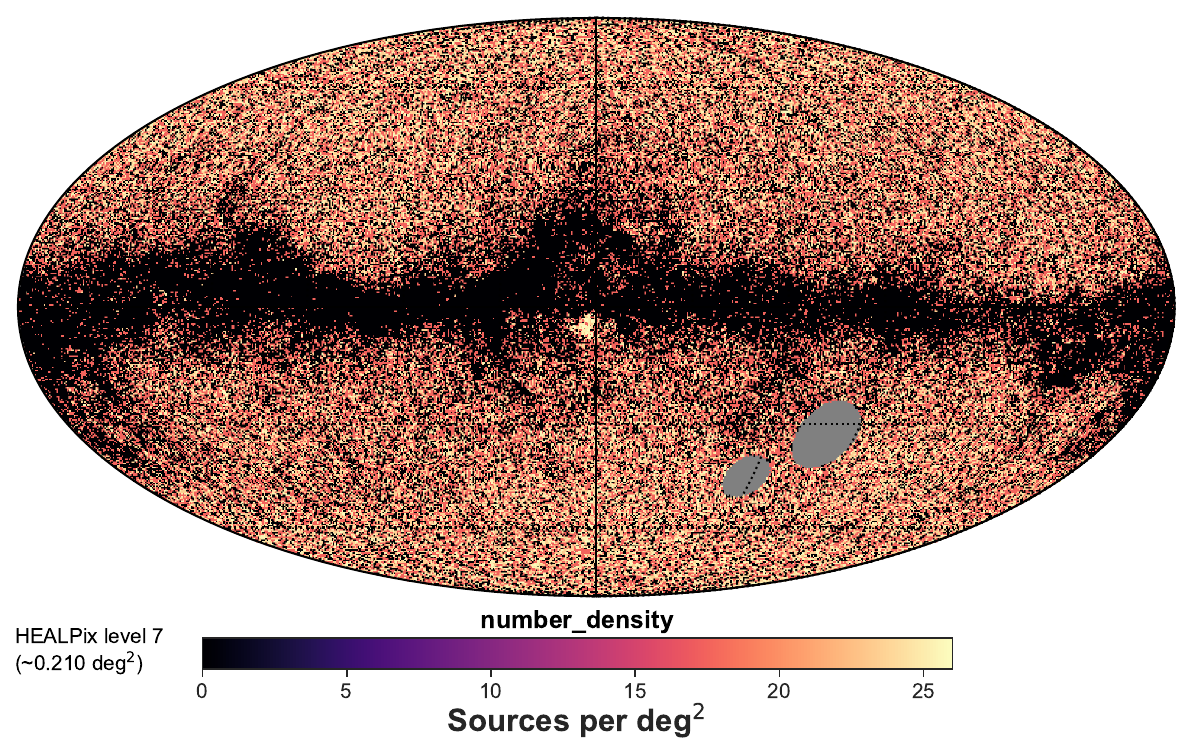}
	  \caption{\num{190265} quasars, 14.5$\leq$$G$<19 and \texttt{phot\_[bp,rp]\_n\_obs}$\geq$10}
	\end{subfigure}
	
	\begin{subfigure}{0.3333\textwidth}
         \centering
	 \includegraphics[scale=0.30]{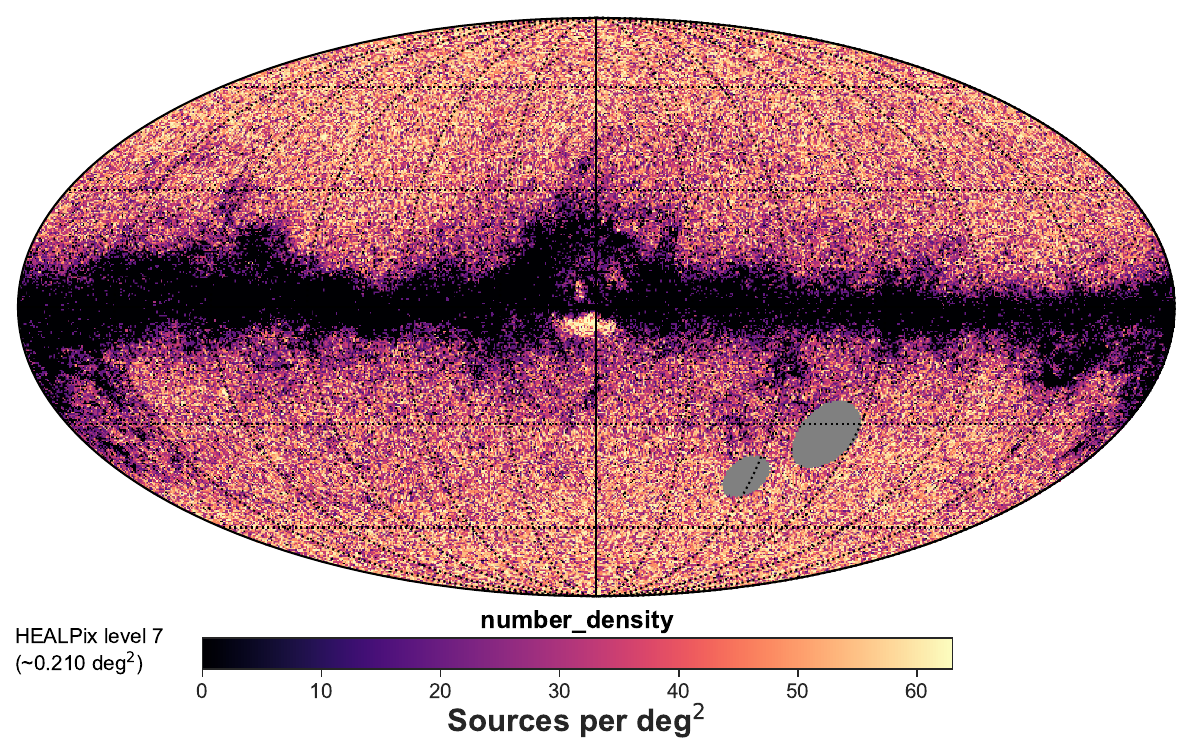}
	  \caption{\num{572266} quasars, 19$\leq$$G$<20\\\,}
	\end{subfigure}
	\hspace{0.1cm}
	\begin{subfigure}{0.3333\textwidth}
         \centering
	 \includegraphics[scale=0.30]{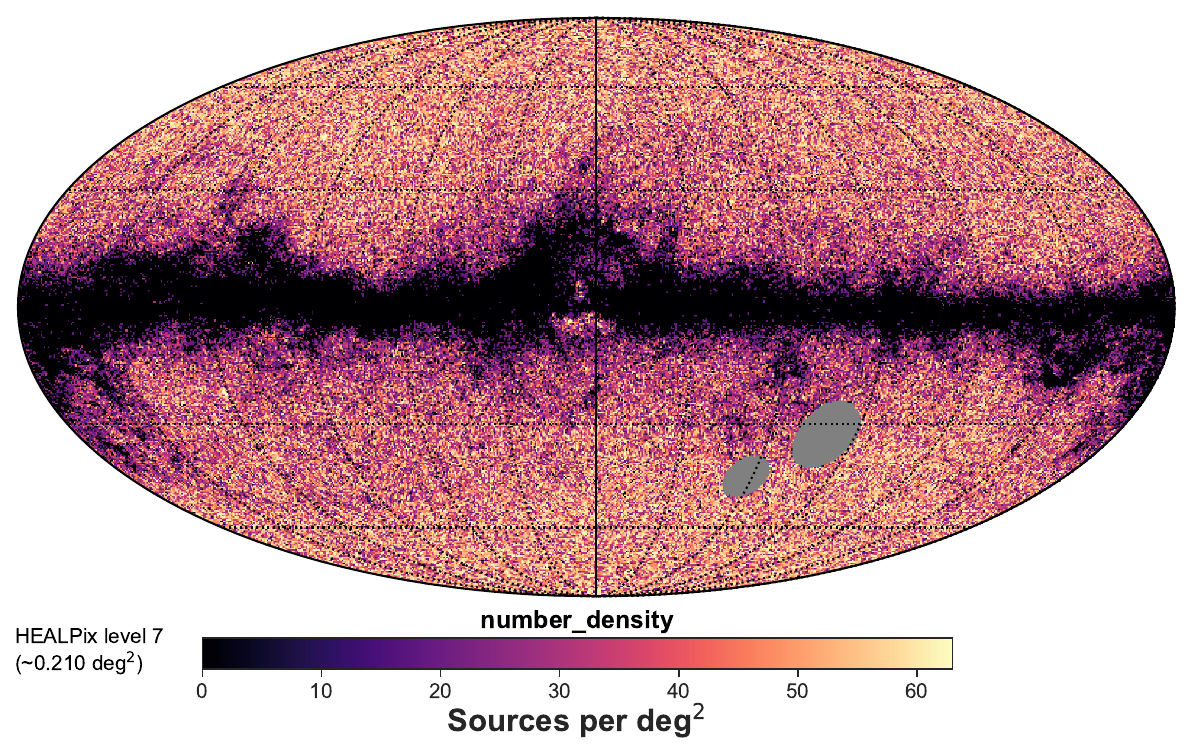}
	  \caption{\num{567858} quasars, 19$\leq$$G$<20 and \texttt{phot\_[bp,rp]\_n\_obs}$\geq$6$\,$}
	\end{subfigure}
	\hspace{0.1cm}
	\begin{subfigure}{0.3333\textwidth}
         \centering
	 \includegraphics[scale=0.30]{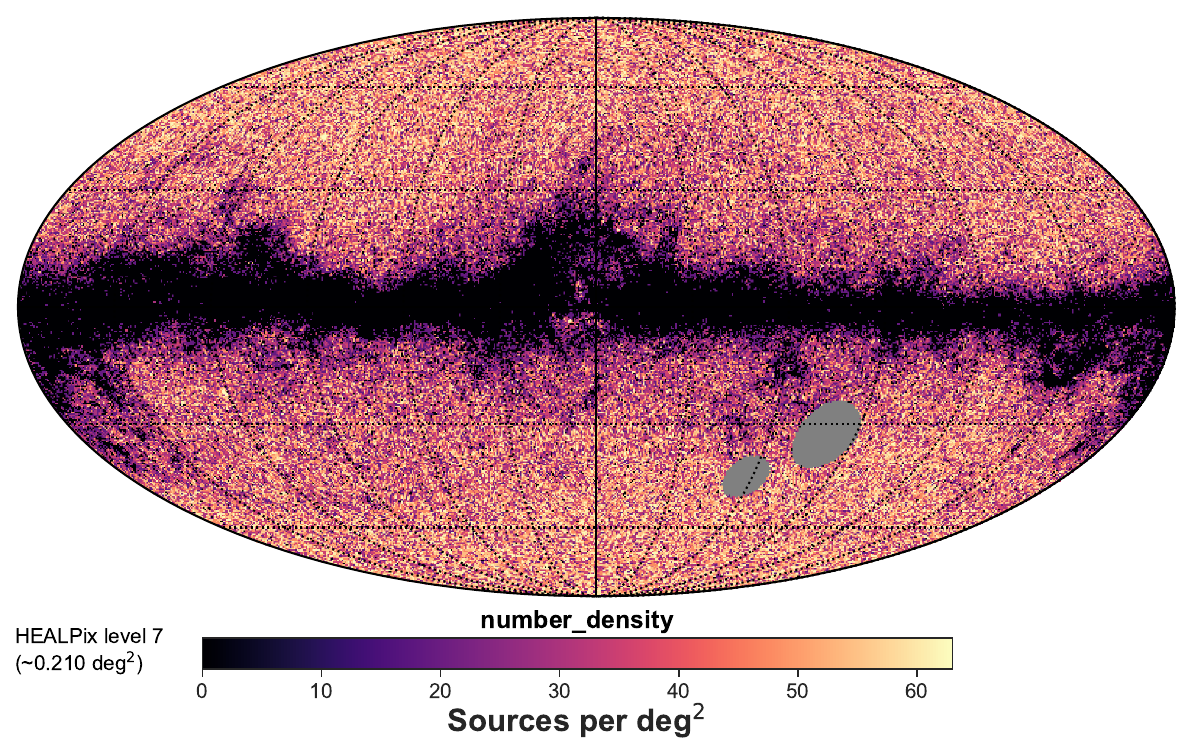}
	  \caption{\num{566334} quasars, 19$\leq$$G$<20 and \texttt{phot\_[bp,rp]\_n\_obs}$\geq$10}
	\end{subfigure}

	\begin{subfigure}{0.3333\textwidth}
         \centering
	 \includegraphics[scale=0.30]{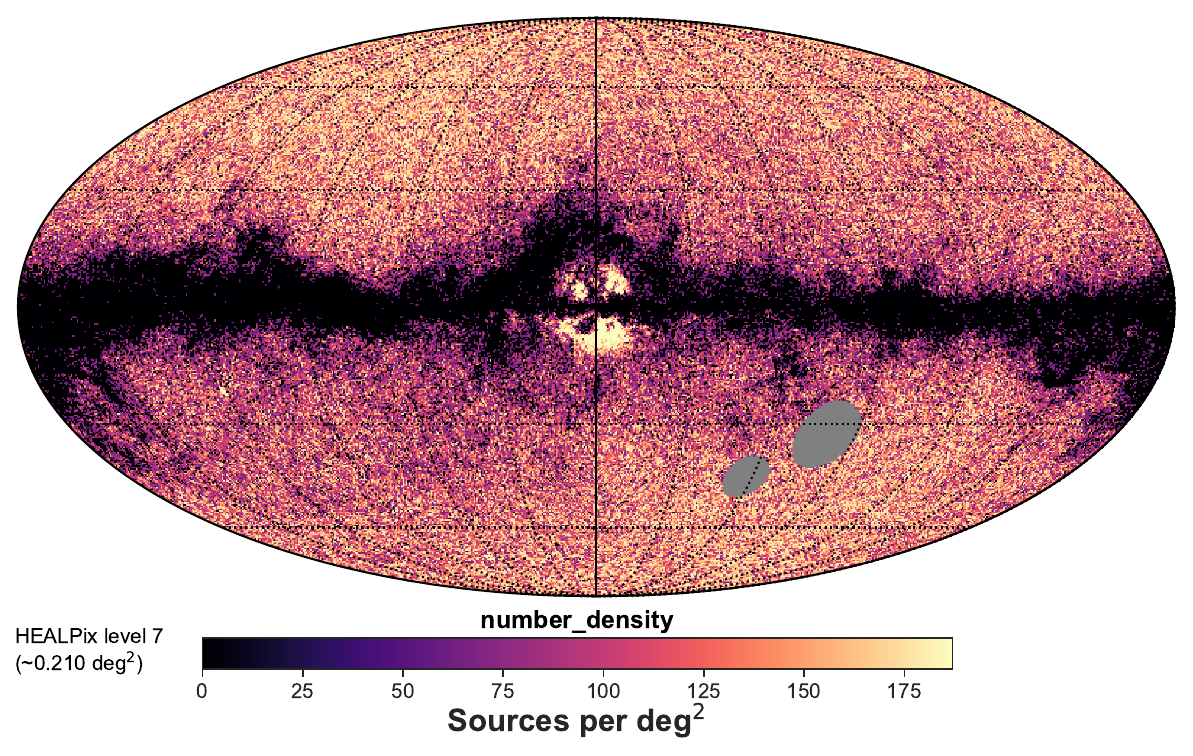}
	  \caption{\num{525260} quasars, 20$\leq$$G$<20.5\\\,}
	\end{subfigure}
	\hspace{0.1cm}
	\begin{subfigure}{0.3333\textwidth}
         \centering
	 \includegraphics[scale=0.30]{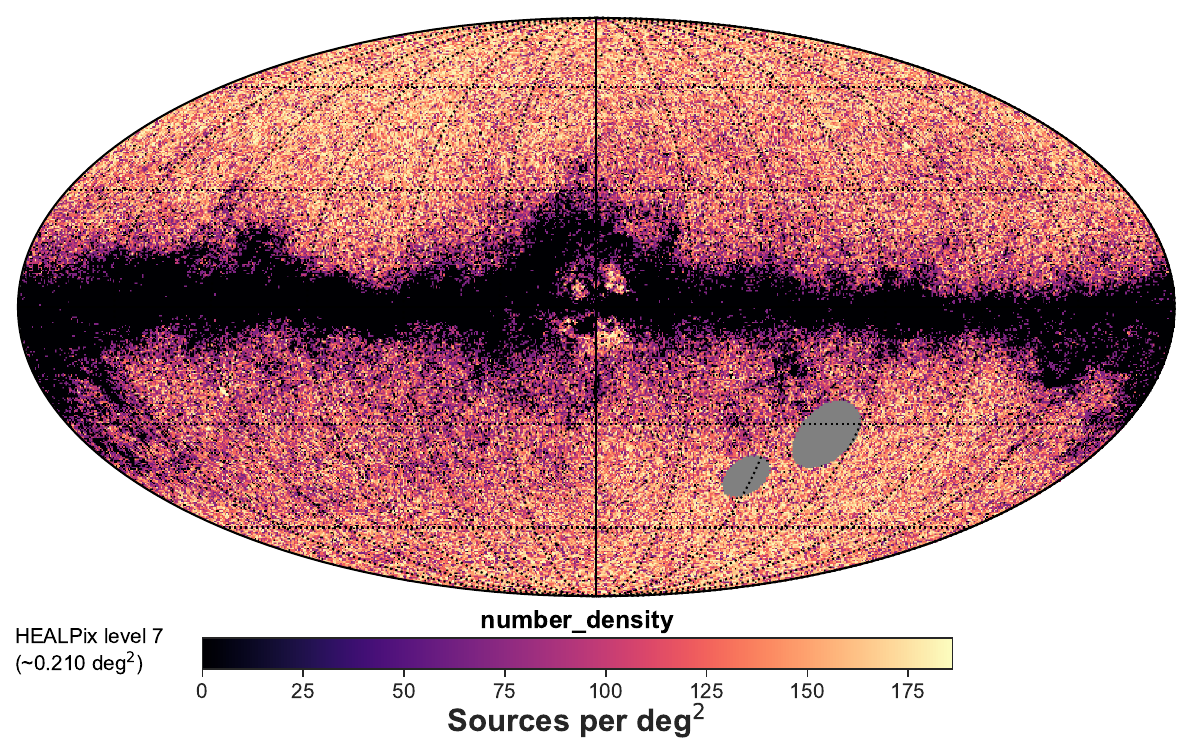}
	  \caption{\num{515391} quasars, 20$\leq$$G$<20.5 and \texttt{phot\_[bp,rp]\_n\_obs}$\geq$6$\,$}
	\end{subfigure}
	\hspace{0.1cm}
	\begin{subfigure}{0.3333\textwidth}
         \centering
	 \includegraphics[scale=0.30]{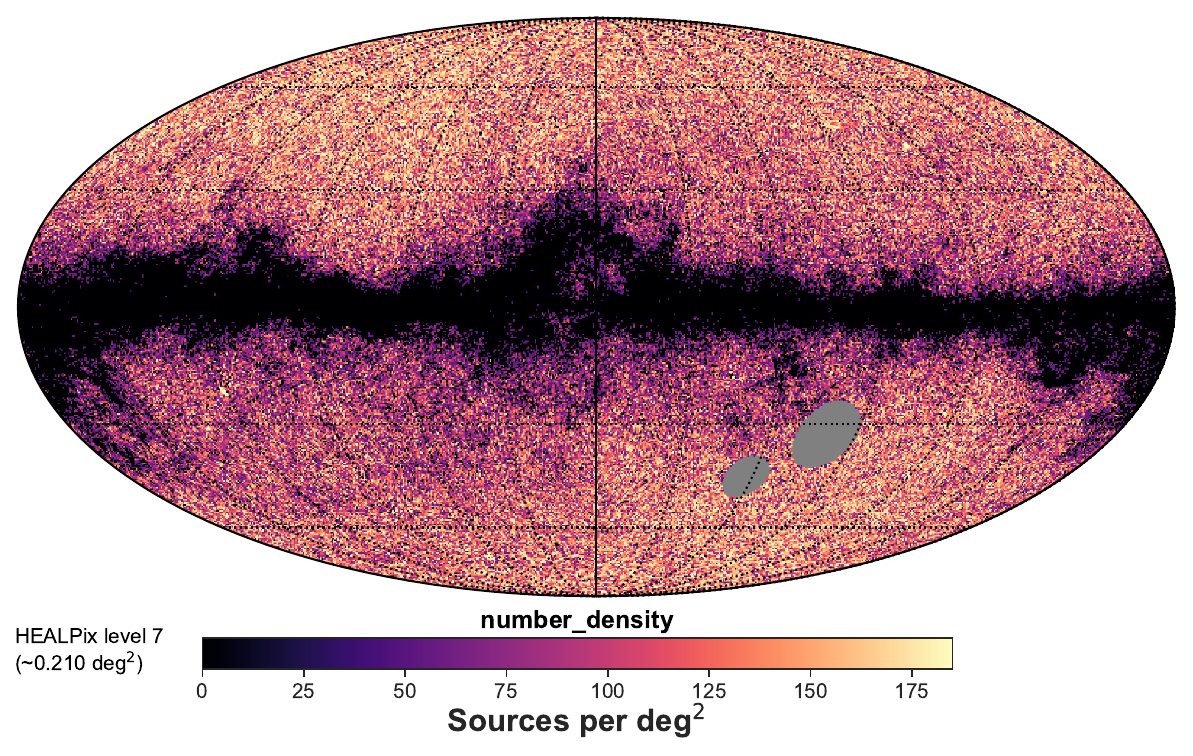}
	  \caption{\num{511486} quasars, 20$\leq$$G$<20.5 and \texttt{phot\_[bp,rp]\_n\_obs}$\geq$10}
	\end{subfigure}
	
	\begin{subfigure}{0.3333\textwidth}
         \centering
	 \includegraphics[scale=0.30]{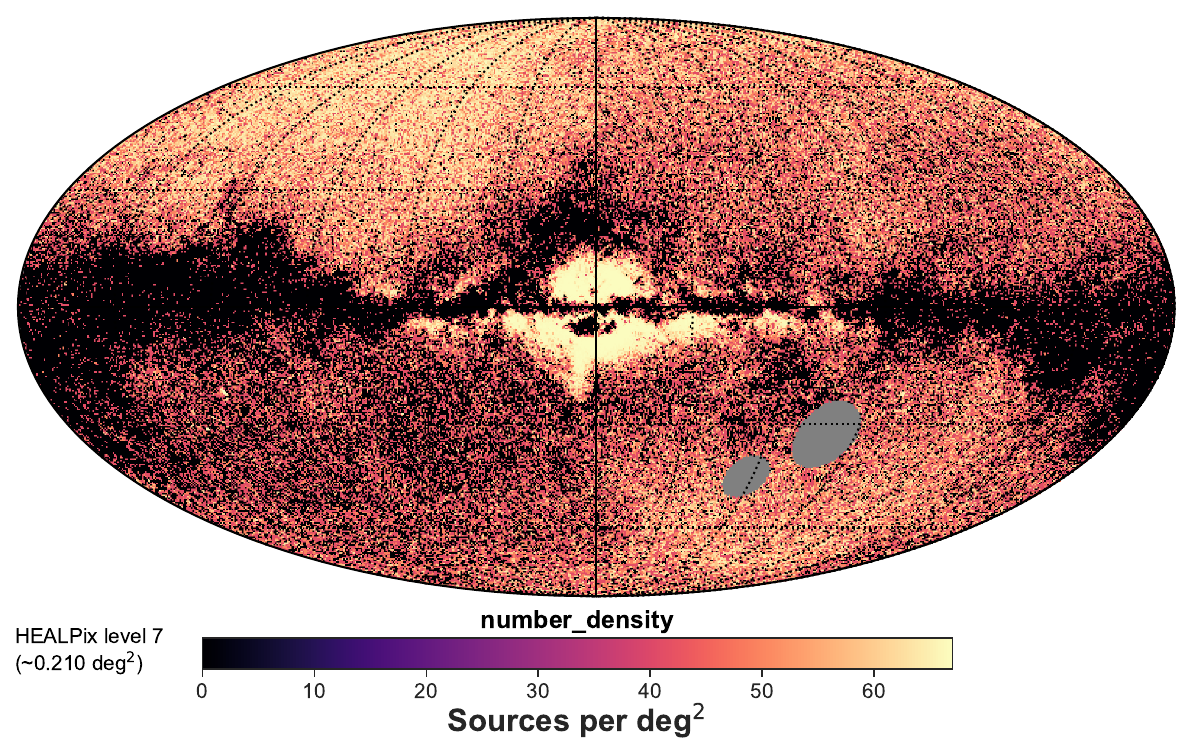}
	  \caption{\num{315559} quasars, 20.5$\leq$$G$<21\\\,}
	\end{subfigure}
	\hspace{0.1cm}
	\begin{subfigure}{0.3333\textwidth}
         \centering
	 \includegraphics[scale=0.30]{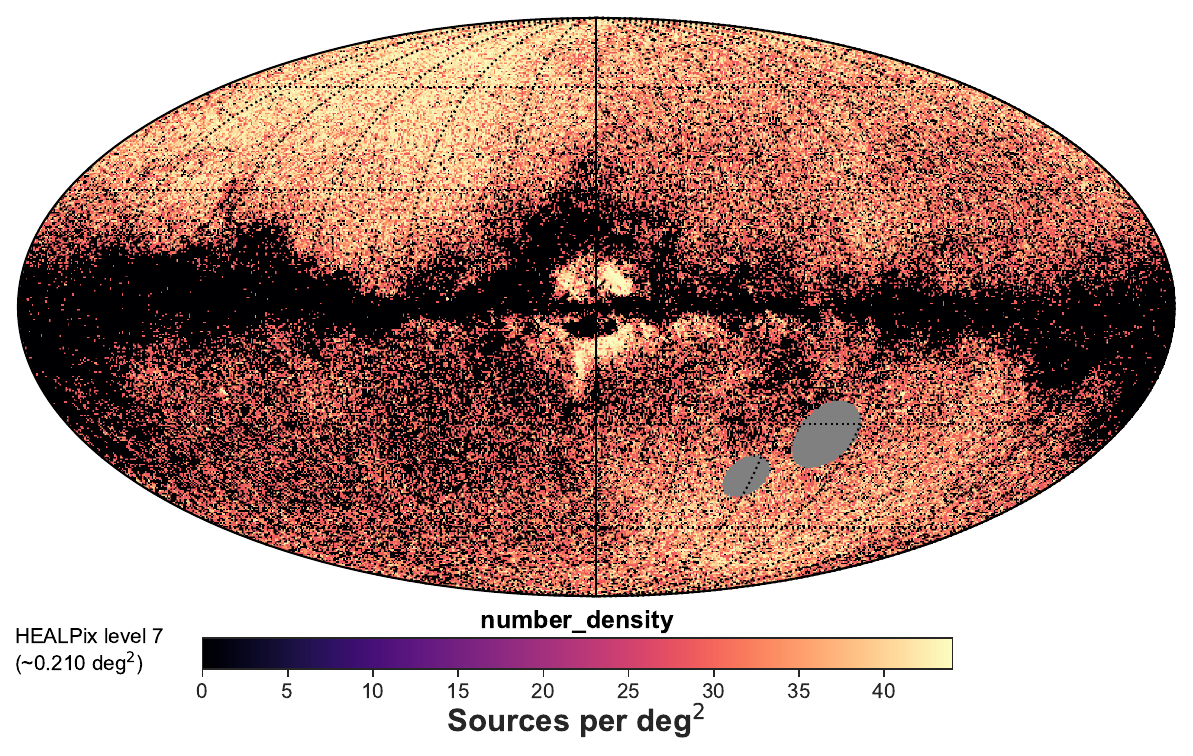}
	  \caption{\num{257208} quasars, 20.5$\leq$$G$<21 and \texttt{phot\_[bp,rp]\_n\_obs}$\geq$6$\,$}
	\end{subfigure}
	\hspace{0.1cm}
	\begin{subfigure}{0.3333\textwidth}
         \centering
	 \includegraphics[scale=0.30]{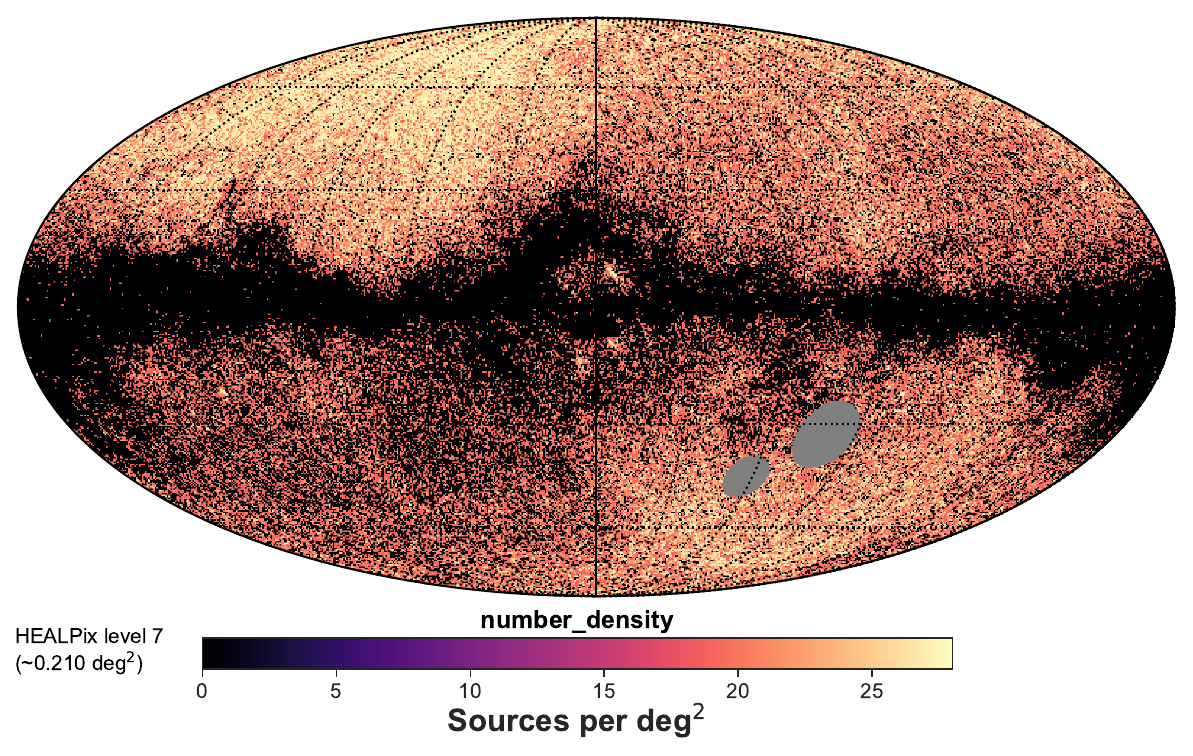}
	  \caption{\num{237665} quasars, 20.5$\leq$$G$<21 and \texttt{phot\_[bp,rp]\_n\_obs}$\geq$10}
	\end{subfigure}

	\begin{subfigure}{0.3333\textwidth}
         \centering
	 \includegraphics[scale=0.30]{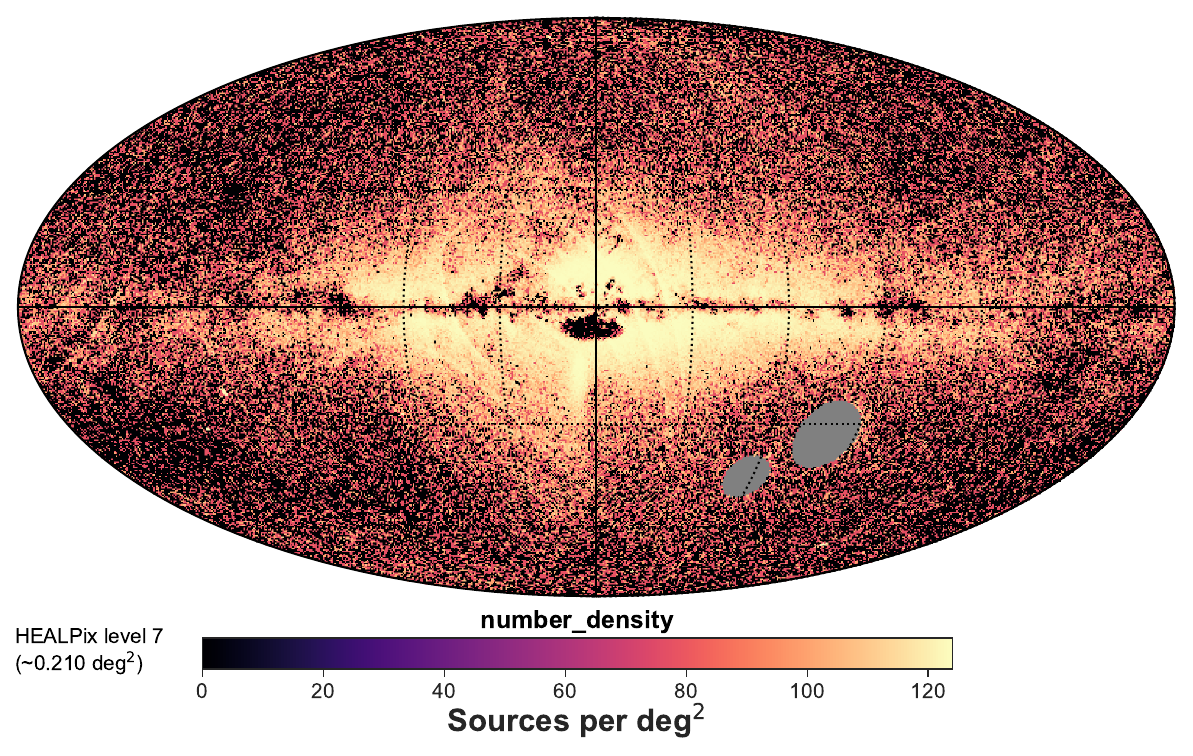}
	  \caption{\num{392863} quasars, 21$\leq$$G$<30\\\,}  
	\end{subfigure}
	\hspace{0.1cm}
	\begin{subfigure}{0.3333\textwidth}
         \centering
	 \includegraphics[scale=0.30]{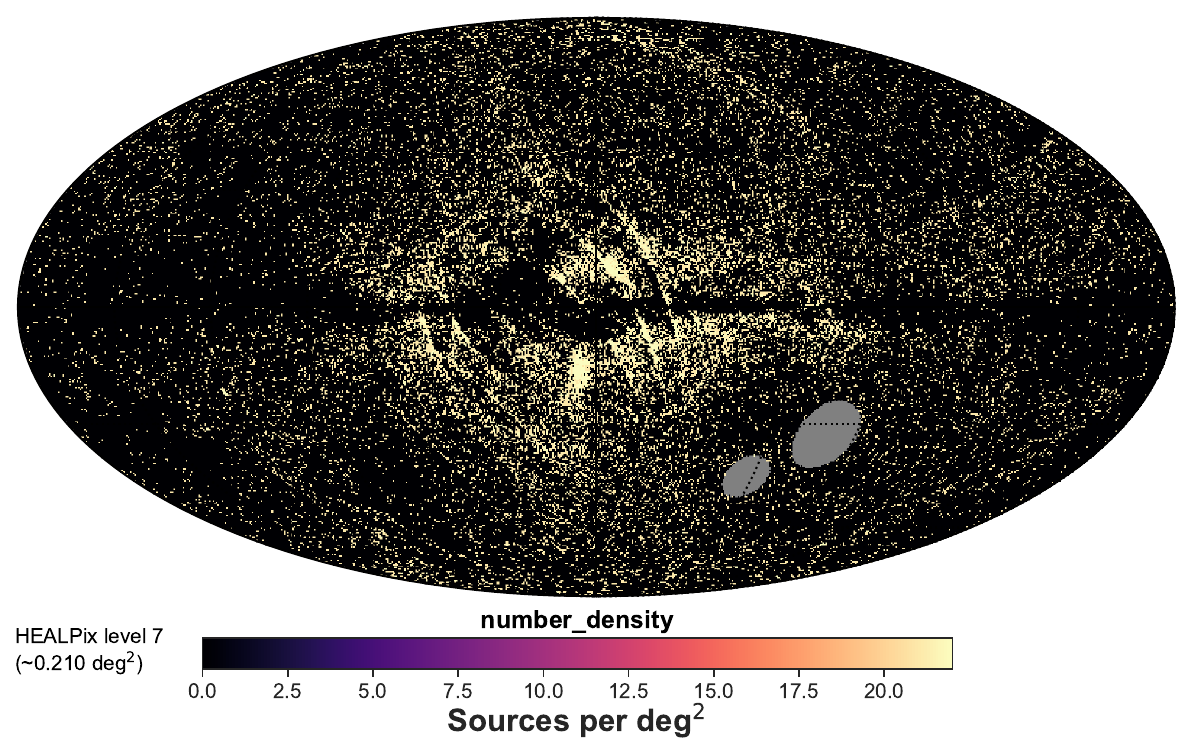}
	  \caption{\num{30105} quasars, 21$\leq$$G$<30 and \texttt{phot\_[bp,rp]\_n\_obs}$\geq$6$\,$}
	\end{subfigure}
	\hspace{0.1cm}
	\begin{subfigure}{0.3333\textwidth}
         \centering
	 \includegraphics[scale=0.30]{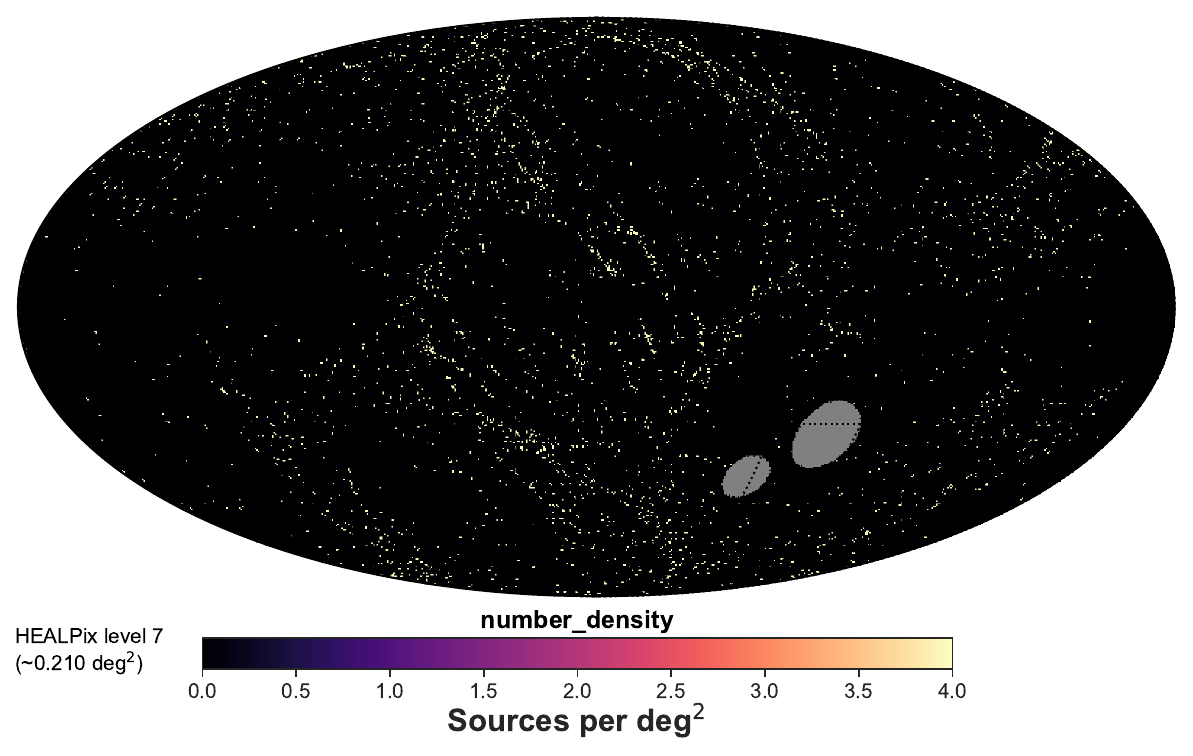}
	  \caption{\num{3379} quasars, 21$\leq$$G$<30 and \texttt{phot\_[bp,rp]\_n\_obs}$\geq$10}
	\end{subfigure}
	
\vspace{-.2cm}
\caption{
	Galactic sky distribution of sources classified as quasars from maximum probabilities 
	by \texttt{Combmod-$\alpha$} using the global prior at HEALpixel level 7 in Mollweide projection.
	The LMC and SMC regions are masked in grey. 
	Sky distributions show predicted sources at different magnitudes after applying a quality cut 
	to the number of photometric observations in the BP- and RP- bands.\\}
\label{fig:perfo_skyplots_fullrun_magbins_combmodalpha_quasars}
\end{minipage}
\end{figure}

\begin{figure}[htp!]
\begin{minipage}[t]{1\textwidth}
	\vspace{-.25cm}  
	 \textbf{\texttt{Combmod-$\alpha$} classifier predictions of galaxy candidates}\\
	\begin{subfigure}{0.3333\textwidth}
         \centering
	\includegraphics[scale=0.30]{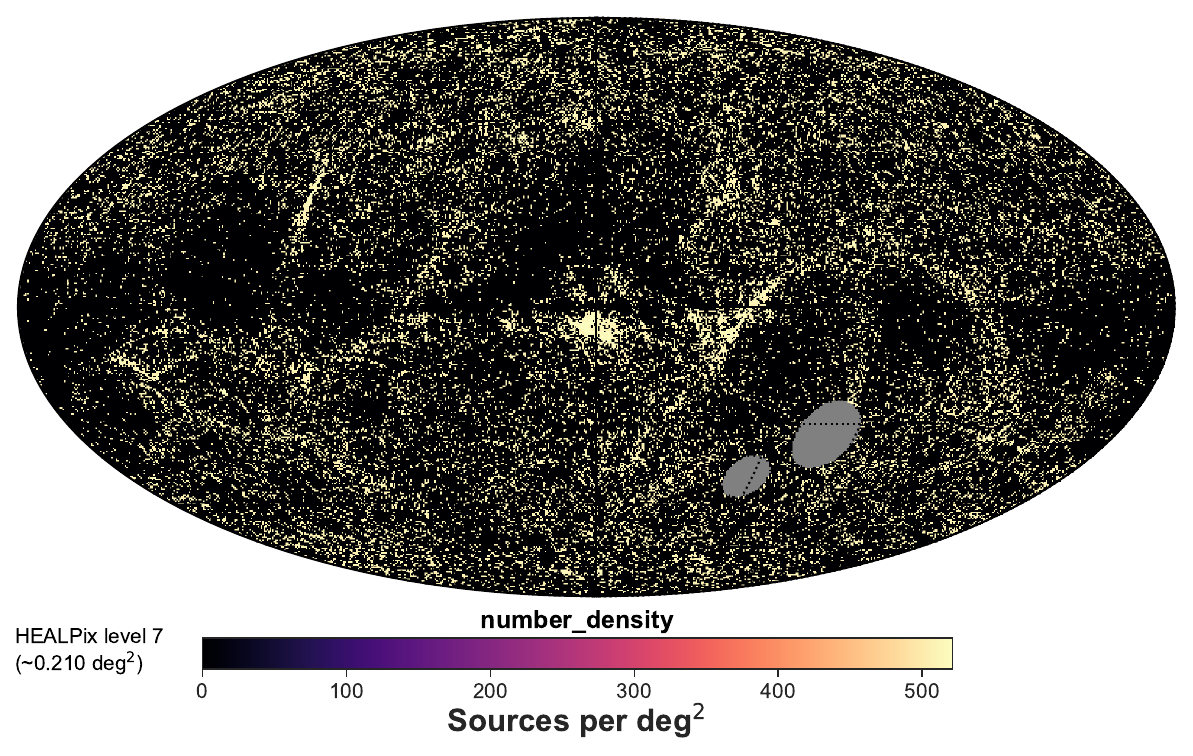}
	\caption{\num{39642} galaxies, 14.5$\leq$$G$<19\\\,}
	\end{subfigure}
	\hspace{0.1cm}
	\begin{subfigure}{0.3333\textwidth}
         \centering
	\includegraphics[scale=0.30]{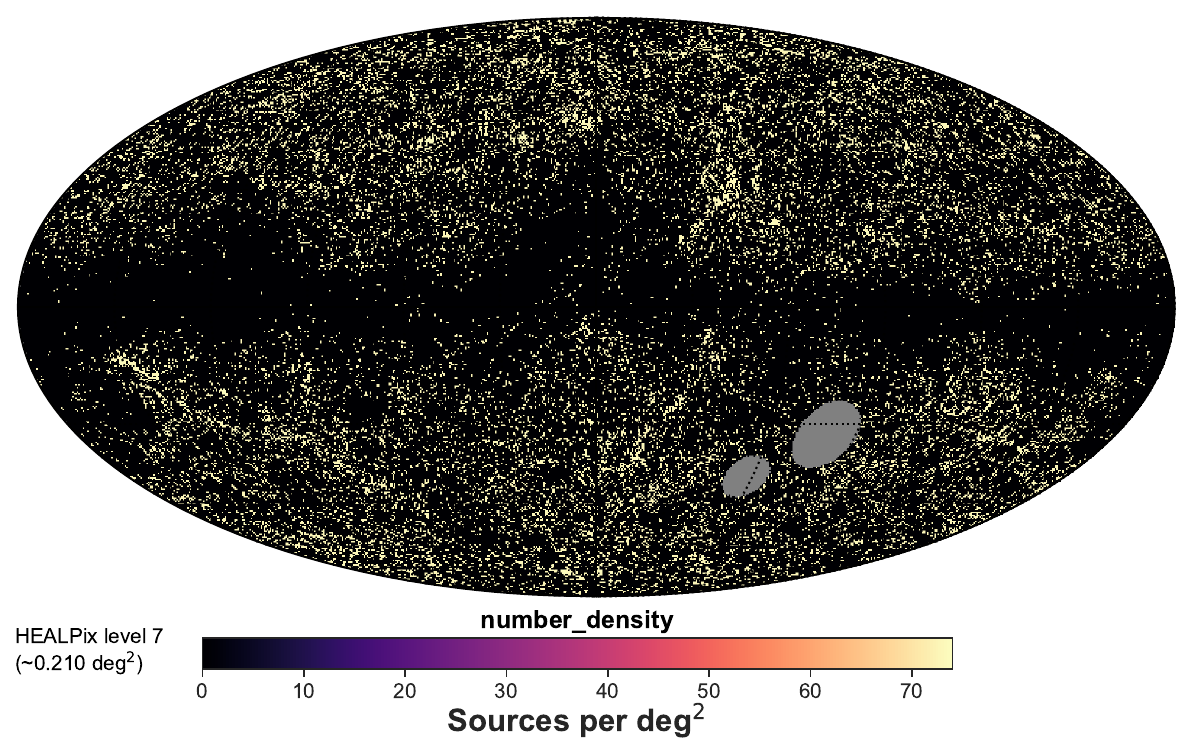}
	\caption{\num{29603} galaxies, 14.5$\leq$$G$<19 and \texttt{phot\_[bp,rp]\_n\_obs}$\geq$6$\,$}
	\end{subfigure}
	\hspace{0.1cm}
	\begin{subfigure}{0.3333\textwidth}
         \centering
	\includegraphics[scale=0.30]{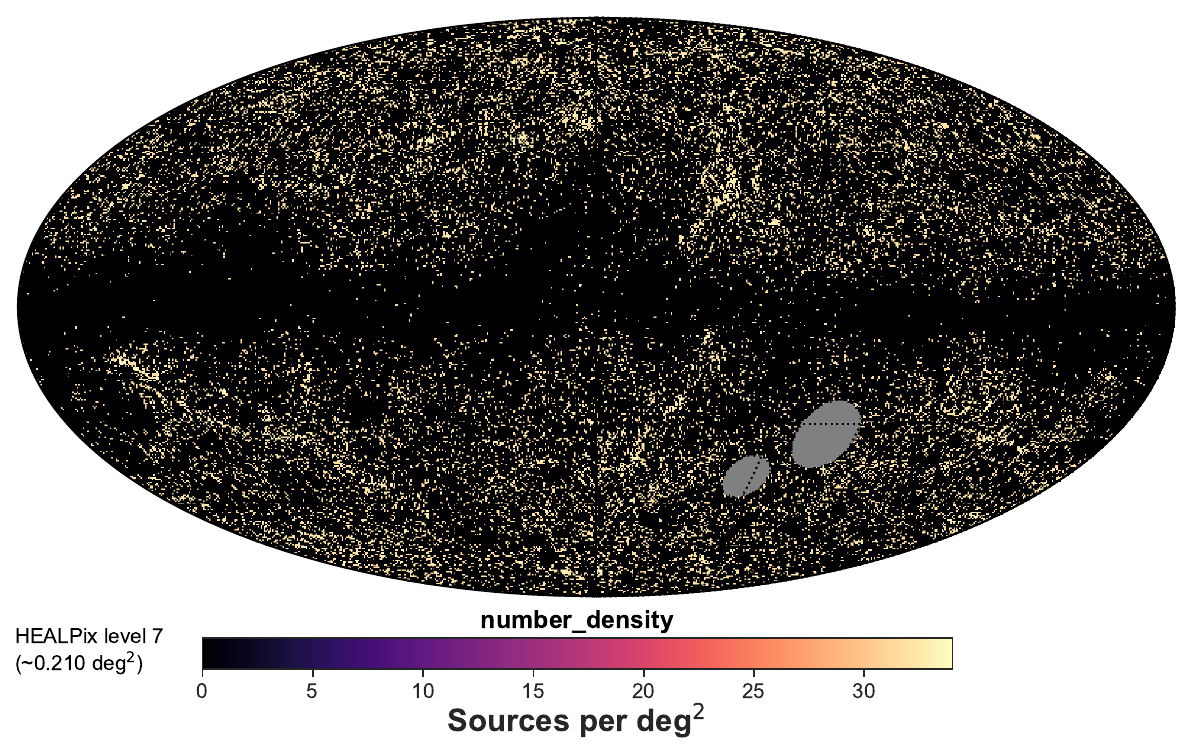}
	\caption{\num{29023} galaxies, 14.5$\leq$$G$<19 and \texttt{phot\_[bp,rp]\_n\_obs}$\geq$10}
	\end{subfigure}
	
	\begin{subfigure}{0.3333\textwidth}
         \centering
	\includegraphics[scale=0.30]{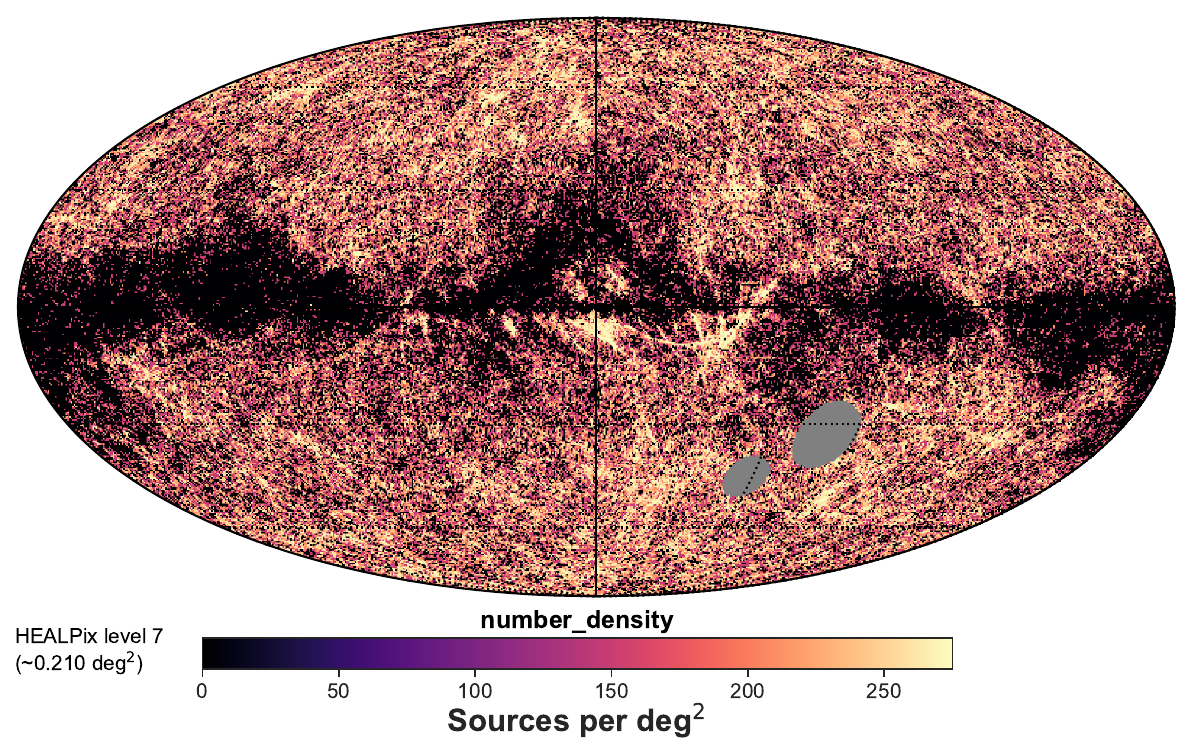}
	\caption{\num{299315} galaxies, 19$\leq$$G$<20\\\,}
	\end{subfigure}
	\hspace{0.1cm}
	\begin{subfigure}{0.3333\textwidth}
         \centering
	\includegraphics[scale=0.30]{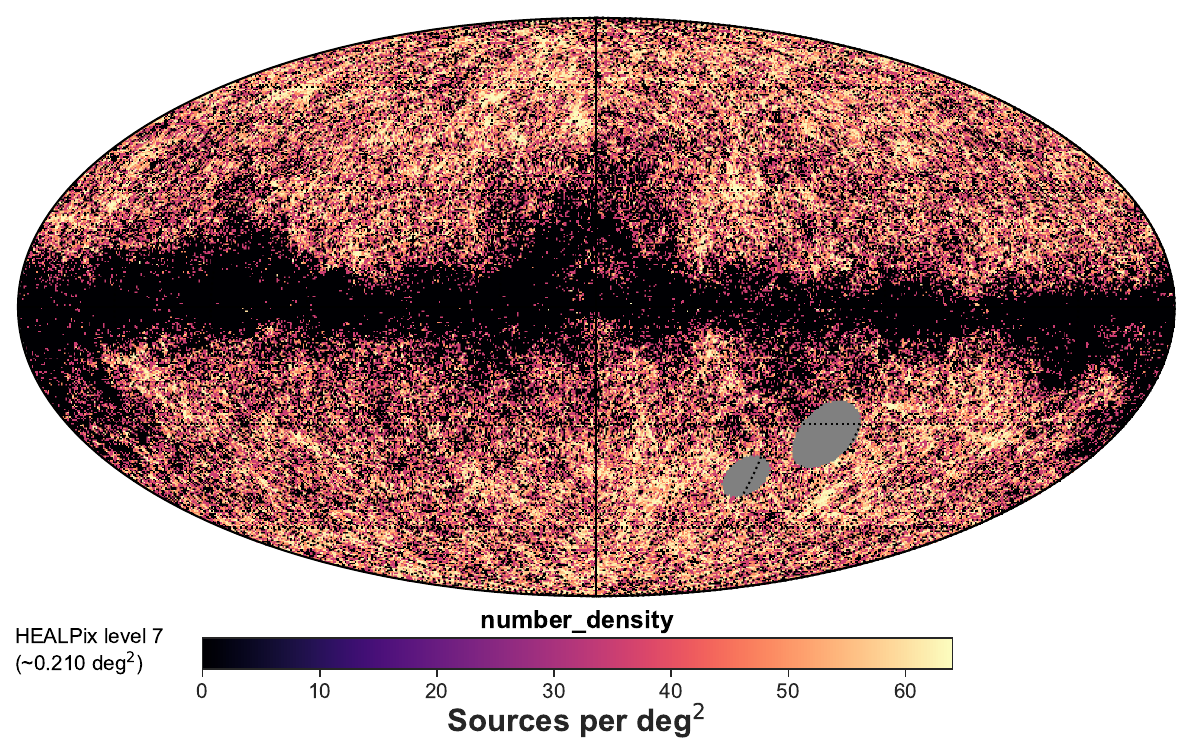}
	\caption{\num{279898} galaxies, 19$\leq$$G$<20 and \texttt{phot\_[bp,rp]\_n\_obs}$\geq$6$\,$}
	\end{subfigure}
	\hspace{0.1cm}
	\begin{subfigure}{0.3333\textwidth}
         \centering
	\includegraphics[scale=0.30]{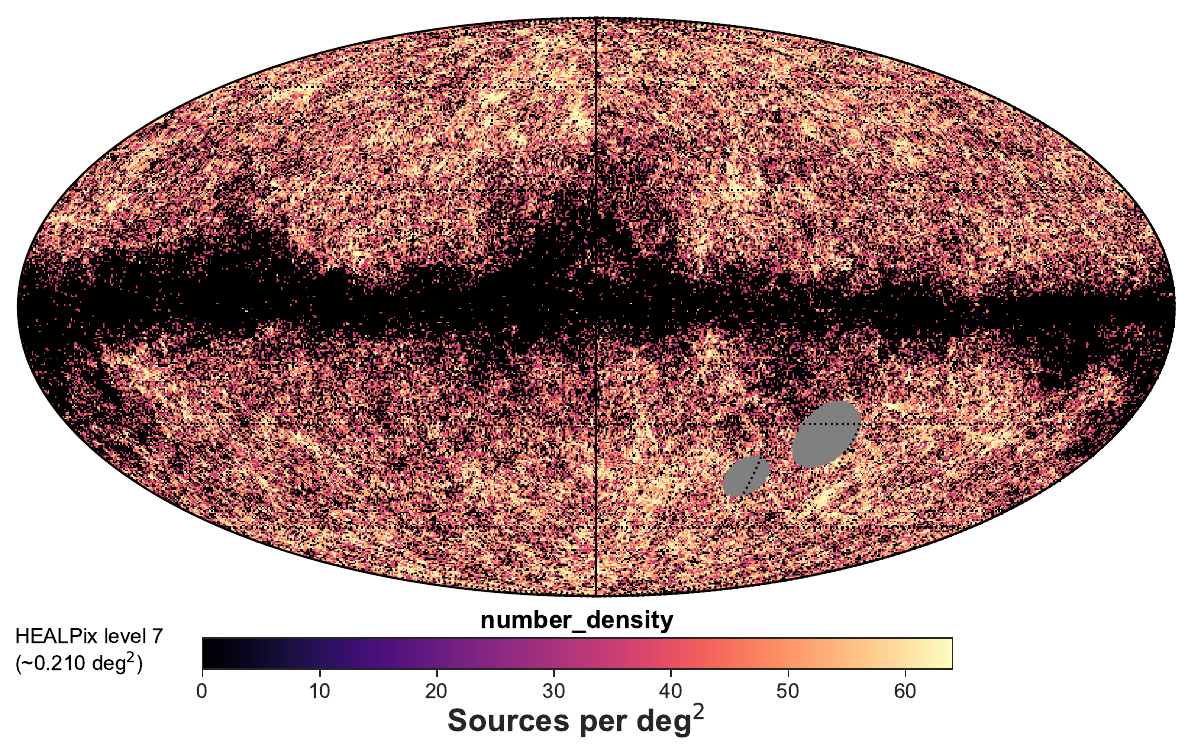}
	\caption{\num{278549} galaxies, 19$\leq$$G$<20 and \texttt{phot\_[bp,rp]\_n\_obs}$\geq$10}
	\end{subfigure}
	
	\begin{subfigure}{0.3333\textwidth}
         \centering
	\includegraphics[scale=0.30]{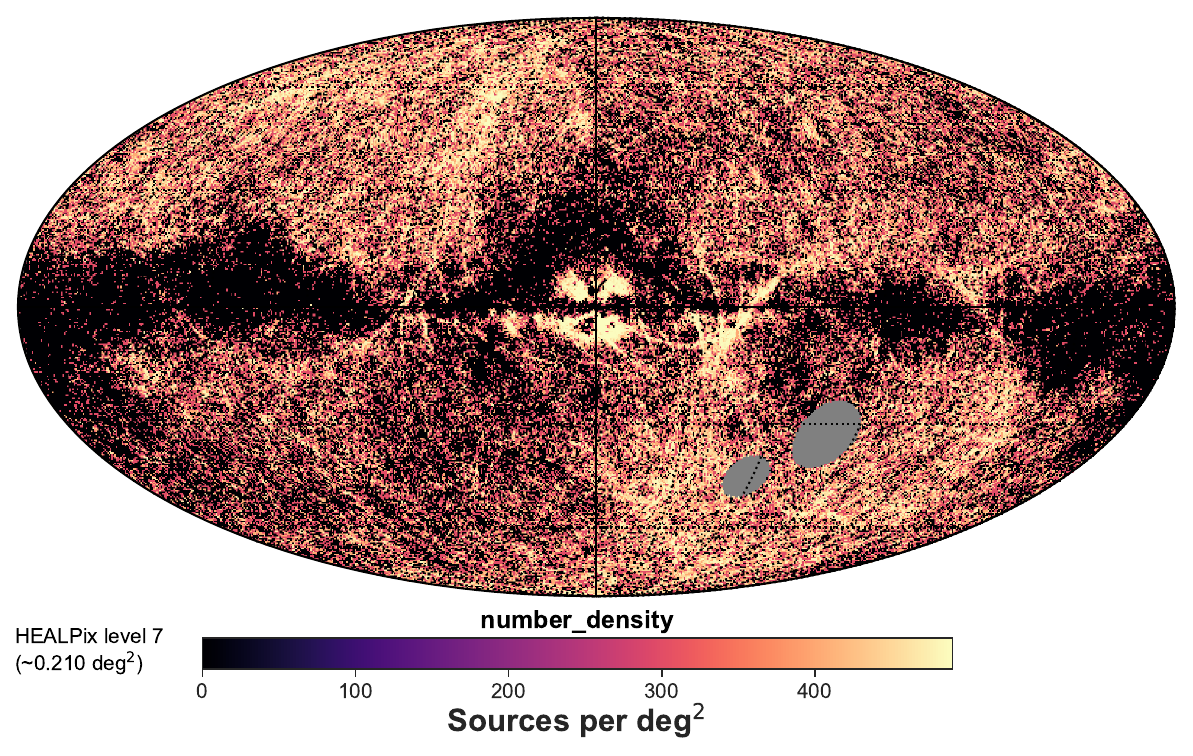}
	\caption{\num{250590} galaxies, 20$\leq$$G$<20.5\\\,}
	\end{subfigure}
	\hspace{0.1cm}
	\begin{subfigure}{0.3333\textwidth}
         \centering
	\includegraphics[scale=0.30]{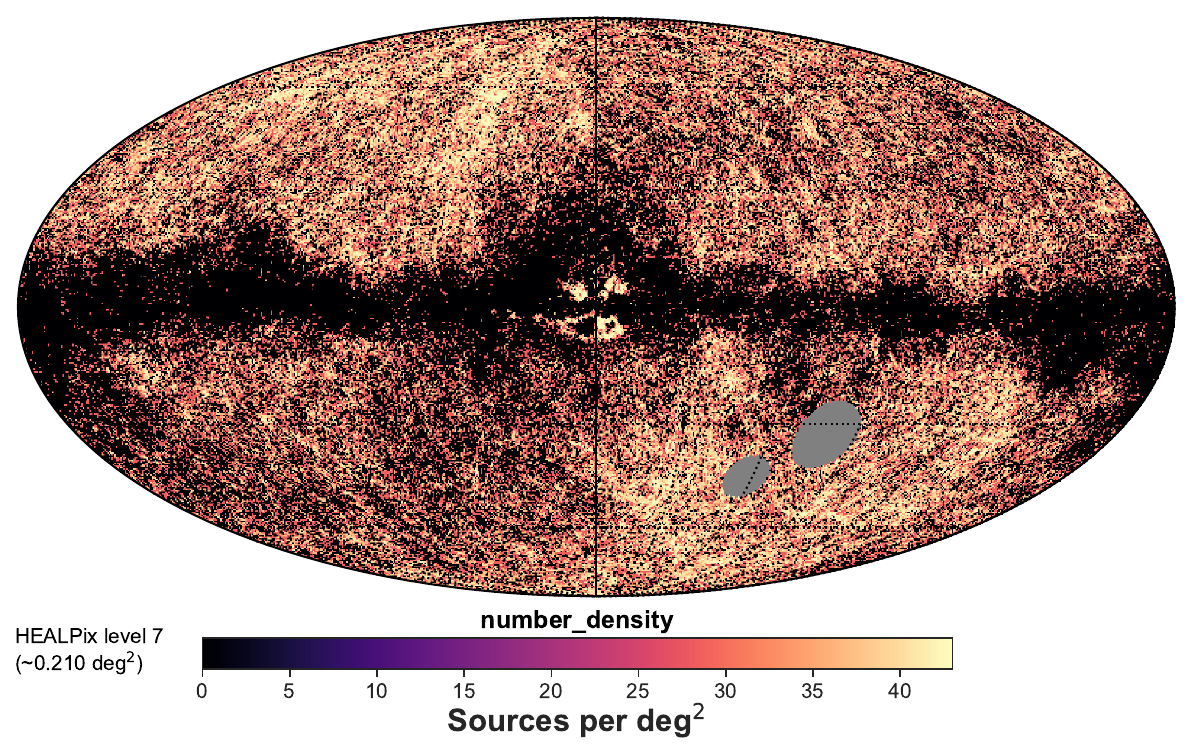}
	\caption{\num{222869} galaxies, 20$\leq$$G$<20.5 and \texttt{phot\_[bp,rp]\_n\_obs}$\geq$6$\,$}
	\end{subfigure}
	\hspace{0.1cm}
	\begin{subfigure}{0.3333\textwidth}
         \centering
	\includegraphics[scale=0.30]{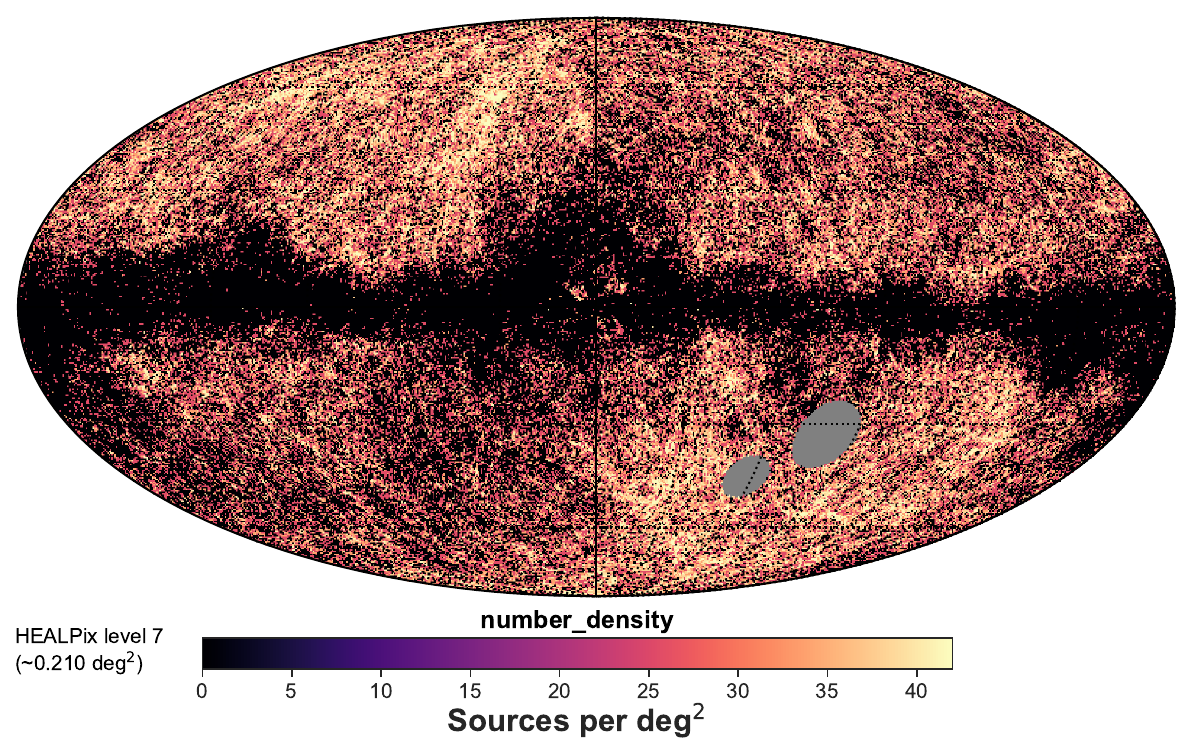}
	\caption{\num{219017} galaxies, 20$\leq$$G$<20.5 and \texttt{phot\_[bp,rp]\_n\_obs}$\geq$10}
	\end{subfigure}
		
	\begin{subfigure}{0.3333\textwidth}
         \centering
	\includegraphics[scale=0.30]{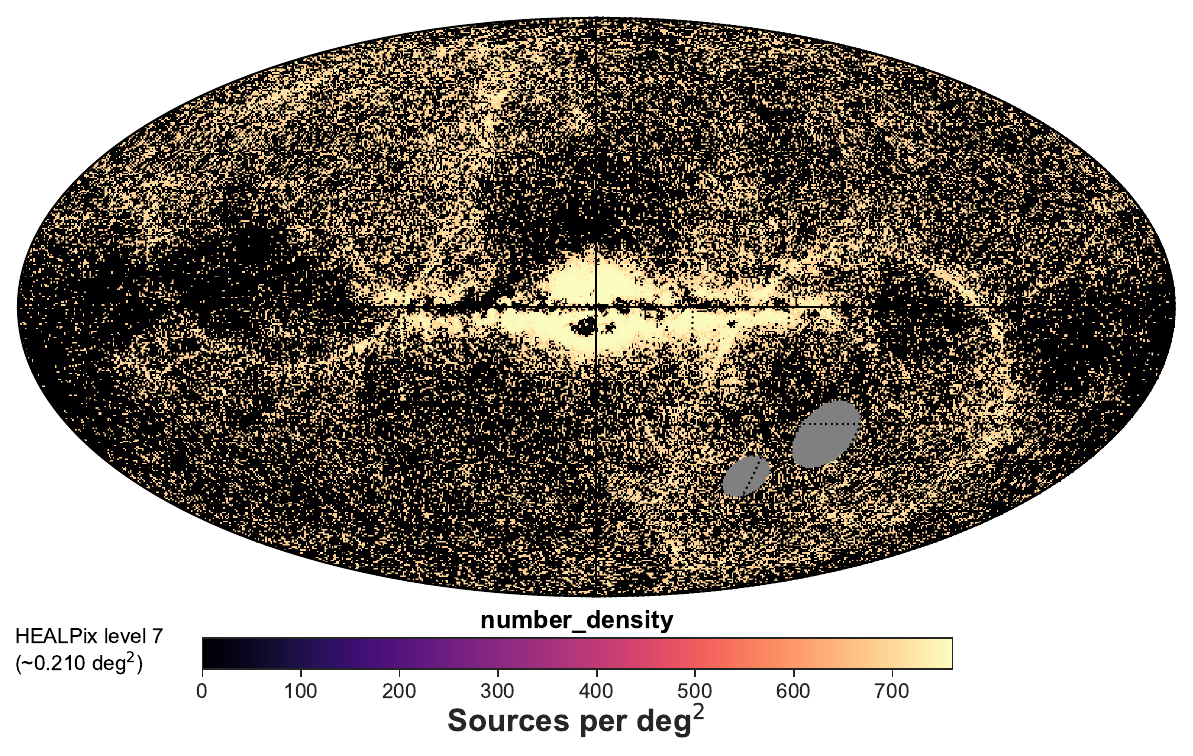}
	\caption{\num{189843} galaxies, 20.5$\leq$$G$<21\\\,}
	\end{subfigure}
	\hspace{0.1cm}
	\begin{subfigure}{0.3333\textwidth}
         \centering
	\includegraphics[scale=0.30]{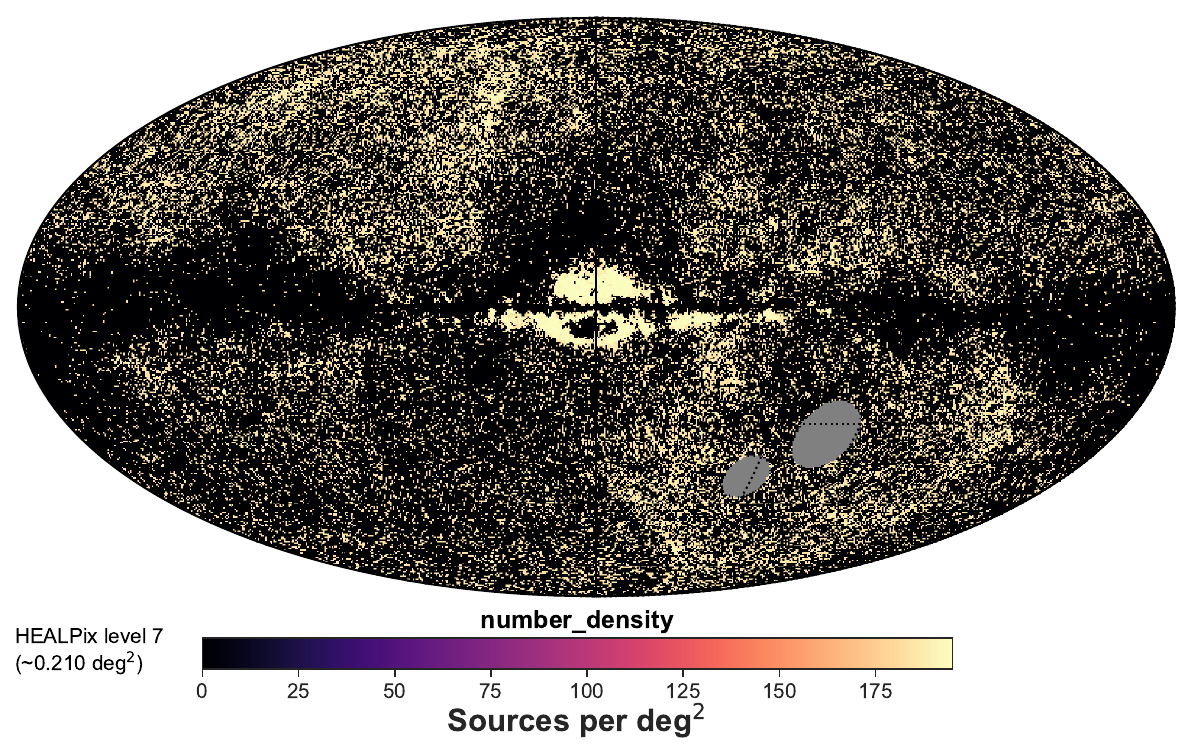}
	\caption{\num{80159} galaxies, 20.5$\leq$$G$<21 and \texttt{phot\_[bp,rp]\_n\_obs}$\geq$6$\,$}
	\end{subfigure}
	\hspace{0.1cm}
	\begin{subfigure}{0.3333\textwidth}
         \centering
	\includegraphics[scale=0.30]{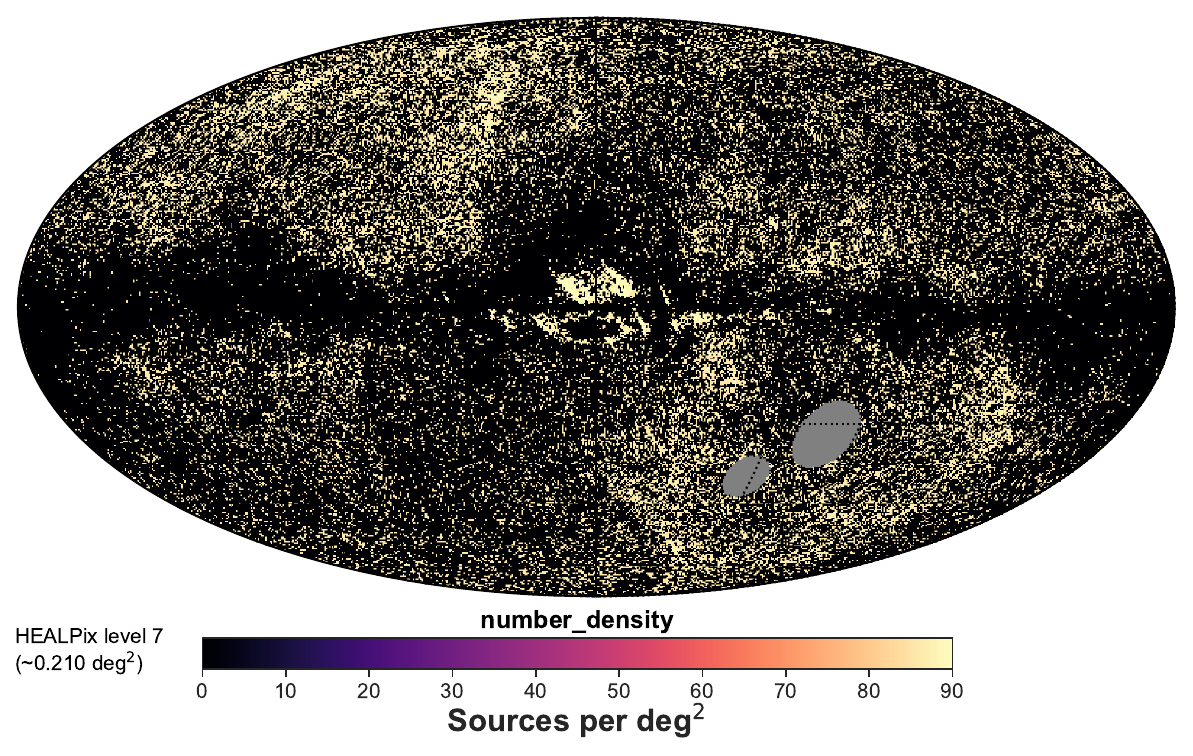}
	\caption{\num{56856} galaxies, 20.5$\leq$$G$<21 and \texttt{phot\_[bp,rp]\_n\_obs}$\geq$10}
	\end{subfigure}
	
	\begin{subfigure}{0.3333\textwidth}
         \centering
	\includegraphics[scale=0.30]{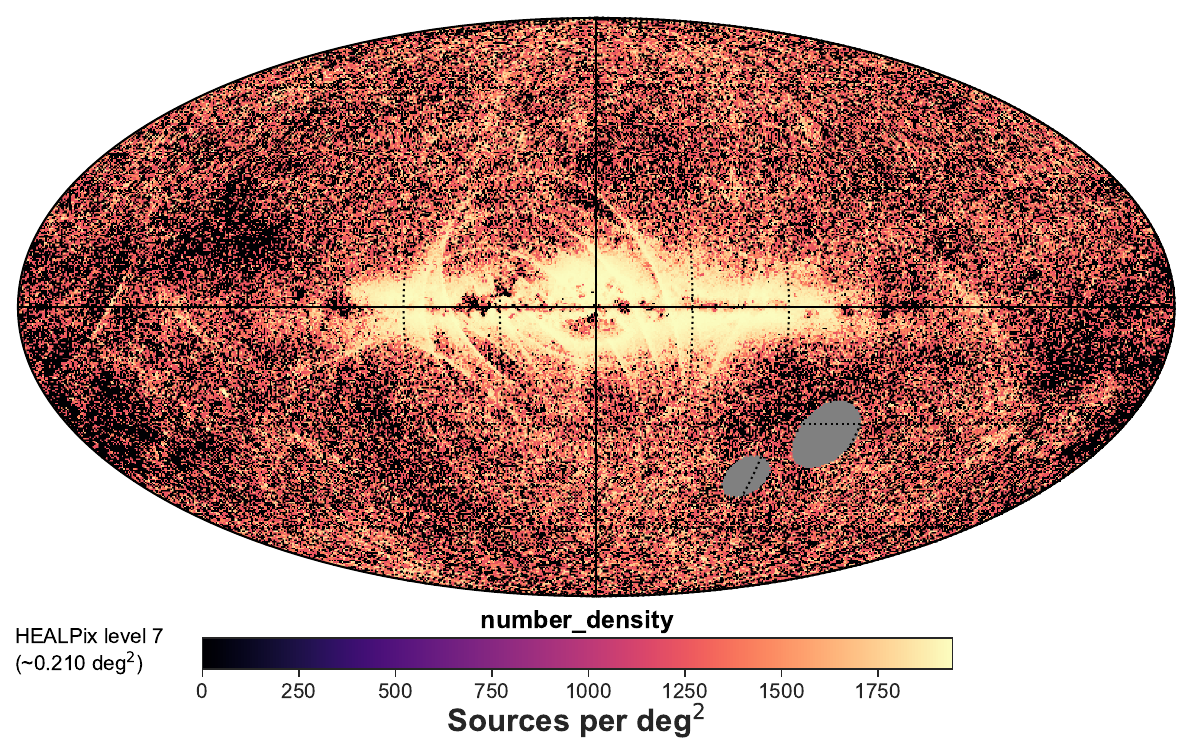}
	\caption{\num{527316} galaxies, 21$\leq$$G$<30\\\,}
	\end{subfigure}
	\hspace{0.1cm}
	\begin{subfigure}{0.3333\textwidth}
         \centering
	\includegraphics[scale=0.30]{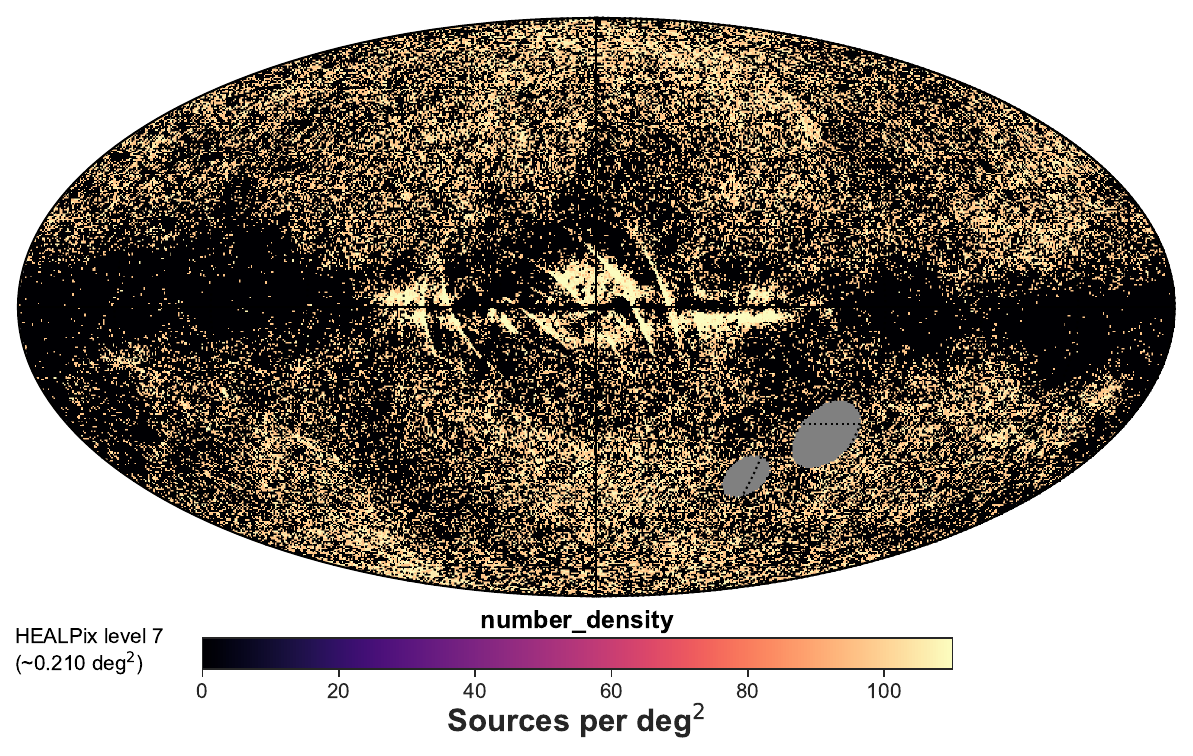}
	\caption{\num{108431} galaxies, 21$\leq$$G$<30 and \texttt{phot\_[bp,rp]\_n\_obs}$\geq$6$\,$}
	\end{subfigure}
	\hspace{0.1cm}
	\begin{subfigure}{0.3333\textwidth}
         \centering
	\includegraphics[scale=0.30]{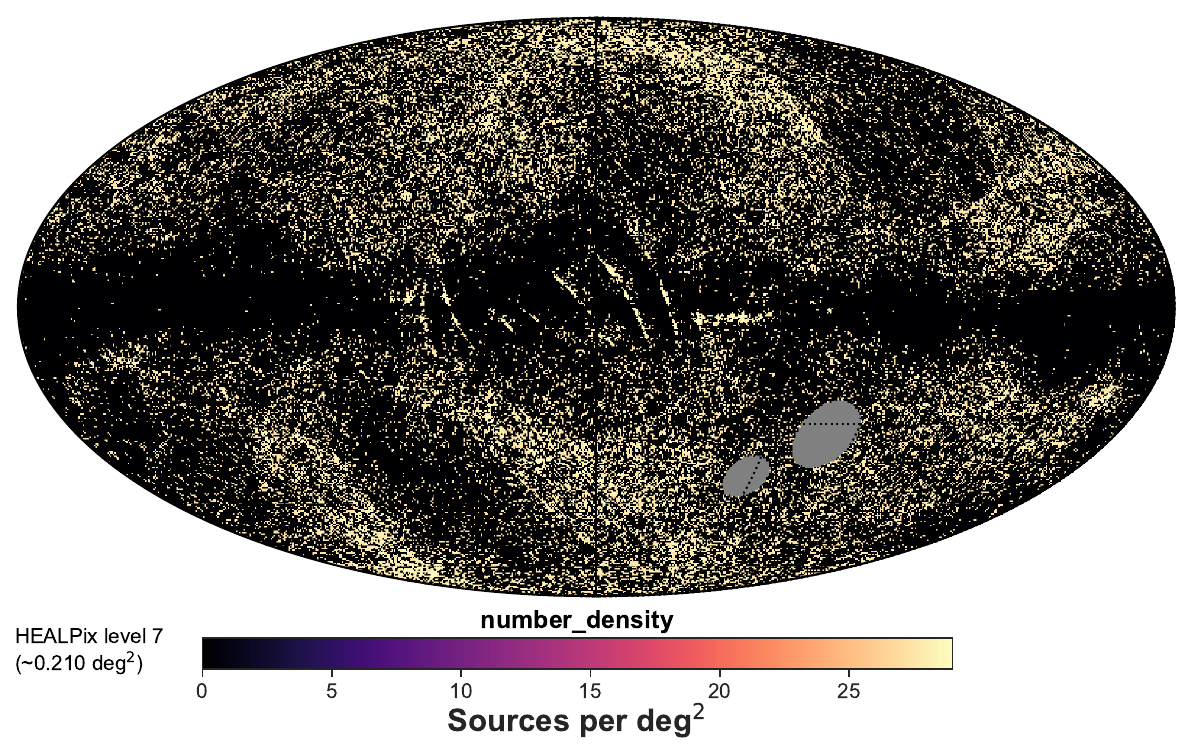}
	\caption{\num{51586} galaxies, 21$\leq$$G$<30 and \texttt{phot\_[bp,rp]\_n\_obs}$\geq$10}
	\end{subfigure}
	
\vspace{-.2cm}
\caption{
	Same as Fig. \ref{fig:perfo_skyplots_fullrun_magbins_combmodalpha_quasars}, 
	but for sources classified as galaxies from maximum probabilities by \texttt{Combmod-$\alpha$} 
	using the global prior at HEALpixel level 7 in Mollweide projection.\\}
\label{fig:perfo_skyplots_fullrun_magbins_combmodalpha_galaxies}
\end{minipage}
\end{figure}

\begin{minipage}{0.95\textwidth}
\section{Quasar and galaxy candidates identified by DSC \texttt{Combmod-$\alpha$} in the GDR4 per magnitude bins}
This section shows how the extragalactic candidates identified by the \texttt{Combmod-$\alpha$} classifier are distributed across the sky in different magnitude ranges. 
Quasar and galaxy candidates are shown in Figures \ref{fig:perfo_skyplots_fullrun_magbins_combmodalpha_quasars} and \ref{fig:perfo_skyplots_fullrun_magbins_combmodalpha_galaxies}, respectively. 
\end{minipage}

\newpage

\newpage~\newpage

\begin{figure}[htp!]
 \begin{minipage}{1\textwidth}
	 \begin{subfigure}{1\textwidth} \centering
         \includegraphics[scale=0.80]{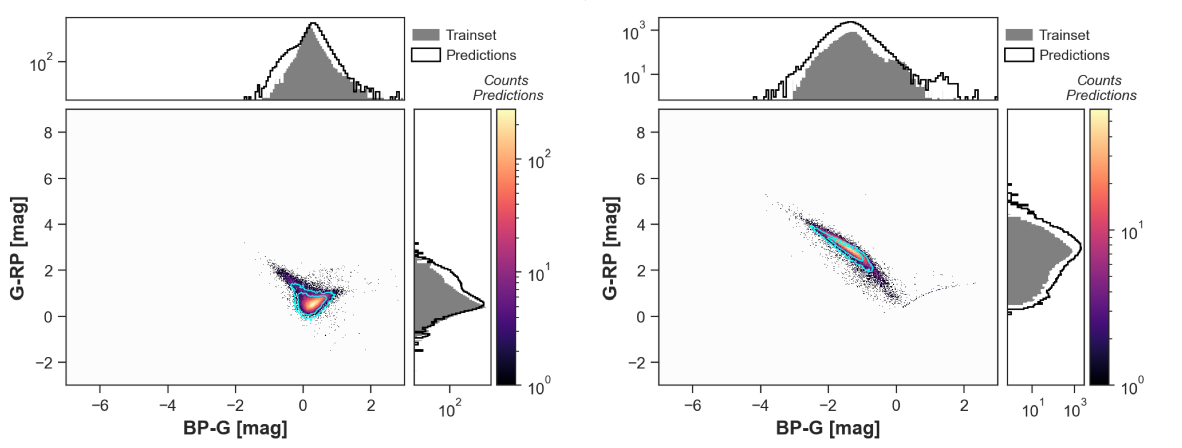}
        \vspace{-0.15cm} \caption{}
    	\end{subfigure}
	
	 \begin{subfigure}{1\textwidth} \centering
         \includegraphics[scale=0.80]{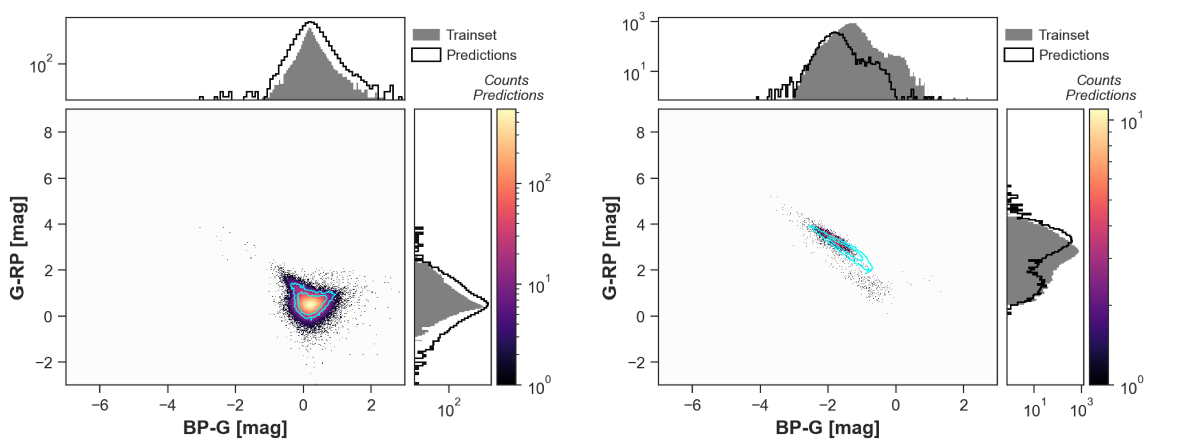}
	\vspace{-0.15cm}\caption{}
    	\end{subfigure}
	
	 \begin{subfigure}{1\textwidth} \centering
         \includegraphics[scale=0.80]{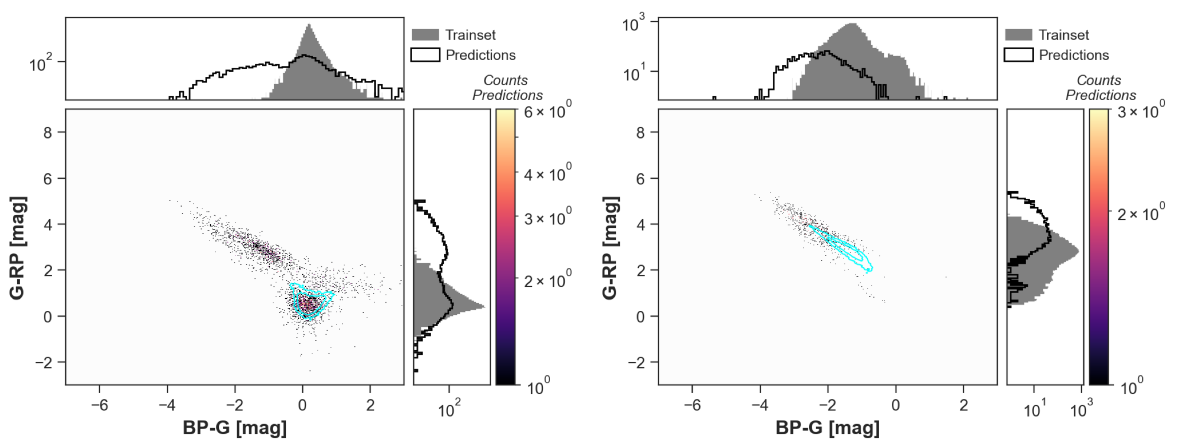}
	\vspace{-0.15cm} \caption{}
    	\end{subfigure}
	\caption{Colour-colour distributions of sources, exclusive to the \textit{Gaia}-IR mode, identified by \texttt{Combmod-$\alpha$}
		and classified from maximum probabilities using the global prior. 
		\textit{Left}: quasars. \textit{Right}: galaxies.
		Sources located in the LMC and SMC are excluded.
		The quality cut is applied for sources at higher latitudes $|\sin b|$>0.20 and \texttt{phot\_[bp,rp]\_n\_obs}$\geq$10.
		Top-to-bottom panels show the distributions at different magnitude limits.
		Contours (cyan lines) show the normalised density on a log scale of the highest-density regions for the trainset.
		$(a) $ \num{113636} quasars and \num{34185} galaxies at magnitudes $G$<20.5.
		$(b)$ \num{215330} quasars and \num{5290} galaxies at magnitudes 20.5$\leq$$G$<21.
		$(c)$ \num{3198} quasars and \num{1290} galaxies at magnitudes 21$\leq$$G$<30.
		}
	\label{fig:perfo_fullrun_distribution_ir_combmodalpha}
	\end{minipage} 	
\end{figure}

\begin{figure*}[htp!]
	 \begin{subfigure}{1\textwidth} \centering
         \includegraphics[scale=0.33]{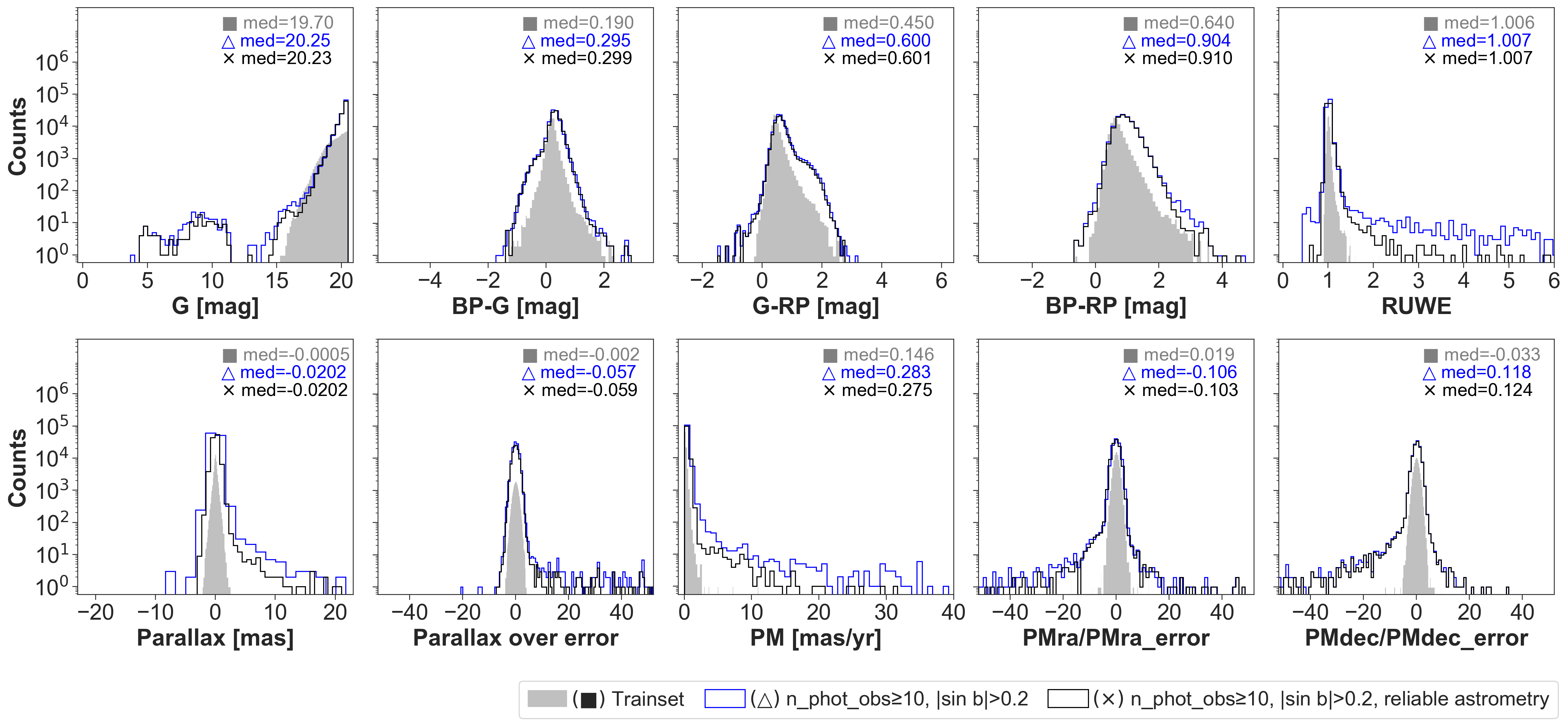}
         \caption{}
    	\end{subfigure}
	
	\begin{subfigure}{1\textwidth}
         \centering
         \includegraphics[scale=0.33]{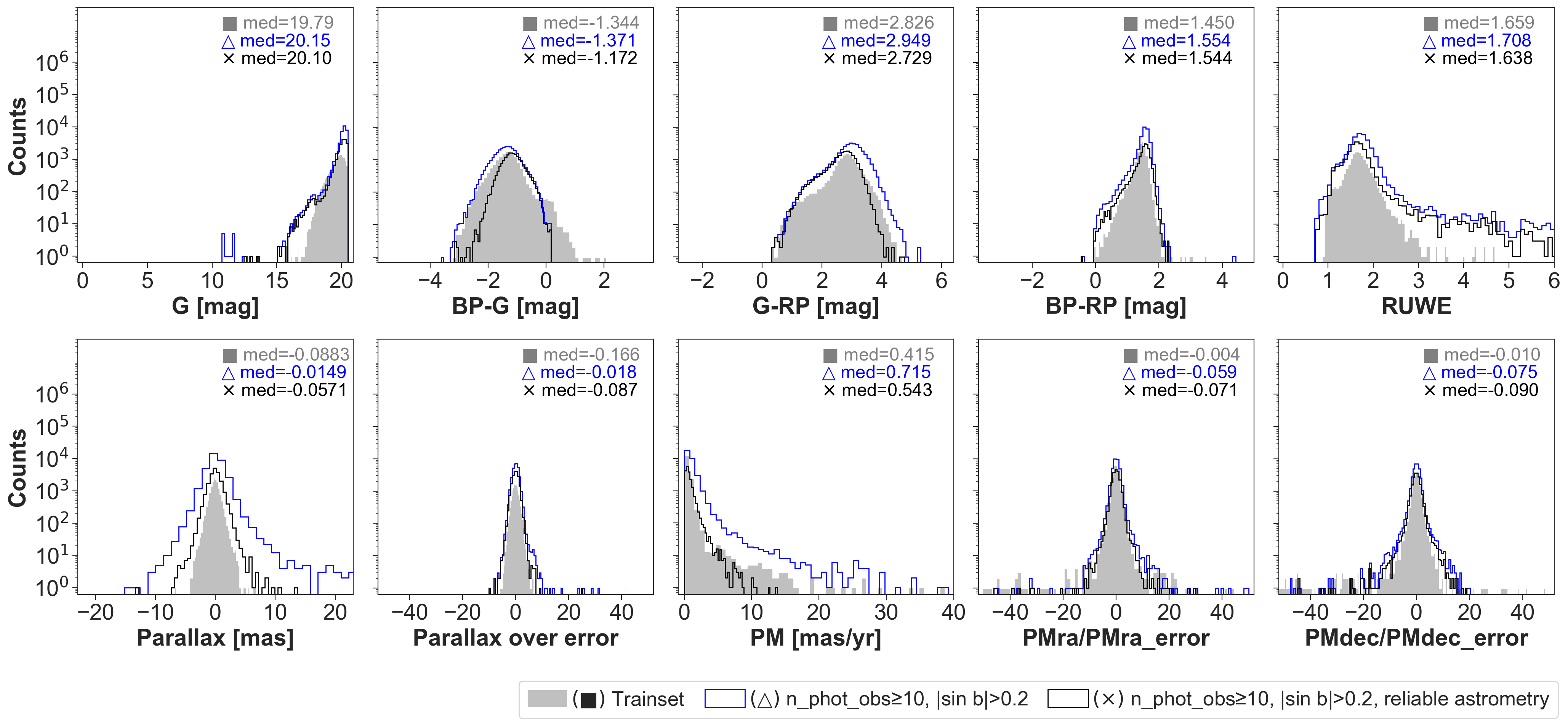}
         \caption{}
    	\end{subfigure}
	
	\caption{
		Same as Fig. \ref{fig:perfo_fullrun_distribution_features_combmodalpha_0-G-20.5_Pmax} but for the extragalactic candidates
		identified exclusively in the \textit{Gaia}-IR mode by \texttt{Combmod-$\alpha$} using the global prior, at magnitudes $G$<20.5.
		$(a)$ Quasar candidates. $(b)$ Galaxy candidates. 
		In $(a)$, the classifier identifies as quasars, \num{113636} sources ($\bigtriangleup$) and \num{108497} sources ($\times$).
		In $(b)$, the classifier identifies as galaxies, \num{34185} sources ($\bigtriangleup$) and \num{19434} sources ($\times$). 
		}
	\label{fig:perfo_fullrun_distribution_features_ir_combmodalpha_0-G-20.5}
\end{figure*}

\newpage~\newpage~\newpage
\begin{minipage}{0.95\textwidth}
\section{Quasar and galaxy candidates identified by DSC \texttt{Combmod-$\alpha$} using \textit{Gaia} data and infrared photometry from CatWISE}
This section illustrates how the quasar and galaxy candidates, as identified by the \texttt{Combmod-$\alpha$} classifier in \textit{Gaia}-IR mode, are distributed in the feature space across different magnitude ranges (Figures \ref{fig:perfo_fullrun_distribution_ir_combmodalpha} and \ref{fig:perfo_fullrun_distribution_features_ir_combmodalpha_0-G-20.5}).
\end{minipage}

\newpage

\end{appendix}

\end{document}